\documentclass[preprint,aps,color,superscriptaddress]{revtex4-1}
\usepackage{graphicx}
\usepackage{epsfig}

\usepackage{color,amsmath,graphicx,latexsym}
\usepackage{amssymb}

\usepackage[toc,page]{appendix}

\begin{document}

\title{Dzyaloshinskii-Moriya coupling  in 3d insulators}
\author{A.S.\ Moskvin}
\affiliation{Ural Federal University, 620083 Ekaterinburg,  
Russia}
\date{\today}

\begin{abstract}
We present an  overview of the microscopic theory of the Dzyaloshinskii-Moriya (DM) coupling in strongly correlated 3d compounds. 
Most attention in the paper centers around the derivation of the Dzyaloshinskii vector, its value, orientation, and sense (sign)  under different types of the (super)exchange interaction and crystal field. We consider both the Moriya mechanism of the antisymmetric interaction and novel contributions, in particular, that of spin-orbital coupling on the intermediate ligand ions.   We have predicted a novel magnetic phenomenon, {\it weak ferrimagnetism} in mixed weak ferromagnets with competing signs of the Dzyaloshinskii vectors.      
 We revisit a problem of  the DM coupling for a single bond in cuprates specifying the local spin-orbital contributions to Dzyaloshinskii vector focusing on the oxygen term. We predict a novel puzzling effect of the on-site staggered spin polarization to be a result of the on-site spin-orbital coupling and the the cation-ligand spin density transfer. The intermediate ligand NMR measurements are shown to be an effective tool to inspect the effects of the DM coupling in an external magnetic field. We predict the effect of a $strong$ oxygen weak antiferromagnetism in edge-shared CuO$_2$ chains due to uncompensated oxygen Dzyaloshinskii vectors. We revisit the effects of symmetric spin anisotropy directly induced by the DM coupling. A critical analysis will be given of different approaches to exchange-relativistic coupling based on the  cluster and the DFT based calculations. Theoretical results are applied to different classes of 3d compounds from conventional weak ferromagnets ($\alpha$-Fe$_2$O$_3$, FeBO$_3$, FeF$_3$, RFeO$_3$, RCrO$_3$,.. ) to unconventional systems such as weak ferrimagnets (e.g., RFe$_{1-x}$Cr$_x$O$_3$), helimagnets (e.g., CsCuCl$_3$), and  parent cuprates (La$_2$CuO$_4$,...).
\end{abstract} 


\maketitle

\section{Introduction}
More than a hundred years have passed since T. Smith\,\cite{Smith}  in 1916 found a weak, or $parasitic$ ferromagnetism in an "international family line" of different natural hematite $\alpha$-Fe$_2$O$_3$ single crystalline samples from Italy, Hungary, Brasil, and Russia (Schabry, Ural Mountains, small settlement near Ekaterinburg) that was first assigned to ferromagnetic impurities. Later the phenomenon was observed in many other 3d compounds, such as fluoride NiF$_2$ with rutile structure,  orthorhombic orthoferrites RFeO$_3$ (where R is a rare-earth element or Y),  rhombohedral antiferromagnets MnCO$_3$, NiCO$_3$, CoCO$_3$, and FeBO$_3$. However, only in 1954 L.M. Matarrese and J.W. Stout for NiF$_2$\,\cite{Matarrese} and in 1956 A.S. Borovik-Romanov and M.P. Orlova for very pure synthesised  carbonates  MnCO$_3$ and CoCO$_3$\,\cite{Borovik} have firmly established that  the connexion between the weak ferromagnetism and any impurities or inhomogeneities seems very unlikely as weak ferromagnetism is observed in chemically pure antiferromagnetic materials and therefore it is a specific intrinsic property of some antiferromagnets. Furthermore, Borovik-Romanov and Orlova assigned the uncompensated moment in MnCO$_3$ and CoCO$_3$ to an overt canting of the two magnetic sublattices in almost antiferromagnetic matrix.
 The model of a canted  antiferromagnet   became generally adopted model of the weak ferromagnet.



\begin{figure}[t]
\centering
\includegraphics[width=8.5cm,angle=0]{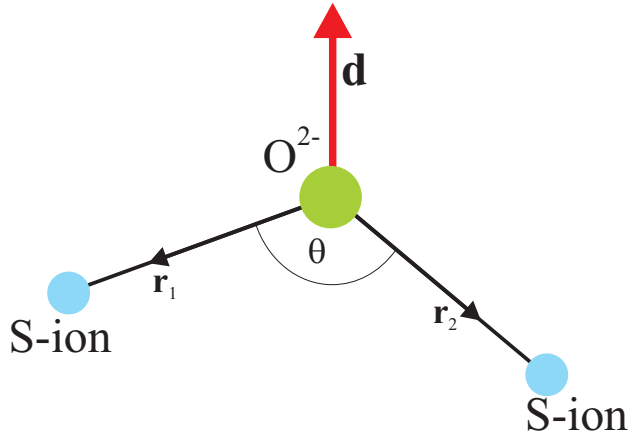}
\caption{Superexchange geometry and the Dzyaloshinskii vector.}
\label{fig1}
\end{figure}

A theoretical explanation and first thermodynamic theory for weak ferromagnetism in $\alpha$-Fe$_2$O$_3$, MnCO$_3$, and CoCO$_3$ was provided by Igor Dzyaloshinskii (Dzialoshinskii, Dzyaloshinsky)\,\cite{Dzyaloshinskii} in 1957 on the basis of symmetry considerations and Landau's theory of the second kind phase transitions.  Free energy of the two-sublattice uniaxial weak ferromagnet such as $\alpha$-$Fe_2O_3$, $MnCO_3$, $CoCO_3$, $FeBO_3$ was shown to be written as follows
\[
F=MH_E({\bf m}_1\cdot {\bf m}_2)-M{\bf H}_0({\bf m}_1+{\bf m}_2)+E_D+E_A
\]
\begin{equation}
	= MH_E({\bf m}^2-{\bf l}^2)-M{\bf H}_0{\bf m}+E_D+E_A \, .
\end{equation}
In this expression ${\bf m}_1$ and ${\bf m}_2$ are unit vectors in the directions of the sublattice moments, $M$ is the sublattice
magnetization, ${\bf m}=\frac{1}{2}({\bf m}_1+{\bf m}_2)$ and  ${\bf l}=\frac{1}{2}({\bf m}_1-{\bf m}_2)$ are the ferro- and antiferromagnetic vectors, respectively, $H_0$ is the applied field, $H_E$ is the exchange field, 
\begin{equation}
	E_D=-MH_D[{\bf m}_1\times {\bf m}_2]_z=+2MH_D[{\bf m}\times {\bf l}]_z=+2MH_D(m_xl_y-m_yl_x)
\end{equation}
is now called the Dzyaloshinskii interaction, $H_D>0$ is the Dzyaloshinskii field.   
The anisotropy energy $E_A$ is assumed to have the form: $E_A = H_A/2M(m_{1z}^2+m_{2z}^2)=2H_A/2M(m_{z}^2+l_{z}^2)$, where $H_A$ is the anisotropy field. The choice of sign for the anisotropy field $H_A$ assumes that the $c$ axis is a hard direction of magnetization. 
In a general sense the Dzyaloshinskii interaction implies the terms that are linear both on  ferro- and antiferromagnetic vectors. For instance, in orthorhombic orthoferrites and orthochromites the Dzyaloshinskii interaction consists of the antisymmetric and symmetric terms
\[
	E_D= d_1m_zl_x+d_2m_xl_z=\frac{d_1-d_2}{2}(m_zl_x-m_xl_z)+\frac{d_1+d_2}{2}(m_zl_x+m_xl_z)=
\]	
	\begin{equation}
	-2MH_D[{\bf m}\times {\bf l}]_y+\frac{d_1+d_2}{2}(m_zl_x+m_xl_z)\, ,
\end{equation}
while for tetragonal fluorides NiF$_2$ and CoF$_2$ the Dzyaloshinskii interaction consists of the only symmetric term. 
Despite Dzyaloshinskii supposed that weak ferromagnetism is due to relativistic spin-lattice and magnetic dipole interaction,  the theory was phenomenological one and did not clarify the microscopic nature of the Dzyaloshinskii interaction that does result in the canting.
 Later on, in 1960, Toru Moriya\,\cite{Moriya} suggested a model microscopic theory of the exchange-relativistic antisymmetric exchange interaction to be a main contributing mechanism of weak ferromagnetism
\begin{equation}
	V_{DM}=\sum_{mn}({\bf d}_{mn}\cdot\left[{\bf S}_m\times{\bf S}_n\right]) \, ,
	\label{DM}
\end{equation}
now called Dzyaloshinskii-Moriya (DM) spin coupling. Here,  ${\bf d}_{mn}$ is the axial
 Dzyaloshinskii vector.
Presently Keffer\,\cite{Keffer} proposed a simple phenomenological expression for the Dzyaloshinskii vector for two magnetic ions M$_i$ and M$_j$ interacting by the superexchange mechanism via intermediate ligand O (see Fig.\ref{fig1}):
\begin{equation}
	\mathbf{d}_{ij} \propto [\mathbf{r}_i \times \mathbf{r}_j] \, ,
\end{equation}
where ${\bf r}_{i,j}$ are unit radius vectors for O\,-\,M$_{i,j}$ bonds with presumably equal bond lenghts. Later on Moskvin\,\cite{1971} derived a microscopic formula for Dzyaloshinskii vector 
 \begin{equation}
	\mathbf{d}_{ij} = d_{ij}(\theta) [\mathbf{r}_i \times \mathbf{r}_j] \, ,
	\label{d12}
\end{equation} 
where
\begin{equation}
	d_{ij}(\theta)=d_1(R_i,R_j)+d_2(R_i,R_j)cos\theta_{ij} \, ,
\end{equation}
with $\theta_{ij}$ being the M$_{i}$\,-\,O\,-\,M$_{j}$ bonding angle, $R_{i,j}$ are the O\,-\,M$_{i,j}$ separations.
The sign of the scalar parameter $d_{ij}(\theta)$ can be addressed to be the sign, or sense of the Dzyaloshinskii vector.  The formula (\ref{d12}) was shown to work only for S-type magnetic ions with orbitally nondegenerate ground state, e.g. for 3d ions with half-filled shells (3d$^5$, $t_{2g}^3$, $t_{2g}^3e_g^2$, $t_{2g}^6e_g^2$).

It should be noted that sometimes instead of (\ref{d12}) one may use another form of the structural factor for the Dzyaloshinskii vector:
\begin{equation}
	\left[{\bf r}_1\times{\bf r}_2\right]=\frac{1}{2}\left[({\bf r}_1-{\bf r}_2)\times({\bf r}_1+{\bf r}_2)\right]=\frac{1}{2l^2}\left[{\bf R}_{12}\times{\pmb \rho}_{12
}\right] \, ,
\label{Rrho}
\end{equation}
where ${\bf R}_{12}={\bf R}_1-{\bf R}_2$, ${\pmb \rho}_{12}=({\bf R}_1+{\bf R}_2)$, $l$\,=\,$|{\bf R}_{1,2}|$,  ${\bf R}_{1,2}$ are radius vectors for O\,-\,M$_{1,2}$ bonds, respectively. 

Starting with the pioneering papers by Dzyaloshinskii\,\cite{Dzyaloshinskii} and Moriya\,\cite{Moriya} the DM coupling was extensively investigated in 60-80ths in connection with the weak ferromagnetism focusing on  hematite $\alpha$-Fe$_2$O$_3$ and orthoferrites RFeO$_3$\,\cite{1972,1975,1977,thesis}. Typical values of the canting angle $\alpha_D$ turned out to be on the order of 0.001-0.01, in particular, $\alpha_D=1.1\cdot 10^{-3}$ in $\alpha$-Fe$_2$O$_3$\,\cite{Fe2O3}, 2.2-2.9$\cdot 10^{-3}$ in $La_2CuO_4$\,\cite{Thio}, $5.5\cdot 10^{-3}$ in $FeF_3$\,\cite{Prozorova},  $1.1\cdot 10^{-2}$ in $YFeO_3$\,\cite{Jacobs}, $1.7\cdot 10^{-2}$ in $FeBO_3$\,\cite{Kotyuzhanskii}. Main exchange and DM coupling parameters for these weak ferromagnets are given in the Table\,1.
\begin{table}
\caption{Main exchange and DM coupling parameters in weak ferromagnets (WFMs), $I$ is the exchange integral, $\alpha_D$ is the canting angle. See text for detail.}
\begin{tabular}{|c|c|c|c|c|c|c|c|c|}
\hline
 WFM & R$_{FeO}$, \AA & $\theta$ & T$_N$, K & $I$, K (MFA) & H$_E$, Tesla & $\alpha_D$ & H$_D$, Tesla & d($\theta$), K \\ \hline
$\alpha$-Fe$_2$O$_3$\,\cite{Fe2O3} & 2.111  & 145$^{\circ}$ &  948 & 54.2 & 870-920 & $1.1\cdot 10^{-3}$ & 1.9-2.2 & 2.3\\ \hline
YFeO$_3$ & 2.001 (x2) & 145$^{\circ}$ &  640 & 36.6 & 640 & $1.1\cdot 10^{-2}$ & 14 & 3.2\\ \hline
FeBO$_3$ & 2.028 & 126$^{\circ}$ &  348 & 19.9 & 300 & $1.7\cdot 10^{-2}$ & 10 & 2.3\\ \hline
FeF$_3$ & 1.914  & 153$^{\circ}$ &  363 & 20.7 & 440 & $5.5\cdot 10^{-3}$ & 4.88 & 1.1\\ \hline
 \end{tabular}
\label{AFC}
\end{table}

Valerii Ozhogin {\it et al}.\,\cite{Ozhogin} in 1968 first  raised the issue of the sign of the Dzyaloshinskii vector, however, only in 1990 the reliable local information on the sign of the Dzyaloshinskii vector, or to be exact, that of the Dzyaloshinskii parameter $d_{12}$, was first extracted from the $^{19}F$ ligand NMR data in weak ferromagnet $FeF_3$\,\cite{sign}. 
In 1977 we have shown that the Dzyaloshinskii vectors can be of opposite sign for different pairs of $S$-type ions\,\cite{1977} that allowed us to uncover a novel magnetic phenomenon, {\it weak ferrimagnetism}, and a novel class of magnetic materials, {\it weak ferrimagnets}, which are systems such as solid solutions YFe$_{1-x}$Cr$_x$O$_3$ with competing signs of the Dzyaloshinskii vectors and the very unusual concentration and temperature dependence of the magnetization\,\cite{WFIM-1}. 
The relation between Dzyaloshinskii vector and the superexchange geometry (\ref{d12}) allowed us to find numerically all the overt and hidden canting angles in the rare-earth orthoferrites\,\cite{1975} that was nicely confirmed in $^{57}Fe$ NMR\,\cite{Luetgemeier} and neutron diffraction\,\cite{Plakhtii} measurements.

The stimulus to a renewed interest to the subject was given by the cuprate problem, in particular, by the weak ferromagnetism observed in the parent cuprate La$_2$CuO$_4$\,\cite{Thio} and  many other interesting effects for the DM systems, in particular, the "field-induced gap"\, phenomena\,\cite{Affleck}.  At variance with typical 3D systems such as orthoferrites, the cuprates  are characterised by a low-dimensionality, large diversity of Cu-O-Cu bonds including corner- and edge-sharing, different ladder configurations, strong quantum effects for $s=1/2$ Cu$^{2+}$ centers, and a particularly strong Cu-O covalency resulting in a comparable magnitude of hole charge/spin densities on copper and oxygen sites. 
Several groups (see, e.g., Refs.\,\cite{Coffey,Koshibae,Shekhtman}) developed the microscopic model approach by Moriya for different 1D and 2D cuprates, making use of different perturbation schemes, different types of the low-symmetry crystalline field, different approaches to the intra-atomic electron-electron repulsion. However, despite a rather large number of publications and hot debates (see, e.g., Ref.\,\cite{debate}) the problem of exchange-relativistic effects, that is of the DM coupling and related problem of spin anisotropy in cuprates remains to be open (see, e.g., Refs.\,\cite{Tsukada,Kataev} for experimental data and discussion). Common shortcomings of current approaches to DM coupling in 3d oxides concern a problem of allocation of the Dzyaloshinskii vector and respective "weak" (anti)ferromagnetic moments, and  full neglect of spin-orbital effects for "nonmagnetic"\, oxygen O$^{2-}$ ions, which are usually believed to play only  indirect intervening role. From the other hand, the oxygen $^{17}$O NMR-NQR studies of weak ferromagnet La$_2$CuO$_4$\,\cite{Walstedt}  seem  to evidence unconventional local oxygen "weak-ferromagnetic" polarization whose origin cannot be explained in frames of current models.

In recent years interest has shifted towards other manifestation of the DM coupling, such as the magnetoelectric effect\,\cite{Dagotto,SLD}, so-called flexoelectric effect in multiferroic bismuth ferrite $BiFeO_3$ with coexisting spin canting and the spin cycloidal ordering\,\cite{Pyatakov}, and skyrmion states\,\cite{Bogdanov}, where reliable theoretical predictions have been lacking. 

It was shown the particular importance of this interaction for magnetic nanostructures, e.g. ultrathin films on surfaces, where it can give rise to cycloidal spiral spin density waves with a unique sense of rotation. Despite a clear weakness of the typical DM coupling as compared with typical isotropic exchange interactions the DM coupling  can be a central ingredient in the stabilization of complex magnetic textures.

In fact, it is known for a long time that the DM coupling can produce long-period magnetic spiral structures in ferromagnetic and antiferromagnetic crystals lacking inversion symmetry. This effect was suggested for MnSi and other crystals with B20 structure and it has been carefully proved that the sign of the DM coupling, hence the sign of the spin helix, is determined by the crystal handedness.

Phenomenologically antisymmetric DM coupling in a continual approximation  gives rise to so-called Lifshitz invariants, or energy contributions linear in first spatial derivatives of the magnetization ${\bf m}({\bf r})$\,\cite{Lifshits}
\begin{equation}
	m_i\frac{\partial m_j}{\partial x_l}-m_j\frac{\partial m_i}{\partial x_l}
\end{equation}
($x_l$ is a spatial coordinate). These chiral interactions derived from the DM coupling stabilize localized (vortices) and spatially modulated structures with a fixed rotation sense of the magnetization\,\cite{Bogdanov}. In fact, these are the only mechanism to induce nanosize skyrmion structures in condensed matter.

In this paper we present an  overview of the microscopic theory of the DM coupling in strongly correlated compounds such as 3d oxides. 
The rest part of the paper is organized as follows. In Sec.\,2 we shortly address main results of the microscopic theory of the isotropic superexchange interactions for so-called $S$-type ions focusing on the angular dependence of the exchange integrals.
Most attention in Sec.\,3 centers around the derivation of the Dzyaloshinskii vector, its value, orientation, and sense (sign)  under different types of the (super)exchange interaction and crystal field. Theoretical predictions of this section are compared in Sec.\,4 with experimental data for the overt and hidden canting in orthoferrites.  Here, too, we consider a {\it weak ferrimagnetism}, a novel type of magnetic ordering in systems with competing signs of the Dzyaloshinskii vectors. The ligand NMR in weak ferromagnet $FeF_3$ and first reliable determination of the sign of the Dzyaloshinskii vector are considered in Sec.\,5. An alternarive method to derive DM coupling is discussed in Sec.\,6 by the example of the three-center two-electron/hole system such as a triad $Cu^{2+}-O^{2-}-Cu^{2+}$ in cuprates. Here we emphasize specific features of the ligand contribution to the DM coupling and some inconsistencies of its traditional form. As a direct application of the theory we address in Sec.\,7 the $^{17}O$ NMR in $La_2CuO_4$ and argue that the field-induced staggered magnetization due to DM coupling does explain a puzzling Knight shift anomaly. In Sec.\,8 we consider features of the DM coupling in helimagnetic cuprate $CsCuCl_3$. Short Sec.\,9 is devoted to puzzling features of the exchange-relativistic  two-ion symmetric spin anisotropy due to DM coupling in quantum s=1/2 magnets. So-called "first-principles" calculations of the exchange interactions and DM coupling are critically discussed in Sec.\,10.   Short summary is presented in Sec.\,11. 

\section{Microscopic theory of the isotropic superexchange coupling}

DM coupling is derived from the off-diagonal (super)exchange coupling and does usually accompany a conventional (diagonal) Heisenberg type isotropic (super)exchange coupling:
\begin{equation}
	{\hat V}_{ex}=I_{12}({\bf S}_1\cdot{\bf S}_2)	\, .
\end{equation}
 The modern microscopic theory of the (super)exchange coupling had been elaborated by many physicists starting with well-known papers by P. Anderson\,\cite{PWA}, especially  intensively in 1960-70th (see review articles\,\cite{superexchange}). Numerous papers devoted to the problem pointed to existence  of many hardly estimated exchange mechanisms, seemingly comparable in value, in particular, for superexchange via intermediate ligand ion to be the most interesting for strongly correlated systems such as 3d oxides. Unfortunately, up to now we have no reliable estimations of the exchange parameters, though from the other hand we have no reliable experimental information about their magnitudes. To that end, many efforts were focused on the fundamental points such as many-electron theory and orbital dependence\,\cite{1968,Levy,1971,Veltrusky}, crystal-field effects\,\cite{Sidorov}, off-diagonal exchange\,\cite{1970}, exchange in excited states\,\cite{Luk}, angular dependence of the superexchange coupling\,\cite{1971}. The irreducible tensor operators (the Racah algebra) were shown to be very instructive tool both for description and  analysis of the exchange coupling in the 3d- and 4f-systems\,\cite{1968,Levy,1971,Veltrusky,Sidorov}.

First poor man's microscopic derivation for the  dependence of the superexchange integral on the bonding angle (see Fig.\,\ref{fig1}) was performed by the author in 1970\,\cite{1971}  under simplified assumptions. As a result, for $S$-ions with configuration $3d^5$ ($Fe^{3+}, Mn^{2+}$)
\begin{equation}
	I_{12}(\theta )=a+b\cdot cos\theta_{12} + c\cdot cos^2\theta_{12} \, ,
	\label{angle}
	\end{equation}
where  parameters $a, b, c$ depend on the cation-ligand separation. A more comprehensive analysis  has supported validity of the expression. Interestingly, the second term in (\ref{angle}) is determined by the ligand inter-configurational $2p$-$ns$ excitations, while other terms are related with intra-configurational $2p$-, $2s$-contributions.  

Later on the derivation had been generalized for the $3d$ ions in a strong cubic crystal field\,\cite{thesis}.
Orbitally isotropic contribution to the exchange integral for pair of 3d-ions with configurations $t_{2g}^{n_1}e_g^{n_2}$ can be written as follows
\begin{equation}
I=\sum_{\gamma_i,\gamma_j}I(\gamma_i\gamma_j)\,(g_{\gamma_i}-1)\, (g_{\gamma_j}-1)	\, ,
\end{equation}
where $g_{\gamma_i}, g_{\gamma_j}$ are effective "$g$-factors" of the $\gamma_i, \gamma_j$ subshells of ion 1 and 2, respectively:
\begin{equation}
	g_{\gamma_i} =1+\frac{S(S+1)+S_i(S_i+1)-S_j(S_j+1)}{2S(S+1)}\,  .
\end{equation}
 Kinetic exchange contribution to partial exchange parameters $I(\gamma_i\gamma_j)$ related with the electron transfer to partially filled shells can be written as follows\,\cite{Sidorov,thesis} 
\begin{equation}
I(e_ge_g)=\frac{(t_{ss}+t_{\sigma\sigma}cos\theta)^2}{2U};\,
I(e_gt_{2g})=\frac{t_{\sigma\pi}^2}{3U}sin^2\theta;\,
I(t_{2g}t_{2g})=\frac{2t_{\pi\pi}^2}{9U}(2-sin^2\theta )\, ,
\label{kinetic}	
\end{equation}
where $t_{\sigma\sigma}>t_{\pi\sigma}>t_{\pi\pi}>t_{ss}$ are positive definite d-d transfer integrals, $U$ is a mean d-d transfer energy (correlation energy). 
All the partial exchange integrals appear to be positive or "antiferromagnetic", irrespective of the bonding angle value,  though the combined effect of the $ss$ and $\sigma\sigma$ bonds $\propto cos\theta$ in $I(e_ge_g)$ yields a ferromagnetic contribution given bonding angles $\pi /2<\theta <\pi$. It should be noted that the "large" ferromagnetic potential contribution\,\cite{Freeman} has a similar angular dependence\,\cite{Luk}.
\begin{figure}[t]
\centering
\includegraphics[width=8.5cm,angle=0]{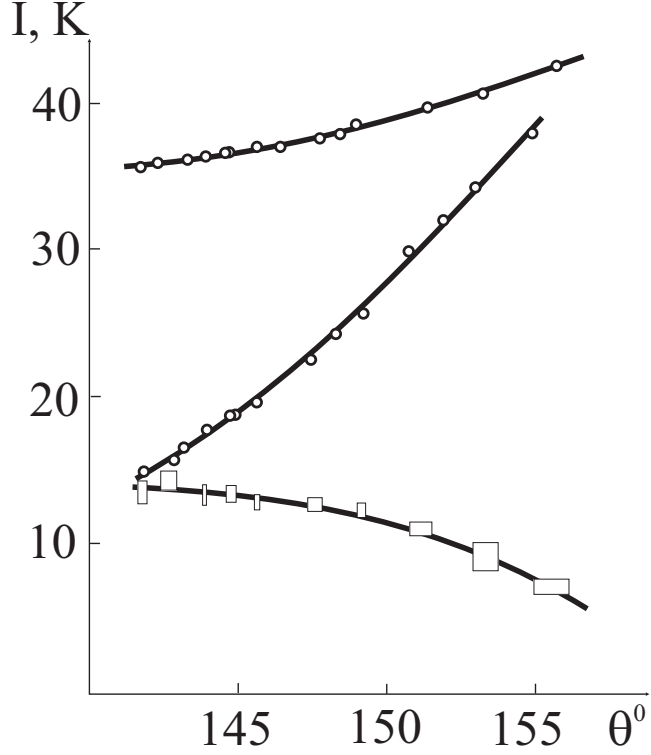}
\caption{Dependence of the $Fe^{3+}-Fe^{3+}$, $Cr^{3+}-Cr^{3+}$, $Fe^{3+}-Cr^{3+}$ exchange integrals (in K) on the superexchange bond angle in orthoferrites-orthocromites\,\cite{Ovanesyan}.}
\label{fig2}
\end{figure}

Some predictions regarding the relative magnitude of the $I(\gamma_i\gamma_j)$ exchange  parameters can be made using the relation among different d-d transfer integrals as follows
\begin{equation}
	t_{\sigma\sigma}\,:\,t_{\pi\sigma}\,:\,t_{\pi\pi}\,:\,t_{ss}\approx \lambda_{\sigma}^2\,:\,\lambda_{\pi}\lambda_{\sigma}\,:\,\lambda_{\pi}^2\,:\,\lambda_{s}^2 \, ,
	\label{t-lambda}
\end{equation}
where $\lambda_{\sigma}, \lambda_{\pi}, \lambda_{s}$ are covalency parameters.
The simplified kinetic exchange contribution (\ref{kinetic}) related with the electron transfer to partially filled shells does not account for intra-center correlations which are of a particular importance for the  contribution related with the electron transfer to empty shells. For instance, appropriate contributions related with the transfer to empty $e_g$ subshell for the $Cr^{3+}-Cr^{3+}$ and $Fe^{3+}-Cr^{3+}$ exchange integrals are
\begin{equation}
\Delta I_{CrCr}=-\frac{\Delta E(35)}{6U}\frac{t_{\sigma\pi}^2}{U}sin^2\theta \, ;\,\,\,
	\Delta I_{FeCr}=-\frac{\Delta E(35)}{10U}\left[\frac{(t_{ss}+t_{\sigma\sigma}cos\theta)^2}{U}+\frac{t_{\sigma\pi}^2}{U}sin^2\theta\right] \, ,
\end{equation}
where $\Delta E(35)$ is the energy separation between $^3E_g$ and $^5E_g$ terms for $t_{2g}^3e_g$ configuration ($Cr^{2+}$ ion). Obviously, these contributions have a ferromagnetic sign. Furthermore, the exchange integral $I(CrCr)$ can change sign  at $\theta$\,=\,$\theta_{cr}$:
\begin{equation}
	sin^2\theta_{cr}=\frac{1}{\left(\frac{1}{2}+\frac{3}{8}\frac{\Delta E(35)}{U}\frac{t^2_{\sigma\pi}}{t^2_{\pi\pi}}\right)} \, .
\end{equation}
Microscopically derived angular dependence of the superexchange integrals does nicely describe the experimental data for exchange integrals $I(FeFe)$, $I(CrCr)$, and $I(FeCr)$ in orthoferrites, orthochromites, and orthoferrites-orthochromites\,\cite{Ovanesyan} (see Fig.\,\ref{fig2}). The fitting allows us to predict the sign change for  $I(CrCr)$ and $I(FeCr)$ at $\theta_{12}$\,$\approx$\,133$^{\circ}$ and 170$^{\circ}$, respectively. In other words, the $Cr^{3+}-O^{2-}-Cr^{3+}$ ($Fe^{3+}-O^{2-}-Cr^{3+}$) superexchange coupling becomes ferromagnetic at $\theta_{12}\leq 133^{\circ}$ ($\theta_{12}\geq 170^{\circ}$). 
However, it should be noted that too narrow (141-156$^{\circ}$) range of the superexchange bonding angles we used for the fitting with  assumption of the same Fe(Cr)-O bond separations and mean superexchange bonding angles for  all the systems gives rise to a sizeable parameter's uncertainty, in particular, for $I(FeFe)$ and $I(FeCr)$. In addition, it is necessary to note a large uncertainty regarding what is here called the "experimental"\, value of the exchange integral. The fact is that the "experimental"\, exchange integrals we have just used above are calculated using simple MFA relation
\begin{equation}
	T_N=\frac{zS(S+1)}{3k_B}I \, ,
\end{equation}
however, this relation yields the exchange integrals that can be one and a half or even twice less than the values obtained by other methods\,\cite{thesis,LuCrO3}.

Above we addressed only typically antiferromagnetic kinetic (super)exchange contribution as a result of the second order perturbation theory. However, actually this contribution does compete with typically ferromagnetic potential (super)exchange contribution, or Heisenberg exchange, which is a result of the first order perturbation theory. The most important contribution to the potential superexchange can be related with the intra-atomic ferromagnetic Hund exchange interaction of unpaired electrons on orthogonal ligand orbitals hybridized with the d-orbitals of the two nearest magnetic cations.

Strong dependence of the $d-d$ superexchange integrals  on the cation-ligand-cation separation is usually described by the Bloch's rule\,\cite{Bloch}:  
\begin{equation}
	\frac{\partial\ln I}{\partial\ln R}=\frac{\partial I}{\partial R}/\frac{I}{R} \approx -\,10 \, .
\end{equation}

\section{Microscopic theory of the DM coupling}

\subsection{Moriya's microscopic theory}

First microscopic theory of weak ferromagnetism, or theory of anisotropic superexchange interaction was provided by Moriya\,\cite{Moriya}, who extended the Anderson theory of superexchange to include spin-orbital coupling $V_{so}=\sum_i \xi({\bf l}_i\cdot{\bf s}_i)$.
Moriya started with the one-electron Hamiltonian for $d$-electrons as follows  
\begin{equation}
	{\hat H}=\sum_{fm\sigma}\epsilon_m{\hat d}^{\dagger}_{fm\sigma}{\hat d}_{fm\sigma}+\sum_{m\not= m^{\prime}, \sigma}t_{fmf^{\prime}m^{\prime}}{\hat d}^{\dagger}_{fm\sigma}{\hat d}_{f^{\prime}m^{\prime}\sigma}+
\sum_{fm\not= f^{\prime}m^{\prime}, \sigma\sigma^{\prime}}{\hat d}^{\dagger}_{fm\sigma}({\bf C}_{fmf^{\prime}m^{\prime}}\cdot{\pmb \sigma}){\hat d}_{f^{\prime}m^{\prime}\sigma^{\prime}} \, ,
\end{equation}
where
\begin{equation}
{\bf C}_{fmf^{\prime}m^{\prime}}=-\frac{\xi}{2}\sum_{m^{\prime\prime}}\left(\frac{{\bf l}_{fmfm^{\prime\prime}}t_{fm^{\prime\prime}f^{\prime}m^{\prime}}}{\epsilon_{m^{\prime\prime}}-\epsilon_m}+\frac{t_{fmf^{\prime}m^{\prime\prime}}{\bf l}_{f^{\prime}m^{\prime\prime}f^{\prime}m^{\prime}}}{\epsilon_{m^{\prime\prime}}-\epsilon_{m^{\prime}}}\right)	
\label{C}
\end{equation}
is a spin-orbital correction to transfer integral, $m$ and $m^{\prime}$ are orbitally nondegenerate ground states  on sites $f$ and $f^{\prime}$, respectively. Then Moriya did calculate the generalized Anderson kinetic exchange that contains both conventional isotropic exchange and anisotropic symmetric and antisymmetric terms, that is quasidipole anisotropy and DM coupling, respectively. 
We emphasize that the expression for the Dzyaloshinskii vector  
\begin{equation}
	{\bf d}_{ff^{\prime}}=\frac{4i}{U}\sum_{m\not= m^{\prime}}\left[t_{fmf^{\prime}m^{\prime}}{\bf C}_{f^{\prime}m^{\prime}fm}-{\bf C}_{fmf^{\prime}m^{\prime}}t_{f^{\prime}m^{\prime}fm}\right] \, .
\end{equation}
has been obtained by Moriya assuming orbitally nondegenerate ground states $m$ and $m^{\prime}$ on sites $f$ and $f^{\prime}$, respectively.
 It is worth noting that the spin-operator form of the DM coupling  follows from the relation: 
\begin{equation}
	{\bf S}_1({\bf S}_1\cdot{\bf S}_2)+ ({\bf S}_1\cdot{\bf S}_2) {\bf S}_2=-i[{\bf S}_1\times{\bf S}_2]\, ,
\end{equation}
which is a simple consequence of the spin algebra, in particular, of the commutation relations for the spin projection operators.

Moriya found the symmetry constraints on the orientation of the Dzyaloshinskii vector $\mathbf{d}_{ij}$. 
Let two ions 1 and 2 are located at the points A and B, respectively, with C point bisecting the AB line:

1. When C is a center of inversion: d=0.

2. When a mirror plane $\perp$AB passes through C, 
     ${\bf d} \parallel$ mirror plane or  ${\bf d} \perp$ AB.

3. When there is a mirror plane including A and B, 
     ${\bf d} \perp$ mirror plane.

4. When a twofold rotation axis $\perp$ AB passes through C, 
    ${\bf d} \perp$ twofold axis.

5. When there is an n-fold axis (n$
\geq$2) along AB,  ${\bf d} \parallel$ AB.


Despite its seeming simplicity the operator form of the DM coupling (\ref{DM})  raises some questions and doubts. First, at variance with the scalar product  $\left({\bf S}_1\cdot{\bf S}_2\right)$  the vector product of the  spin operators $\left[{\bf S}_1\times{\bf S}_2\right]$ changes the spin multiplicity, that is the net spin ${\bf S}_{12}={\bf S}_1+{\bf S}_2$, that underscores the need for quantum description. Spin nondiagonality of the DM coupling implies very unusual features of the ${\bf d}$-vector somewhat resembling vector orbital operator whose transformational properties cannot be isolated from the lattice\,\cite{Dmitrienko-2010}. It seems  the ${\bf d}$-vector  does not transform as a vector at all. 

Another issue that causes some concern is the structure and location of the ${\bf d}$ vector and corresponding spin cantings. Obviously, the ${\bf d}_{12}$ vector should be related in one or another way to spin-orbital contributions localized on sites 1 and 2, respectively. These components  may differ in their magnitude and direction, while the operator form (\ref{DM}) implies some  averaging both for ${\bf d}_{12}$ vector and spin canting between the two sites. 

Moriya did not take into account the effects of the crystal field symmetry and strength and did not specify the character of the (super)exchange coupling, that, as we'll see below, can crucially affect the direction and value of the Dzyaloshinskii vector up to its vanishing. Furthermore, he made use of a very simplified form (\ref{C}) of the spin-orbital perturbation correction to  the transfer integral (see Exp.\,(2.5) in Ref.\,\cite{Moriya}). The fact is that the structure of the charge transfer matrix elements implies the  involvement of several different on-site configurations ($t_{kn}\propto \langle N_1-1N_2+1|{\hat H}|N_1N_2\rangle$). Hence, the  perturbation correction has to be more complicated than (\ref{C}), at least, it should involve the spin-orbital matrix elements (and excitation energies!) for one- and two-particle configurations. As a result, it does invalidate the author's conclusion about the equivalence of the two perturbation schemes, based on the $V_{SO}$ corrections to the transfer integral and to the exchange coupling, respectively.
Another limitation of the Moriya's theory is related to a full neglect of the ligand spin-orbital contribution to DM coupling. 
Despite these shortcomings the Moriya's estimation for the ratio between the  magnitudes of the Dzyaloshinskii vector $d=|{\bf d}|$ and isotropic exchange $J$: $d/J\approx \Delta g/g$, where $g$ is the gyromagnetic ratio, $\Delta g$ is its deviation from the free-electron value, respectively, in some cases may be helpful, however, only for a very  rough estimation.  

\subsection{Microscopic theory of the DM coupling: direct exchange interaction of the S-type ions}

We start with a derivation of the DM coupling in the pair of the exchange coupled free ions with valent  $n_1l_1^{N_{1}}$  and $n_2l_2^{N_{2}}$ 
shells to be a result of the second-order perturbation theory as a combined effect of the exchange and spin-orbital couplings when schematically
\begin{equation}
	{\hat V}_{DM}=\sum_{ES}\frac{\langle GS|(V_{so}(1)V_{ex}(12)+ V_{ex}(12) V_{so}(2)+h.c.)|GS\rangle}{\Delta E_{ES}}\, ,
\end{equation}
where excited states $|ES\rangle$ are the terms which are allowable one by the spin-orbital selection rules $\Delta L\leq 1$, $\Delta S\leq 1$. Spin-orbit interaction has a fairly simple form   $V_{so}=\sum_i \xi_{nl}({\bf l}_i\cdot{\bf s}_i)$, whereas for the exchange interaction Hamiltonian one has to use a complex expression in terms of irreducible tensor operators\,\cite{1971,1977,1968,Levy,Veltrusky,Atoms}.
 The task seems to be more limited to academic interest, however,  it is of a great importance from methodological point of view. 
After some routine though rather intricate procedure we arrive at the Dzyaloshinskii vector to be a complicated "multistory" irreducible orbital operator as follows\,\cite{1977} 
$$
{\hat D}_q=-\frac{i\xi_{n_1l_1}}{\sqrt{2}} \sum_{b_1b_2b} \sum_{b_{1}^{'}b^{'}}\sum_{S_{1}^{''}L_{1}^{''}} (-1)^{2S_{1}+b+b^{'}+b_2} (2b_{1}^{'}+1)[(2b+1)(2b^{'}+1)]^{1/2}\times
$$
$$
\left\{
   \begin{array}{ccc}
    b_1 & b_{1}^{'}& 1 \\  
		b^{'} & b    & b_2
   \end{array}
  \right\} 
	\left\{
   \begin{array}{ccc}
    b_1 & 1& b_{1}^{'} \\  
		L_1 & L_1    & L_{1}^{''}
   \end{array}
  \right\}                    (\langle S_1\|S\|S_1\rangle \langle S_2\|S\|S_2\rangle )^{-1}\Delta E^{-1}_{S_1^{''}L_1^{''}}
$$
$$
W^{(1b_2)}_{S_2L_2;S_2L_2}\left( W^{(1b_1)}_{S_1L_1;S_{1}^{''}L_{1}^{''}}W^{(11)}_{S_{1}^{''}L_{1}^{''};S_1L_1}+(-1)^{b_1+b_{1}^{'}}W^{(11)}_{S_1L_1;S_{1}^{''}L_{1}^{''}}W^{(1b_1)}_{S_{1}^{''}L_{1}^{''};S_1L_1}\right)
$$
\begin{equation}
\left[I(b_1b_2b)\times \left[{\hat V}^{b_{1}^{'}}(L_1)\times {\hat V}^{b_{2}}(L_2)\right]^{b^{'}}\right]^1_q \, ,
\label{DM1}
\end{equation}
where we make use of standard notations for $6j$-symbols, irreducible matrix elements, spectroscopic coefficients, and irreducible tensor products\,\cite{Sobelman,Var,Atoms,MRS}. Matrix elements of irreducible tensor operators ${\hat V}^{b}(L)$ are defined by the Wigner-Eckart theorem\,\cite{Sobelman,Var}  as follows
$$
\langle LM|{\hat V}^b_{\beta}(L)|LM^{\prime}\rangle = (-1)^{L-M} \left( \begin{array}{ccc}
L & b & L \\
-M & \beta & M^{\prime}
\end{array} \right)\, .
$$
  For the exchange parameters we have a simple dependence on the pair radius-vector: $I(b_1b_2b\beta)=J(b_1b_2b)C^b_{\beta}({\bf R}_{12})$, where $C^b_{\beta}$ is the tensor  spherical harmonics ($C^k_q=\sqrt{\frac{4\pi}{2k+1}}Y_{kq}$).
Here in (\ref{DM1}) we took into account $V_{so}(1)$ while the contribution of $V_{so}(2)$ has the same expression with the minus sign and the 1$\leftrightarrow$2 permutation. In addition, we restrict ourselves by the DM coupling operator which is diagonal on the spin and orbital moments. Obviously, nonzero DM coupling is only at even value of ($b_1+b_{1}^{'}$) and $|b_1-b_{1}^{'}|\leq 1\leq b_1+b_{1}^{'}$. In addition  ($b_2+b_{1}^{'}$) is also  should be an even number. Thus we should conclude that for the pair of equivalent free $S$-ions ($Fe^{3+}, Mn^{2+}$) when $b_2=b_{1}^{'}$\,=\,0  we have no DM coupling\,\cite{thesis}. We arrive at the same conclusion, if to take into account that the exchange parameters $I(101q)$ and $I(011q)$ specifying the appropriate contribution turn into zero\,\cite{thesis}.
The appearance of the DM coupling in such a case can be driven by the inter-configurational or crystal field effects.  

As the most illustrative example we consider a pair of 3d$^5$ ions such as $Fe^{3+}$, or Mn$^{2+}$ with the ground state $^6S$ in an intermediate octahedral crystal field which does split the $^{2S+1}L$ terms into crystal  $^{2S+1}L\Gamma$ terms and mix the crystal terms with the same octahedral symmetry, that is with the same $\Gamma$'s\,\cite{STK}. Spin-orbital coupling does mix the $^6S$ ground state with the $^4PT_{1g}$ term, 
however the $^4PT_{1g}$ term has been mixed with other $^4T_{1g}$ terms, $^4FT_{1g}$ and  $^4GT_{1g}$. Namely the latter effect is believed to be  a decisive factor for appearance of the DM coupling. The $|4(L)T_{1g}\rangle $ wave functions can be easily calculated by a standard technique\,\cite{STK} as follows\,\cite{thesis}:
$$
|4(P)T_{1g}\rangle = 0.679 |4PT_{1g}\rangle -0.604 |4FT_{1g}\rangle  +0.418 |4GT_{1g}\rangle \, ;
$$
$$
|4(F)T_{1g}\rangle = 0.387 |4PT_{1g}\rangle +0.777 |4FT_{1g}\rangle  +0.495 |4GT_{1g}\rangle \, ;
$$
\begin{equation}
|4(G)T_{1g}\rangle = -0.604 |4PT_{1g}\rangle -0.169 |4FT_{1g}\rangle  +0.737 |4GT_{1g}\rangle \, ,
\label{4T1} 	
\end{equation}
given the crystal field and intra-atomic correlation parameters\,\cite{STK} typical for orthoferrites\,\cite{Kahn}: 10Dq\,=\,12200\,$cm^{-1}$; B\,=\,700\,$cm^{-1}$; C\,=\,2600\,$cm^{-1}$.


The huge expression (\ref{DM1}) reduces to a more compact form as follows:
\begin{equation}
	d_q(12)=-\frac{2\sqrt{2}i\xi_{3d}}{5\sqrt{3}}V^{(14)}_{^6S^4G}\left(\sum_{^4T_{1g}}\frac{\alpha_{^4G}\alpha_{^4P}}{\Delta E_{^4T_{1g}}}\right)(I(404T_1q)-I(044T_1q)) \, ,
\end{equation}
where $V^{(14)}_{^6S^4G}$ is the conventional spectroscopic Racah coefficient\,\cite{Sobelman}, $\alpha_{^4P}$, $\alpha_{^4G}$ are the mixing coefficients for the $^4T_{1g}$ term, $I(404T_1q)=\sum_{\beta}\alpha_{4\beta}^{T_1q}I(404\beta)$ are the $T_1$-symmetry combinations of the exchange parameters. 
It is worth noting the conclusive effect of the ${}^4P-{}^4G$ mixing. 

For the direct exchange we have a simple expression for the parameters
\begin{equation} 
	I(404\beta)=J(404)C^4_{\beta}({\bf R}_{12})\rightarrow I(404T_1q)=J(404)C^{4T_1}_{q}({\bf R}_{12})\, ,
\end{equation}
where  $C^{4T_1}_{q}$ is the $T_1$-symmetry combination, or cubic harmonics. Finally we arrive at a remarkable relation: 
\begin{equation}
	d_q(12)=i\left[d_0(12)C^{4T_1}_{q}({\bf R}_{12})-d_0(21)C^{4T_1}_{q}({\bf R}_{21})\right]\, ,
\label{d_q}
\end{equation}
where the $T_1$-symmetry combinations of spherical harmonics are taken in local coordinate systems for the first and second ions, respectively, $d_0(12)\propto J(404)$ and $d_0(21)\propto J(044)$ are determined by the spin-orbital coupling on the sites 1 and 2, respectively. For locally equivalent $Fe^{3+}$ centers $J(404)=J(044)$ and $d_0(12)=d_0(21)$.
In the coordinate axes with $O_{z}\parallel C_4$
\begin{equation}
	\alpha_{44}^{T_10}=-\alpha_{4-4}^{T_10}=\frac{1}{\sqrt{2}};  
\alpha_{41}^{T_11}=-\alpha_{4-1}^{T_1-1}=\frac{\sqrt{7}}{\sqrt{8}};\alpha_{4-3}^{T_11}=-\alpha_{43}^{T_1-1}=\frac{1}{\sqrt{8}} \, ,
\label{alpha}
\end{equation}
and 
\begin{equation}
	C_0^{4T_1}=i\frac{\sqrt{35}}{8}sin^4\theta sin4\varphi \,;
C_{\pm 1}^{4T_1}=\frac{\sqrt{35}}{16\sqrt{2}}sin2\theta [(3-7cos^2\theta )e^{\pm i\varphi}+sin^2\theta e^{\mp 3i\varphi}] \, ,
\end{equation}
where $\theta$ and $\varphi$ are polar and azimuthal angles of the ${\bf R}_{12}$ vector.
Obviously, the Dzyaloshinskii vector turns into zero, if local crystal field axes coincide for the both ions. In addition, ${\bf d}(12)$\,=\,0, if ${\bf R}_{12}\parallel C_2, C_3, C_4$, that is to any symmetry axis for the first and second site. If ${\bf R}_{12}$ lies in a mirror plane ${\bf d}(12)$ $\perp$ mirror plane.
It should be pointed out a very untypical vector character of the Dzyaloshinskii vector.

\subsection{Microscopic theory of the DM coupling: superexchange interaction of the S-ions}
Hamiltonian for the superexchange  coupling of two ions with electron configurations $n_1l_1^{N_1}$ and $n_2l_2^{N_2}$ via intermediate nonmagnetic ligand ion has the same general expression as for direct exchange\,\cite{thesis}, however, with a specific dependence of the exchange parameters on the superexchange geometry:
\begin{equation}
I(b_{1} b_{2} b \beta ) =	\sum_{k_1k_2}J(b_{1} b_{2}k_1k_2 b)\left[C^{k_1}({\bf R}_{10})\times C^{k_2}({\bf R}_{20})\right]^b_{\beta} \, .
\label{SE}
\end{equation}
In the local coordinate system for the site $1$ with $O_z\parallel {\bf R}_{10}$ we can write out the superexchange parameter $I(404T_1q)$ as follows
\begin{equation}
	I(404T_1q)=\sum_{k_1k_2q_2}J(40k_1k_24)\left[
   \begin{array}{ccc}
    k_1 & k_2& 4 \\  
		0 & q_2    & q_2 
   \end{array}
  \right]  C^{k_2}_{q_2}({\bf r}_{20})\alpha_{4q_2}^{T_1q} \, ,
	\label{k1k2}
\end{equation}
where $\left[
   \begin{array}{ccc}
    k_1 & k_2& 4 \\  
		0 & q_2    & q_2 
   \end{array}
  \right]$ is the Clebsch-Gordan coefficient\,\cite{Sobelman,Var}.
 Obviously, for the superexchange mechanisms related with a particular ligand $2s$ or $2p$ electrons  we have for $k_2$: $k_{2}=0$ or $k_{2}=0;2$, respectively. For mechanisms related with the ligand inter-configurational  $2p\rightarrow 3s$ excitations $k_{2}=1$. Taking into account the properties of the $\alpha_{4q_2}^{T_1q}$ coefficients (\ref{alpha}) we see that since $|q_2|\leq 2$ it follows that the terms with $k_{2}=1$ and $k_{2}=2$ in (\ref{k1k2}) can be expressed in terms of the vector product $\left[C^{1}({\bf R}_{10})\times C^{1}({\bf R}_{20})\right]^1_{q}$\,=\,$\frac{i}{\sqrt{2}}\left[{\bf r}_1\times {\bf r}_2)\right]_{q}$. Indeed
\begin{equation}
	\sum_{q_2}\left[
   \begin{array}{ccc}
    k_1 & 1& 4 \\  
		0 & q_2    & q_2 
   \end{array}
  \right]  C^{1}_{q_2}({\bf r}_{20})\alpha_{4q_2}^{T_1q}=\sqrt{\frac{7}{8}}\left[
   \begin{array}{ccc}
    k_1 & 1& 4 \\  
		0 & 1    & 1 
   \end{array}
  \right]  \left[
   \begin{array}{ccc}
    1 & 1& 1 \\  
		0 & 1    & 1 
   \end{array}
  \right]^{-1}\left[C^{1}({\bf r}_{10})\times C^{1}({\bf r}_{20})\right]^1_{q}
\end{equation}

\begin{equation}
	\sum_{q_2}\left[
   \begin{array}{ccc}
    k_1 & 2& 4 \\  
		0 & q_2    & q_2 
   \end{array}
  \right]  C^{2}_{q_2}({\bf r}_{20})\alpha_{4q_2}^{T_1q}=\sqrt{\frac{3}{8}}\left[
   \begin{array}{ccc}
    k_1 & 2& 4 \\  
		0 & 1    & 1 
   \end{array}
  \right]  \left[
   \begin{array}{ccc}
    1 & 1& 1 \\  
		0 & 1    & 1 
   \end{array}
  \right]^{-1}cos\theta_{12}\left[C^{1}({\bf r}_{10})\times C^{1}({\bf r}_{20})\right]^1_{q}
\end{equation}

Obviously, final expression for the Dzyaloshinskii vector can be written as follows
\begin{equation}
	\mathbf{d}_{12} = d_{12}(\theta_{12}) [\mathbf{r}_1 \times \mathbf{r}_2] \, ,
	\label{d}
\end{equation} 
with
\begin{equation}
	d_{12}(\theta)=d_1(R_{10},R_{20})+d_2(R_{10},R_{20})cos\theta_{12} \, ,
	\label{dcos}
\end{equation}
where the first and the second terms are determined by the superexchange mechanisms related with the ligand inter-configurational  $2p\rightarrow 3s$ excitations and intra-configurational $2p-2p$ effects, respectively. It should be noted that given $\theta$\,=\,$\theta_{cr}$, where 
$cos\theta_{cr}=-d_1/d_2$, the   Dzyaloshinskii vector changes its sign.

\subsection{Microscopic theory of the DM coupling: superexchange interaction of the S-type ions in a strong cubic crystal field}

Hereafter we address the DM coupling for the  S-type magnetic 3d ions with orbitally nondegenerate high-spin ground state in a strong cubic crystal field, that is for the 3d ions with half-filled shells $t_{2g}^3$, $t_{2g}^3e_g^2$, $t_{2g}^6e_g^2$ and ground states $^4A_{2g}$,  $^6A_{1g}$, $^3A_{2g}$, respectively. 
The strong crystal field approximation seems to be more appropriate for the most part of 3d ions in crystals. In particular, for the ${}^{4}T_{1g}$ terms of the 3d$^5$ ion in a strong cubic crystal field approximation instead of expressions (\ref{4T1}) we arrive at a superposition of the wave functions for different $t_{2g}^{n_1}e_g^{n2}$ configurations ($n_1+n_2$\,=\,5)\,\cite{STK}.  Using  the same crystal field and correlation parameters as in Exp.\,(\ref{4T1}) we get a triplet of new functions as follows
$$
|{}^4T_{1g}(41)\rangle = 0.988 |t_{2g}^{4}e_g^{1}{}^4T_{1g}\rangle -0.123 |t_{2g}^{3}e_g^{2}{}^4T_{1g}\rangle  +0.088 |t_{2g}^{2}e_g^{3}{}^4T_{1g}\rangle \,;\,\, E(41)= 0.96\cdot 10^4\,cm^{-1}
$$
$$
|{}^4T_{1g}(32)\rangle = 0.058 |t_{2g}^{4}e_g^{1}{}^4T_{1g}\rangle +0.844 |t_{2g}^{3}e_g^{2}{}^4T_{1g}\rangle  -0.534 |t_{2g}^{2}e_g^{3}{}^4T_{1g}\rangle  \,;\,\, E(32)= 2.96\cdot 10^4\,cm^{-1}
$$
\begin{equation}
|{}^4T_{1g}(23)\rangle = -0.140 |t_{2g}^{4}e_g^{1}{}^4T_{1g}\rangle -0.522 |t_{2g}^{3}e_g^{2}{}^4T_{1g}\rangle  +0.841 |t_{2g}^{2}e_g^{3}{}^4T_{1g}\rangle \,;\,\, E(23)= 3.69\cdot 10^4\,cm^{-1} \, ,
\label{4T1a} 	
\end{equation}
with a more clearly defined contribution of a particular configuration compared with the intermediate crystal field scheme.

Making use of expressions for spin-orbital coupling $V_{so}$\,\cite{Atoms}  and main kinetic contribution to the  superexchange parameters, that define the DM coupling,  after routine algebra we have found that the DM coupling can be written in a standard form (\ref{d}), where  $d_{12}$ can   be written as follows\,\cite{1977,thesis}  
\begin{equation}
	d_{12}=X_1Y_2+X_2Y_1 \, ,
	\label{XY}
\end{equation}
where the $X$ and $Y$ factors do reflect the exchange-relativistic structure of the second-order perturbation theory and details of the electron configuration for S-type ion. The exchange factors $X$ are
\begin{equation}
	X_i=\frac{(g^{(i)}_{e_g}-1)}{2U}t_{\pi\sigma}(t_{ss}+t_{\sigma\sigma}cos\theta )-\frac{(g^{(i)}_{t_{2g}}-1)}{3U}t_{\pi\pi}t_{\sigma\pi}cos\theta  \, ,
	\label{X}
\end{equation}
 where $g^{(i)}_{e_g}$, $g^{(i)}_{t_{2g}}$ are effective $g$-factors for $e_g$, ${t_{2g}}$ subshells, respectively, $t_{\sigma\sigma}>t_{\pi\sigma}>t_{\pi\pi}>t_{ss}$ are positive definite $d$\,-\,$d$ transfer integrals, $U$ is the $d$\,-\,$d$ transfer energy (correlation energy). The dimensionless factors $Y$ are determined by the spin-orbital constants  and excitation energies as follows
$$
Y_i= \sum_{S\Gamma} (-1)^{2S+1}  
  \langle S_i\|S\|S_i\rangle^{-1} \left\{
   \begin{array}{ccc}
    1 & 1& 1 \\  
		S_i & S_i    & S
   \end{array}
  \right\} 
\frac{<e_{g}\|\xi \|t_{2g}>}{\Delta E_{ S\Gamma}}
$$
\begin{equation}
W^{(1T_1)}_{S_i\Gamma_i ;S\Gamma }(e_gt_{2g})\left( W^{(1T_1)}_{S\Gamma ;S_{i}\Gamma_i}(e_gt_{2g}) - W^{(1T_1)}_{S\Gamma ;S_{i}\Gamma_i}( t_{2g}e_g) \right) \, ,
\label{Y}
\end{equation}
where $W^{(1T_1)}$ are spectroscopic coefficients for cubic point group\,\cite{Atoms} and summation runs on all the terms ${}^{2S+1}\Gamma$, mixed by the spin-orbital coupling with the ground state term ${}^{2S_i+1}\Gamma_i$
($\Gamma_i =A_{1,2}$, $\Gamma = \Gamma_i\times T_1 =T_{1,2}$).
It should be noted that the nonzero DM coupling for S-type ions can be obtained only due to inter-configurational $t_{2g}-e_g$ interaction.  
The factors $X$ and $Y$ are presented in Table\,\ref{tableXY} for S-type 3d-ions. There  $\xi_{3d}$ is the spin-orbital parameter, $\Delta E_{^{2S+1}\Gamma}$ is the energy of the $^{2S+1}\Gamma$ crystal term.

The signs for $X$ and $Y$ factors in Table\,\ref{tableXY} are predicted for rather large superexchange bonding angles $|cos\theta_{12}|>t_{ss}/t_{\sigma\sigma}$ which are typical for many 3d compounds such as oxides and a relation $\Delta E_{^4T_{1g}}(41)<\Delta E_{^4T_{1g}}(32)$ which is typical for high-spin $3d^5$ configurations. 

It is worth noting that while working with the paper we have detected and corrected a casual and unintentional error in sign of the $X_i$ parameters having made both in our earlier papers\,\cite{1977,thesis} and very recent paper Ref.\,\cite{2016}. 
Hereafter we present correct signs for $X_i$ in (\ref{X}) and Table\,\ref{tableXY}.

Rather simple expressions (\ref{X}) and (\ref{Y}) for the factors $X_i$ and $Y_i$ do not take into account the mixing/interaction effects for the ${}^{2S+1}\Gamma$  terms with the same symmetry and the contribution of empty subshells to the exchange coupling (see Ref.\,\cite{thesis}). Nevertheless, the data in Table\,\ref{tableXY} allow us to evaluate both the  numerical value and sign of the $d_{12}$ parameters.

It should be noted that for critical angle $\theta_{cr}$, when the Dzyaloshinskii vector changes its sign we have
$	cos\theta_{cr}=-d_1/d_2=\frac{\lambda_s^2}{\lambda_{\sigma}^2}$ for $d^8-d^8$ pairs and $	cos\theta_{cr}=-d_1/d_2=\frac{\lambda_s^2}{\lambda_{\sigma}^2-\lambda_{\pi}^2}$ for $d^5-d^5$ pairs. Making use of different experimental data for covalency parameters (see, e.g., Ref.\,\cite{Tofield}) we arrive at $d_1/d_2\sim \frac{1}{5}-\frac{1}{3}$ and $\theta_{cr}\approx 100^{\circ}-110^{\circ}$ for $Fe^{3+}-Fe^{3+}$ pairs in oxides.

 Relation among different $X$'s given the superexchange geometry and covalency parameters typical for orthoferrites and orthochromites \,\cite{thesis} is
\begin{equation}
	|X_{d^8}|\geq |X_{d^3}|\geq |X_{d^5}| \, ,
\end{equation}
however, it should be underlined its sensitivity both to superexchange geometry and covalency parameters. Simple comparison of the exchange parameters $X$ (see (\ref{X}) and Table \ref{tableXY}) with exchange parameters $I(\gamma_i\gamma_j)$ (\ref{kinetic}) evidences their close magnitudes. Furthermore, the relation (\ref{t-lambda}) allows us to maintain more definite correspondence.

Given typical values of the cubic crystal field parameter $10Dq\approx$\,1.5\,eV we arrive at a relation among different $Y$'s\,\cite{thesis}
\begin{equation}
	|Y_{d^8}|\geq |Y_{d^5}|\geq |Y_{d^3}|
\end{equation}
with $Y_{d^8}\approx 7.0\cdot 10^{-2}$, $Y_{d^5}\approx -2.5\cdot 10^{-2}$, $Y_{d^3}\approx 1.5\cdot 10^{-2}$.

The highest value of the $d_{12}$ factor  is predicted for $d^8-d^8$ pairs, while for $d^5-d^5$ pairs one expects  a much less (may be one order of magnitude) value. The $d_{12}$ factor for  $d^3-d^3$ pairs  is predicted to be somewhat above the value for $d^5-d^5$ pairs. For different pairs: $d_{12}(d^3-d^5)\approx -d_{12}(d^3-d^3)$;\,$d_{12}(d^8-d^5)\approx d_{12}(d^5-d^5)$;\, $d_{12}(d^3-d^8)\geq d_{12}(d^3-d^3)$. Puzzlingly, that despite strong isotropic exchange coupling for $d^5-d^5$ and $d^5-d^8$ pairs, the DM coupling for these pairs is expected to be  the least one among the S-type pairs. For $d^5-d^5$ pairs, in particular, $Fe^{3+}-Fe^{3+}$ we have two compensation effects.
First, the $\sigma$-bonding contribution to the $X$ parameter is partially compensated by the $\pi$-bonding contribution, second, the contribution of the ${}^4T_{1g}$ term of the $t_{2g}^4e_g^1$ configuration is partially compensated by the contribution of the ${}^4T_{1g}$ term of the $t_{2g}^2e_g^3$ configuration.

Theoretical predictions of the corrected sign of the Dzyaloshinskii vector in pairs of the S-type 3d-ions with local octahedral symmetry (the sign rules) are presented in Table\,\ref{tablesign}. The signs for $d^3-d^3$, $d^5-d^5$, and $d^3-d^8$ pairs turn out to be the same but opposite  to signs for $d^3-d^5$ and $d^8-d^8$ pairs. 
In a similar way to how different signs of the conventional exchange integral determine different (ferro-antiferro) magnetic orders the different signs of the Dzyaloshinskii vectors  create a possibility of nonuniform  (ferro-antiferro) ordering of local weak (anti)ferromagnetic moments, or local overt/hidden cantings. Novel magnetic phenomenon and novel class of magnetic materials, which are systems such as solid solutions $YFe_{1-x}Cr_xO_3$ with competing signs of the Dzyaloshinskii vectors will be addressed below (Sec.\,4.3) in more detail. 
\begin{center}
\begin{table}
\caption{Expressions for the $X$ and $Y$ parameters that define the magnitude and the sign of the Dzyaloshinskii vector in pairs of the S-type 3d-ions with local octahedral symmetry. Signs for $X_i$ correspond to the bonding angle $\theta >\theta_{cr}$.}
\begin{tabular}{|c|c|c|c|c|c|}
\hline
   \begin{tabular}{c}
Ground state \\
configuration \\  
\end{tabular}        & $X$ & Sign $X$ &  $Y$ & Sign $Y$ & \begin{tabular}{c}
Excited state \\
configuration \\  
\end{tabular}  \\ \hline
  \begin{tabular}{c}
$3d^3$($t_{2g}^3$):${}^4A_{2g}$ \\
$V^{2+}$, $Cr^{3+}$, $Mn^{4+}$ \\  
\end{tabular}  & $-\frac{1}{3U}t_{\pi\pi}t_{\sigma\pi}cos\theta$ & + &$\frac{2\xi_{3d}}{3\sqrt{3}}(\frac{1}{\Delta E_{^4T_{2g}}}+\frac{2}{\Delta E_{^2T_{2g}}})$ & + & $t_{2g}^2e_g^1$ \\ \hline
 \begin{tabular}{c}
$3d^5$($t_{2g}^3e_g^2$):${}^6A_{1g}$ \\
$Mn^{2+}$, $Fe^{3+}$ \\  
\end{tabular}  & \begin{tabular}{c}$-\frac{1}{5U}(t_{\pi\pi}t_{\sigma\pi}cos\theta$ - \\
$t_{\pi\sigma}\left(t_{ss}+t_{\sigma\sigma}cos\theta )\right)$\\
\end{tabular} & -- & --$\frac{6\xi_{3d}}{5\sqrt{3}}(\frac{1}{\Delta E_{^4T_{1g}}(41)}-\frac{1}{\Delta E_{^4T_{1g}}(23)})$ & -- & $t_{2g}^4e_g^1$, $t_{2g}^2e_g^3$ \\ \hline	
\begin{tabular}{c}
$3d^8$($t_{2g}^6e_g^2$):${}^3A_{2g}$ \\
$Ni^{2+}$, $Cu^{3+}$ \\  
\end{tabular}  & $\frac{1}{2U}t_{\pi\sigma}(t_{ss}+t_{\sigma\sigma}cos\theta )$& -- & $\frac{3\xi_{3d}}{2\sqrt{3}}(\frac{1}{\Delta E_{^3T_{2g}}}+\frac{1}{\Delta E_{^1T_{2g}}})$ & + & $t_{2g}^5e_g^3$\\ \hline       
	\end{tabular}
\label{tableXY}
\end{table}
\end{center}

\begin{center}
\begin{table}
\caption{Sign rules for the Dzyaloshinskii vector in pairs of the S-type 3d-ions with local octahedral symmetry and the bonding angle $\theta >\theta_{cr}$.}
\centering
\begin{tabular}{|c|c|c|c|}
\hline
   $3d^n$      & $3d^3$($t_{2g}^3$) &  $3d^5$($t_{2g}^3e_g^2$)  & $3d^8$($t_{2g}^6e_g^2$)  \\ \hline
  $3d^3$($t_{2g}^3$)  & + &--& + \\ \hline
 $3d^5$($t_{2g}^3e_g^2$) & -- &+ & + \\ \hline	
$3d^8$($t_{2g}^6e_g^2$) & +& +& -- \\ \hline       
	\end{tabular}
\label{tablesign}
\end{table}
\end{center}
\subsection{DM coupling in trigonal hematite $\alpha$-$Fe_2O_3$}
Making use of our theory based on the bare ideal octahedral symmetry of S-type ions to the classical weak ferromagnet $\alpha$-$Fe_2O_3$ we arrive at a little unexpected  disappointment, as the theory does predict that the  contribution of the three equivalent $Fe^{3+}-O^{2-}-Fe^{3+}$ superexchange pathes for the two corner shared $FeO_6^{9-}$ octahedrons  to the net Dzyaloshinskii vector strictly turns into zero.  Exactly the same  result will be obtained, if we consider the direct $Fe^{3+}-Fe^{3+}$ exchange in the system of two ideal $FeO_6^{9-}$ octahedrons bonded through the three common oxygen ions when ${\bf R}_{12}\parallel C_3$. Obviously, it is precisely this fact that caused  a tiny spin canting in hematite being an order of magnitude smaller than, e.g., in orthoferrites $RFeO_3$ or borate $FeBO_3$. So what was the real reason of weak ferromagnetism in  $\alpha$-$Fe_2O_3$ as "opening a new page of weak ferromagnetism"? What is a microscopic origin of nonzero Dzyaloshinskii vector  which should be directed along the $C_3$ symmetry axis according Moriya rules?
First of all we should consider trigonal distortions for the $FeO_6^{9-}$ octahedrons which have a $T_{2}$ symmetry and give rise to a mixing of the ${}^4T_{1g}$ terms with ${}^4A_{2g}$ and ${}^4T_{2g}$ terms. The best way to solve the problem in principle is to proceed with a coordinate system where $O_z$ axis is directed along the $C_3$ symmetry axis rather than with the usually applied $O_{z}\parallel C_4$ geometry.

In the coordinate axes with $O_{z}\parallel C_3$ the nonzero coefficients $\alpha_{4\beta}^{T_1q}$  have another expression\,\cite{C_3}. Instead of (\ref{alpha}) we arrive at
\begin{equation}
	\alpha_{4\pm 3}^{T_10}=-\frac{1}{\sqrt{2}};  
\alpha_{4\pm 4}^{T_1\pm 1}=\mp\frac{\sqrt{2}}{3};\alpha_{4\mp 2}^{T_1\pm 1}=\mp\frac{7}{3\sqrt{2}}; 
\alpha_{4\pm 1}^{T_1\pm 1}=\mp\frac{7}{3\sqrt{2}}\, .
\label{alpha1}
\end{equation}
It is easy to see that for ${\bf R}_{12}\parallel C_3\parallel O_z$ $C^4_{\beta}({\bf R}_{12})=\delta_{\beta 0}$, that means that all the components of the two contributions to Dzyaloshinskii vector (\ref{d_q}) turn into zero.

However, the situation changes under axial (trigonal) distortion of the $FeO_6^{9-}$ octahedrons that can be described  by a simple effective Hamiltonian as follows
\begin{equation}
	{\hat V}_{trig}=B_{trig}{\hat V}^{T_{2g}} \, ,
\end{equation}
where ${\hat V}^{T_{2g}}={\hat V}^{T_{2g}}_2+i(e^{i\frac{\pi}{4}}{\hat V}^{T_{2g}}_1+e^{-i\frac{\pi}{4}}{\hat V}^{T_{2g}}_{-1})$ ($\propto (xy+yz+zx)$) is the only irreducible tensor operator  permitted by the symmetry of the distortion, $B_{trig}$ is a trigonal field parameter. Such a distortion gives rise to a mixing of the ${}^4T_{1g}$ terms with ${}^4A_{2g}$, ${}^4E_{g}$, and ${}^4T_{2g}$ terms. As a result the bare $|{}^4T_{1g}\rangle$ functions are transformed as follows:    
\begin{equation}
	|{}^4T_{1g}q\rangle _0 \rightarrow |{}^4T_{1g}q\rangle = c_0\left(|{}^4T_{1g}q\rangle _0  + c_{A_{2g}2}^{T_{1g}q}|{}^4A_{2g}2\rangle   + c_{T_{2g}\mu}^{T_{1g}q}|{}^4T_{2g}\mu\rangle\right) \, ,
\end{equation}
where $c_0$ is a normalization coefficient, $c_{T_{2g}\mu}^{T_{1g}q}$, $c_{A_{2g}2}^{T_{1g}q}=c_{A_{2g}2}^{T_{1g}0}\delta_{q0}$ are the mixing coefficients. The Dzyaloshinskii vector now has a representation as follows
\begin{equation}
	d_q(12)=i\sum_{\mu}\left[c_0(1)c_{T_{2g}\mu}^{T_{1g}q}(1)d_0(12)C^{4T_2}_{\mu}({\bf R}_{12})-c_0(2)c_{T_{2g}\mu}^{T_{1g}q}(2)d_0(21)C^{4T_2}_{\mu}({\bf R}_{21})\right]\, ,
\label{d_qC_3}
\end{equation}
where we again suggest the possibility of nonequivalent centers 1,2. Cubic harmonics $C^{4T_2}_{\mu}$ one can find, if make use of data from Ref.\,\cite{C_3}  
\begin{equation}
\alpha_{4+0}^{T_22}=-\frac{2\sqrt{5}}{3\sqrt{3}};\alpha_{4\pm 3}^{T_22}=\pm\frac{\sqrt{7}}{3\sqrt{6}};\,\, 
\alpha_{4\pm 4}^{T_2\pm 1}=-\frac{2\sqrt{7}}{3\sqrt{6}};   \alpha_{4\pm 2}^{T_2\mp 1}=\frac{1}{3\sqrt{6}}; 
\alpha_{4\pm 1}^{T_2\pm 1}=\pm\frac{5}{3\sqrt{6}}\, .
\label{alpha2}
\end{equation}
We see that given ${\bf R}_{12}\parallel C_3$ the only nonzero cubic harmonic is $C^{4T_2}_{2}=-\frac{2\sqrt{5}}{3\sqrt{3}}$ that defines the only nonzero $z$-component of Dzyaloshinskii vector as for $\mu =2$ only $c_{T_{2g}2}^{T_{1g}0}\not= 0$. Thus the axial distortion along the $Fe^{3+}-Fe^{3+}$  bond  can induce the DM coupling with Dzyaloshinskii vector directed along the bond, however, only for locally nonequivalent $Fe^{3+}$ centers, otherwise we arrive at an exact compensation of  the contributions of the spin-orbital couplings on sites 1 and 2.

Trigonal hematite  $\alpha$-$Fe_2O_3$ has the same crystal symmetry $R\overline{3}c-D_{3d}^6$ as weak ferromagnet $FeBO_3$. Furthermore, the borate  can be transformed into hematite by the $Fe^{3+}$ ion substitution for $B^{3+}$ with a displacement of both "old" and "new" iron ions along trigonal axis. As a result we arrive at  emergence of an additional strong isotropic (super)exchange coupling of three-corner-shared non-centrosymmetric $FeO_6$ octahedra with short $Fe-O$ separations (1.942\,\AA) that determines very high N\'{e}el temperature $T_N$\,=\,948\,K in hematite as compared with $T_N$\,=\,348\,K in borate. However, the $D_{3h}$ symmetry of these exchange bonds points to a distinct compensation of the two $Fe$-ion's contribution to Dzyaloshinskii vector. In other words,  weak ferromagnetism in hematite $\alpha$-$Fe_2O_3$ is determined by the DM coupling for the  same $Fe-O-Fe$ bonds as in borate $FeBO_3$. However, the $Fe-O$ separations for these bonds in hematite (2.111\,\AA) are markedly longer than in borate (2.028\,\AA) that points to a significantly weaker DM coupling. Combination of weaker DM coupling and stronger isotropic exchange in $\alpha$-$Fe_2O_3$ as compared with $FeBO_3$ does explain the one order of magnitude difference in canting angles.  


\subsection{DM coupling with participation of rare-earth ions}

Spin-orbital interaction for the rare-earth ions with valent $4f^n$ configuration is diagonalized within the $(LS)J$ multiplets hence the conventional DM coupling
 \begin{equation}
	{\hat H}_{DM}^{ff}=\sum_{m>n}({\bf d}_{mn}\cdot[{\bf S}_m\times{\bf S}_n])=\sum_{m>n}(g_m-1)(g_n-1)({\bf d}_{mn}\cdot[{\bf J}_m\times{\bf J}_n])
\end{equation} 
($g_{m,n}$ are the Lande factors) can arise for $f-f$ superexchange only due to a spin-orbital contribution on intermediate ligands. Obviously, for the rare-earth-3d-ion (super)exchange we have an additional contribution of the 3d-ion spin-orbital interaction.   
The rare-earth-3d-ion  DM coupling Gd$^{3+}$\,-\,O$^{2-}$\,-\,Fe$^{3+}$
\begin{equation}
	{\hat H}_{DM}^{fd}=\sum_{m>n}({\bf d}_{mn}\cdot[{\bf J}_m\times{\bf S}_n])
\end{equation}
 has been  theoretically and experimentally considered in Ref.\,\cite{Belov} for GdFeO$_3$.

\section{Theoretical predictions as compared with experiment}		
		
\subsection{Overt and hidden canting in orthoferrites}

\begin{figure}[t]
\centering
\includegraphics[width=12.5cm,angle=0]{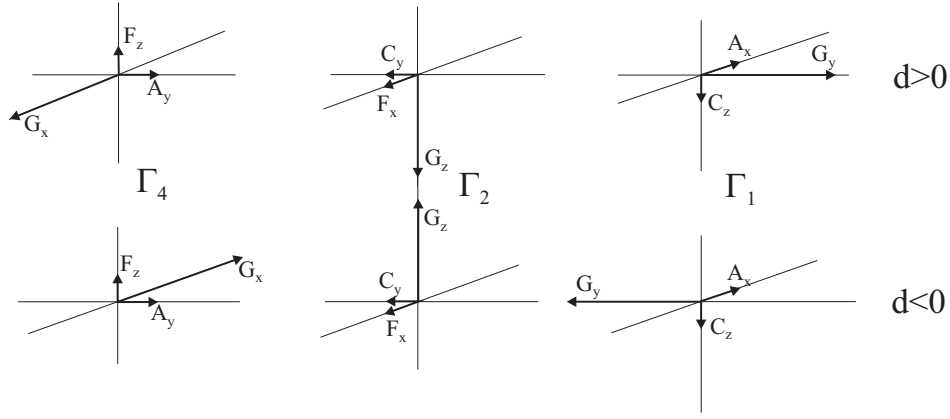}
\caption{Basic vectors of magnetic structure for $3d$ sublattice in orthoferrites and orthochromites}
\label{fig3}
\end{figure}

At variance with isotropic superexchange coupling the DM coupling has a much more complicated structural dependence. In Table\,\ref{tabler1r2}
we present structural factors $\left[{\bf r}_1\times{\bf r}_2\right]_{x,y,z}$ for the superexchange coupled Fe-O-Fe pairs in orthoferrites with numerical values for YFeO$_3$\,\cite{RFeO3}. In all cases, the vector ${\bf r}_1$ is oriented to the Fe ion in the position (1/2,0,0), the vectors ${\bf r}_2$ are oriented to the nearest Fe ions in the $ab$-plane (1a, 1b) or along the $c$-axis (3a). It is easy to see that the weak ferromagnetism in orthoferrites governed by the $y$-component of the Dzyaloshinskii vector does actually make use of only about one-third of its maximal value.
\begin{center}
\begin{table}
\caption{The structural factors $\left[{\bf r}_1\times{\bf r}_2\right]_{x,y,z}$ for the superexchange coupled Fe-O-Fe pairs in orthoferrites with numerical values for YFeO$_3$. See text for detail.}
\centering
\begin{tabular}{|c|c|c|c|}
\hline
         & $\left[{\bf r}_1\times{\bf r}_2\right]_x$ &  $\left[{\bf r}_1\times{\bf r}_2\right]_y$  & $\left[{\bf r}_1\times{\bf r}_2\right]_z$  \\ \hline
  1a  & $-\frac{z_2bc}{2l^2}$\,=\,-0.31 & $-\frac{z_2ac}{2l^2}$\,=\,-0.29 & $\frac{(y_2-x_2+\frac{1}{2})ab}{2l^2}$\,=\,0.41 \\ \hline
 1b & $+\frac{z_2bc}{2l^2}$\,=\,0.31 & $-\frac{z_2ac}{2l^2}$\,=\,-0.29 & $\frac{(y_2-x_2+\frac{1}{2})ab}{2l^2}$\,=\,0.41 \\ \hline	
3a & $\frac{(\frac{1}{2}-y_1)bc}{2l^2}$\,=\,0.20 &-$\frac{x_1ac}{2l^2}$\,=\,-0.55 & 0 \\ \hline       
	\end{tabular}
\label{tabler1r2}
\end{table}
\end{center}
In 1975 we made use of simple formula for the Dzyaloshinskii vector (\ref{d12}) and structural factors from Table\,\ref{tabler1r2} to find a relation between crystallographic and canted magnetic structures for four-sublattice's orthoferrites RFeO$_3$ and orthochromites RCrO$_3$\,\cite{1975,thesis} (see Fig.\ref{fig3}), where main G-type antiferromagnetic order  is accompanied by both overt canting characterized by ferromagnetic vector ${\bf F}$ (weak ferromagnetism!) and two types of a hidden canting, ${\bf A}$ and ${\bf C}$ (weak antiferromagnetism!):
$$
F_z=\frac{(x_1+2z_2)ac}{6l^2}\frac{d}{I}G_x\,;\, F_x=-\frac{(x_1+2z_2)ac}{6l^2}\frac{d}{I}G_z\,;\,A_y=\frac{(\frac{1}{2}+y_2-x_2)ab}{2l^2}\frac{d}{I}G_x\,;\,
$$
\begin{equation}
A_x=-\frac{(\frac{1}{2}+y_2-x_2)ab}{2l^2}\frac{d}{I}G_y\,;\, C_y=\frac{(\frac{1}{2}-y_1)bc}{2l^2}\frac{d}{I}G_z\,;\,C_z=-\frac{(\frac{1}{2}-y_1)bc}{2l^2}\frac{d}{I}G_y\,,
\label{FCA}	
\end{equation}
where $a,b,c$ are unit cell parameters, $x_{1,2}, y_{1,2}, z_2$ are oxygen ($O_{I,II}$) parameters\,\cite{RFeO3}, $l$ is a mean cation-anion separation. These relations imply an averaging on the $Fe^{3+}-O^{2-}-Fe^{3+}$ bonds in $ab$ plane and along $c$-axis. It is worth noting that $|A_{x,y}|>|F_{x,z}|>|C_{y,z}|$.

First of all we arrive at a simple relation between crystallographic parameters and magnetic moment of the Fe-sublattice: in units of $G\cdot g/cm^3$
\begin{equation}
	M_{Fe}=\frac{4g_S\beta_eS}{\rho V}|F_{x,z}|=\frac{2g\beta_eSac}{3l^2\rho V}(x_1+2z_2)\frac{d(\theta)}{I(\theta)}\, ,
\end{equation}
where $\rho$ and $V$ are the unit cell density and volume, respectively.
The overt canting $F_{x,z}$ can be calculated through the ratio of the Dzyaloshinskii ($H_D$) and exchange ($H_E$) fields as follows
\begin{equation}
	F=H_D/2H_E \, .
\end{equation}
If we know the Dzyaloshinskii field we can calculate the $d(\theta )$ parameter in orthoferrites as follows
\begin{equation}
	H_D=\frac{S}{g\mu_B}\sum_i|d_y(1i)|=\frac{S}{g\mu_B}(x_1+2z_2)\frac{ac}{l^2}|d(\theta )|\, ,
\end{equation}
that yields $|d(\theta )|\cong$\,3.2\,K in YFeO$_3$ given $H_D$\,=\,140\,kOe\,\cite{Jacobs}. It is worth noting that despite $F_z\approx$\,0.01 the $d(\theta )$ parameter is only one order of magnitude  smaller than the exchange integral in YFeO$_3$.

Our results have stimulated experimental studies of the  hidden canting, or "weak antiferromagnetism" in orthoferrites. As shown in Table\,\ref{AFC} theoretically predicted relations between overt and hidden canting  nicely agree with the experimental data obtained for different orthoferrites by NMR\,\cite{Luetgemeier} and neutron diffraction \,\cite{Plakhtii,Georgieva}.

\begin{center}
\begin{table}
\caption{Hidden canting in orthoferrites.}
\centering
\begin{tabular}{|c|c|c|c|c|c|c|}
\hline
 Orthoferrite  & A$_y$/F$_z$, theory\,\cite{1975} & A$_y$/F$_z$, exp & A$_y$/C$_y$, theory\,\cite{1975} & A$_y$/C$_y$,  exp  \\ \hline
  YFeO$3$ &  1.10 & \begin{tabular}{c}
1.10\,$\pm$\,0.03\cite{Luetgemeier} \\
1.4\,$\pm$\,0.2\cite{Plakhtii}\\
1.1\,$\pm$\,0.1\cite{Georgieva}\\  
\end{tabular} & 2.04 & ?\\ \hline
  HoFeO$3$ & 1.16
  & 0.85\,$\pm$\,0.10\cite{Georgieva} &  2.00&? \\ \hline
 TmFeO$3$ & 1.10
&  1.25\,$\pm$\,0.05\cite{Luetgemeier}& 1.83 & ?\\ \hline
 YbFeO$3$ & 1.11
&  1.22\,$\pm$\,0.05\cite{Plakhtii} & 1.79 & 2.0\,$\pm$\,0.2\cite{Luetgemeier}\\ \hline
 \end{tabular}
\label{AFC1}
\end{table}
\end{center}

	\subsection{The DM coupling and effective magnetic anisotropy}
	Hereafter we demonstrate a contribution of the DM coupling into effective magnetic anisotropy in orthoferrites.	
	The classical energies of the three spin configurations in orthoferrites $\Gamma_1(A_x,G_y,C_z)$, $\Gamma_2(F_x,C_y,G_z)$, and $\Gamma_4(G_x,A_y,F_z)$ given  $|F_x|=|F_z|=F$, $|C_y|=|C_z|=C$, $|A_x|=|A_z|=A$ can be written as follows\,\cite{DManiso}	
	\begin{eqnarray}
E_{\Gamma_1}= I_{G}-48IS^2F^2\left[\frac{1}{3}(\frac{C}{F})^2+\frac{2}{3}(\frac{A}{F})^2\right] \, ; \\
E_{\Gamma_2}= I_{G}-48IS^2F^2\left[1+\frac{1}{3}(\frac{C}{F})^2\right] \, ; \\
E_{\Gamma_4}= I_{G}-48IS^2F^2\left[1+\frac{2}{3}(\frac{A}{F})^2\right] \, ,	
	\end{eqnarray}
	with obvious relation $E_{\Gamma_4} < E_{\Gamma_1}\leq E_{\Gamma_2}$.
	The energies allow us to find the constants of the in-plane magnetic anisotropy $E_{an}=k_1\,cos2\theta$ ($ac$, $bc$ planes), $E_{an}=k_1\,cos2\varphi$ ($ab$ plane): $k_1(ac)=\frac{1}{2}(E_{\Gamma_2}-E_{\Gamma_4})$;  $k_1(bc)=\frac{1}{2}(E_{\Gamma_2}-E_{\Gamma_1})$;  
$k_1(ab)=\frac{1}{2}(E_{\Gamma_4}-E_{\Gamma_1})$. Detailed analysis of different mechanisms of the magnetic anisotropy of the orthoferrites\,\cite{DManiso} points to a leading contribution of the DM coupling. Indeed, for all the orthoferrites RFeO$_3$ (R = Y, or rare-earth ion) this mechanism does  predict a minimal energy for $\Gamma_4$ configuration which is actually realized as a ground state for all the orthoferrites, if one neglects the R-Fe interaction. Furthermore, predicted value of the constant  of the magnetic anisotropy in $ac$-plane for YFeO$_3$ $k_1(ac)$\,=2.0$\cdot 10^5$\,erg/cm$^3$ is close enough to experimental value of 2.5$\cdot 10^5$\,erg/cm$^3$\,\cite{Jacobs}.
Interestingly, the model predicts a close energy for $\Gamma_1$ and $\Gamma_2$ configurations so that $|k_1(bc)|$ is about one order of magnitude less than $|k_1(ac)|$ and  $|k_1(ab)|$ for most orthoferrites\,\cite{DManiso}. It means the anisotropy in $bc$-plane will be determined by a competition of the DM coupling with relatively weak contributors such as magneto-dipole interaction and single-ion anisotropy. It should be noted that the sign and value of the $k_1(bc)$ is of a great importance for the determination of the type of the domain walls for orthoferrites in their basic $\Gamma_4$ configuration (see, e.g., Ref.\,\cite{DyFeO3}).

\subsection{Weak ferrimagnetism as a novel type of magnetic ordering in systems with competing signs of the Dzyaloshinskii vector.}

\begin{figure}[t]
\begin{minipage}[h]{0.40\linewidth}
\center{\includegraphics[width=0.5\linewidth]{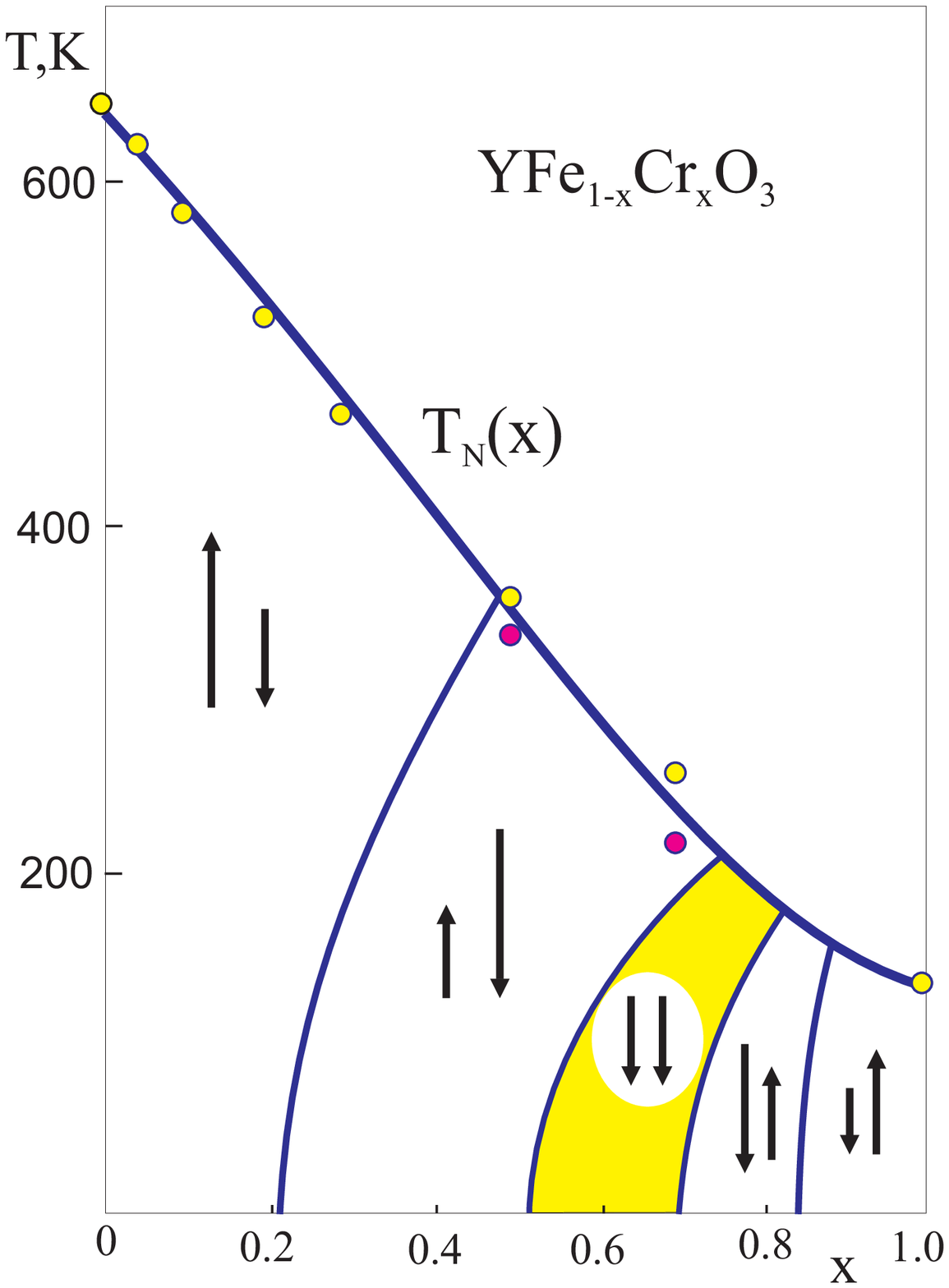} \\ a)}
\end{minipage}
\begin{minipage}[h]{0.40\linewidth}
\center{\includegraphics[width=0.5\linewidth]{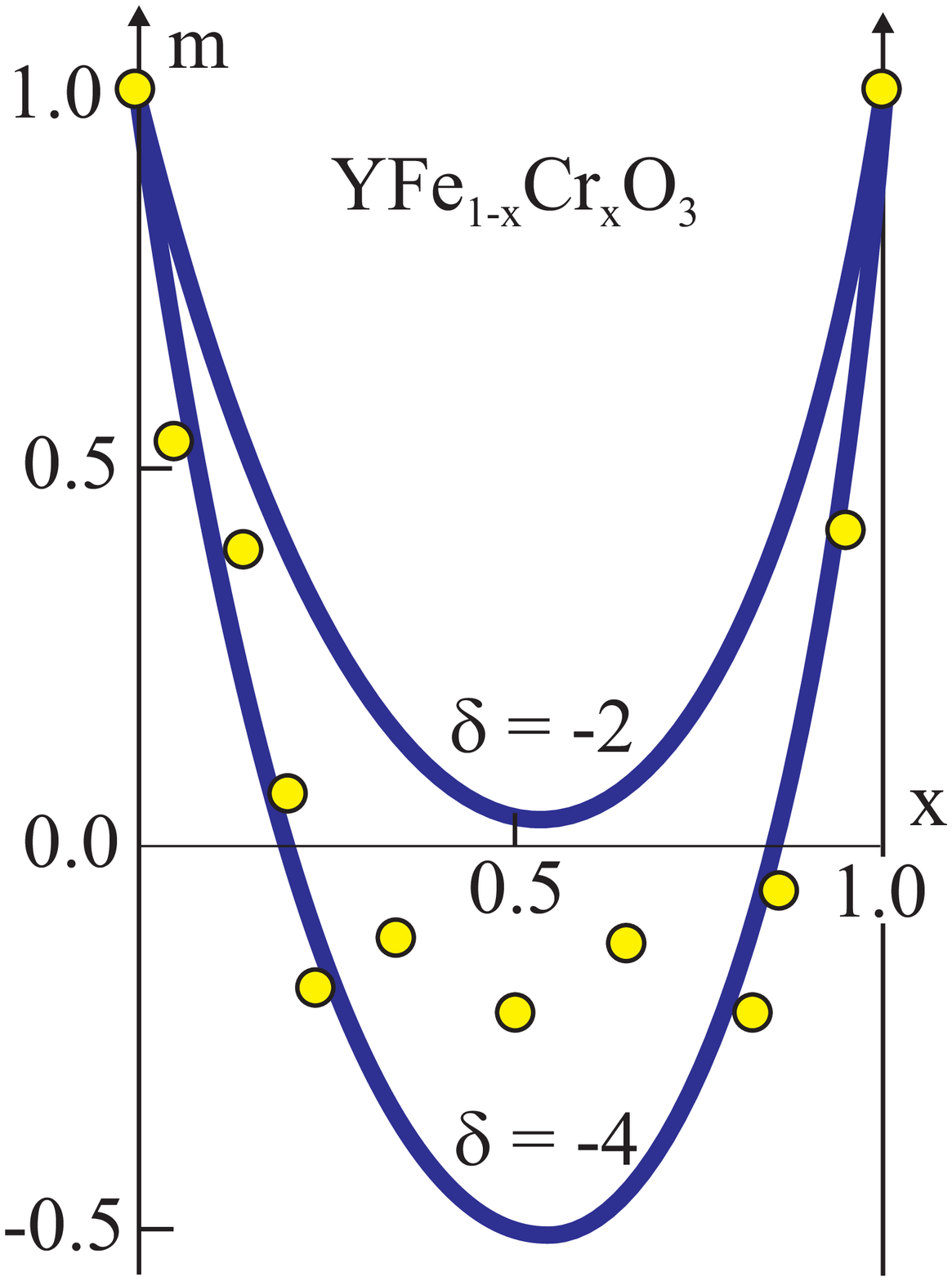} \\ b)}
\end{minipage}
\begin{minipage}[h]{0.40\linewidth}
\center{\includegraphics[width=0.5\linewidth]{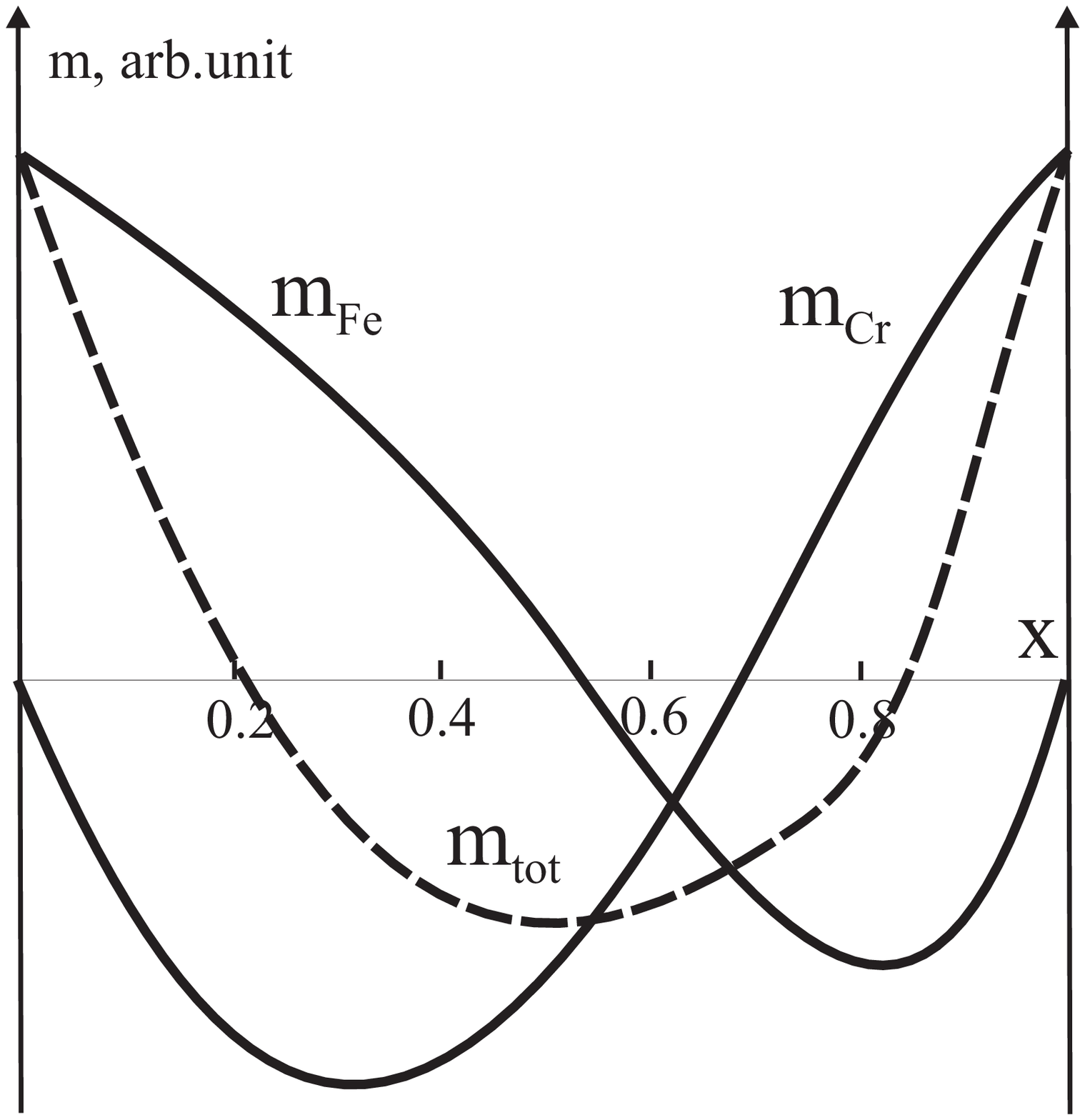} \\ c)}
\end{minipage}
\begin{minipage}[h]{0.40\linewidth}
\center{\includegraphics[width=0.5\linewidth]{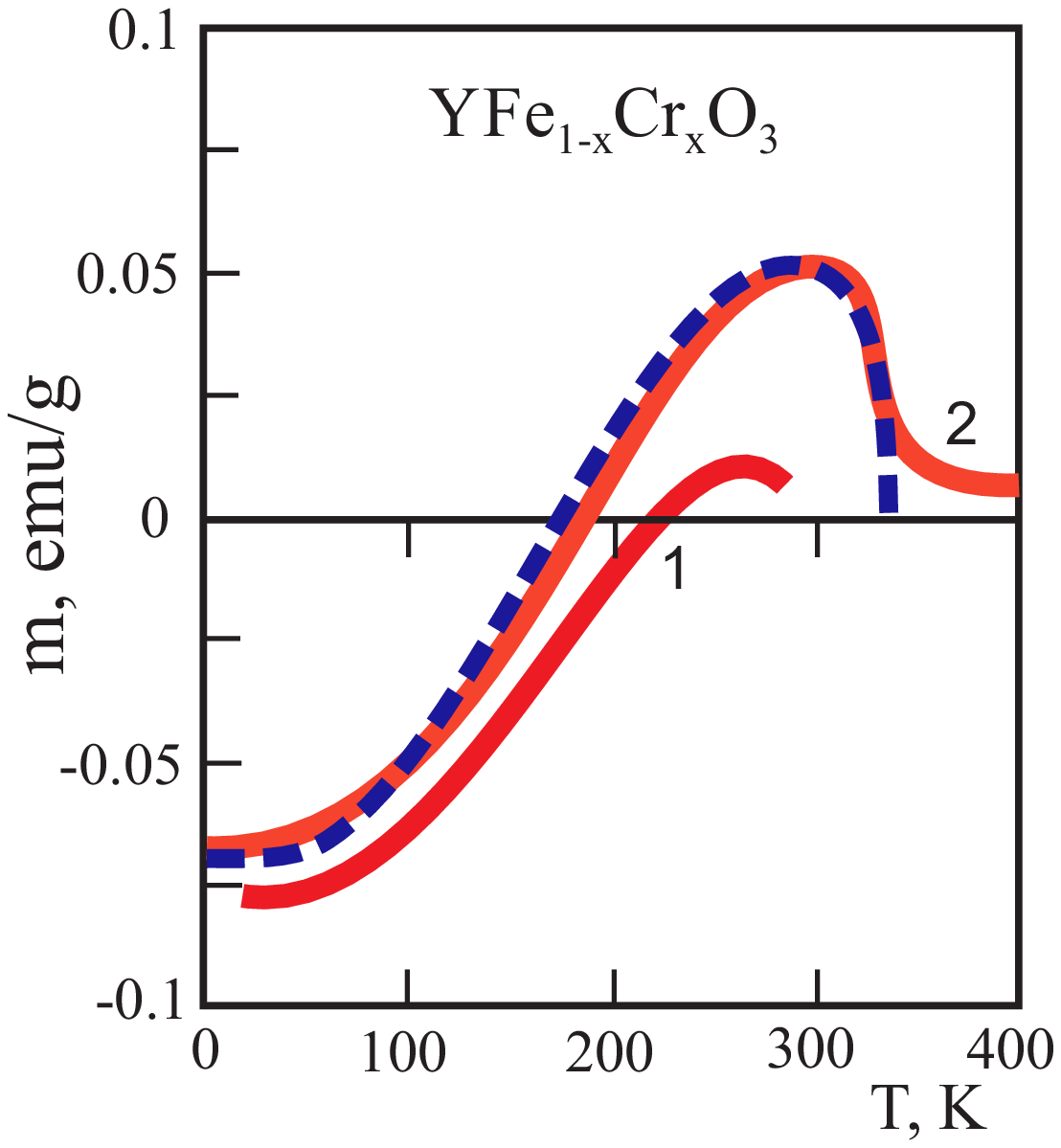} \\ d)}
\end{minipage}
\caption{ a) The MFA phase diagram of weak ferrimagnet $YFe_{1-x}Cr_xO_3$ given $\delta$\,=\,-4; left and right arrows demonstrate the orientation and magnitude of the magnetization for $Fe$- and $Cr$-sublattices, respectively.  The outer and inner thin curves mark the compensation points for the net and partial ($Fe$, $Cr$) magnetization, respectively. Experimental values of $T_N$ for single crystalline and polycrystalline samples are marked by light  and dark circles, respectively.  
b) Concentration dependence of the low-temperature magnetization in $YFe_{1-x}Cr_xO_3$: experimental data (circles)\,\cite{WFIM-1}, the MFA calculations given $\delta$\,=\,-2 and -4; c) Concentration dependence of the magnetization and Fe-, Cr- partial contributions in $YFe_{0.5}Cr_{0.5}O_3$;  d) Temperature dependence of magnetization in $YFe_{1-x}Cr_xO_3$:  solid curves -- experimental data for x\,=\,0.38 (Kadomtseva {\it et al.}, 1977\,\cite{WFIM-1} -- curve 1) and for x\,=\,0.4 (Dasari {\it et al.}, 2012\,\cite{Dasari} -- curve 2), dotted curve -- the MFA calculation for x\,=\,0.4\,\cite{Dasari} given $d_{FeCr}$\,=\,-0.39\,K }
\label{fig4}
\end{figure}

\begin{figure}[t]
\begin{minipage}[h]{0.30\linewidth}
\center{\includegraphics[width=0.5\linewidth]{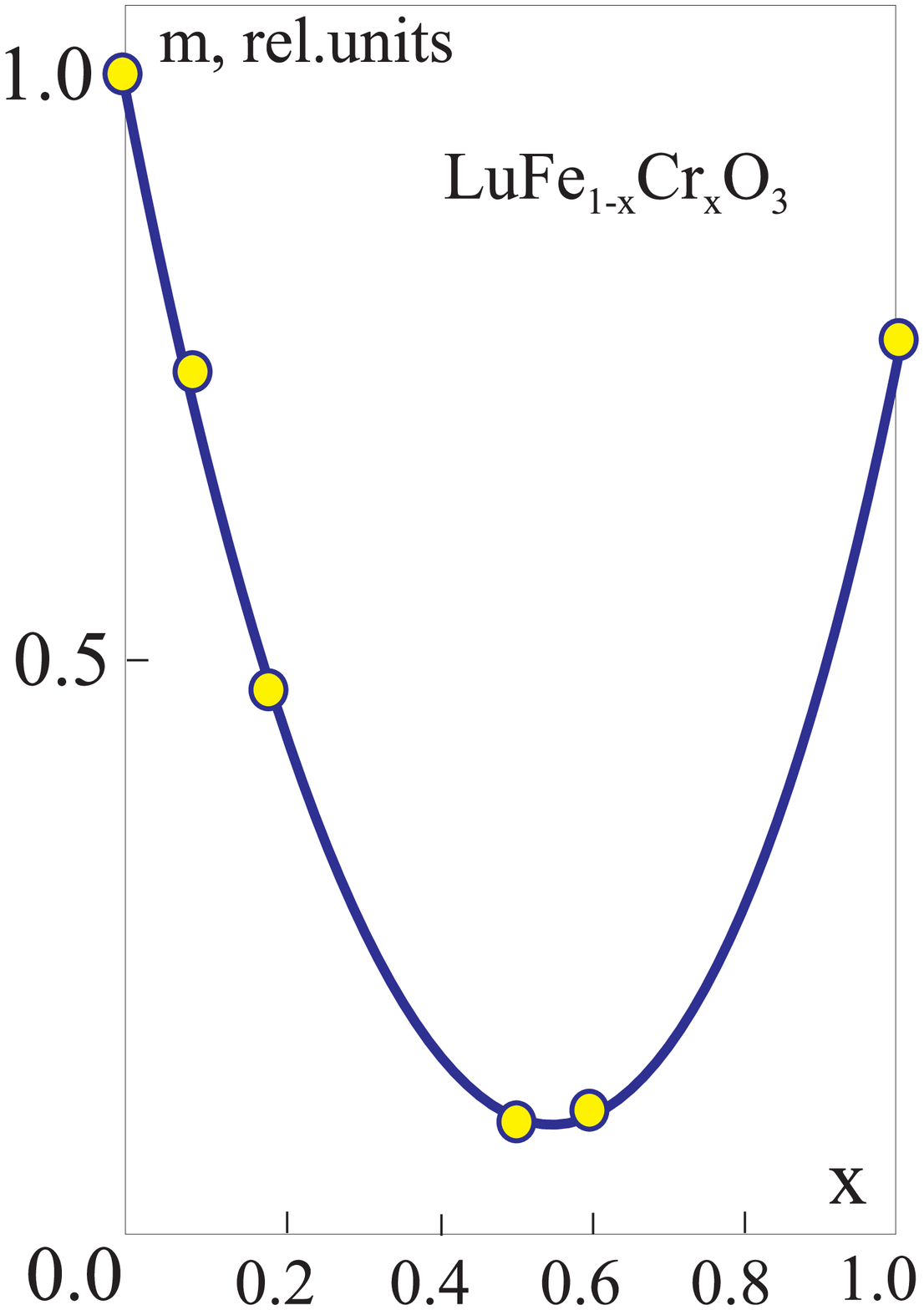} \\ a)}
\end{minipage}
\begin{minipage}[h]{0.40\linewidth}
\center{\includegraphics[width=0.5\linewidth]{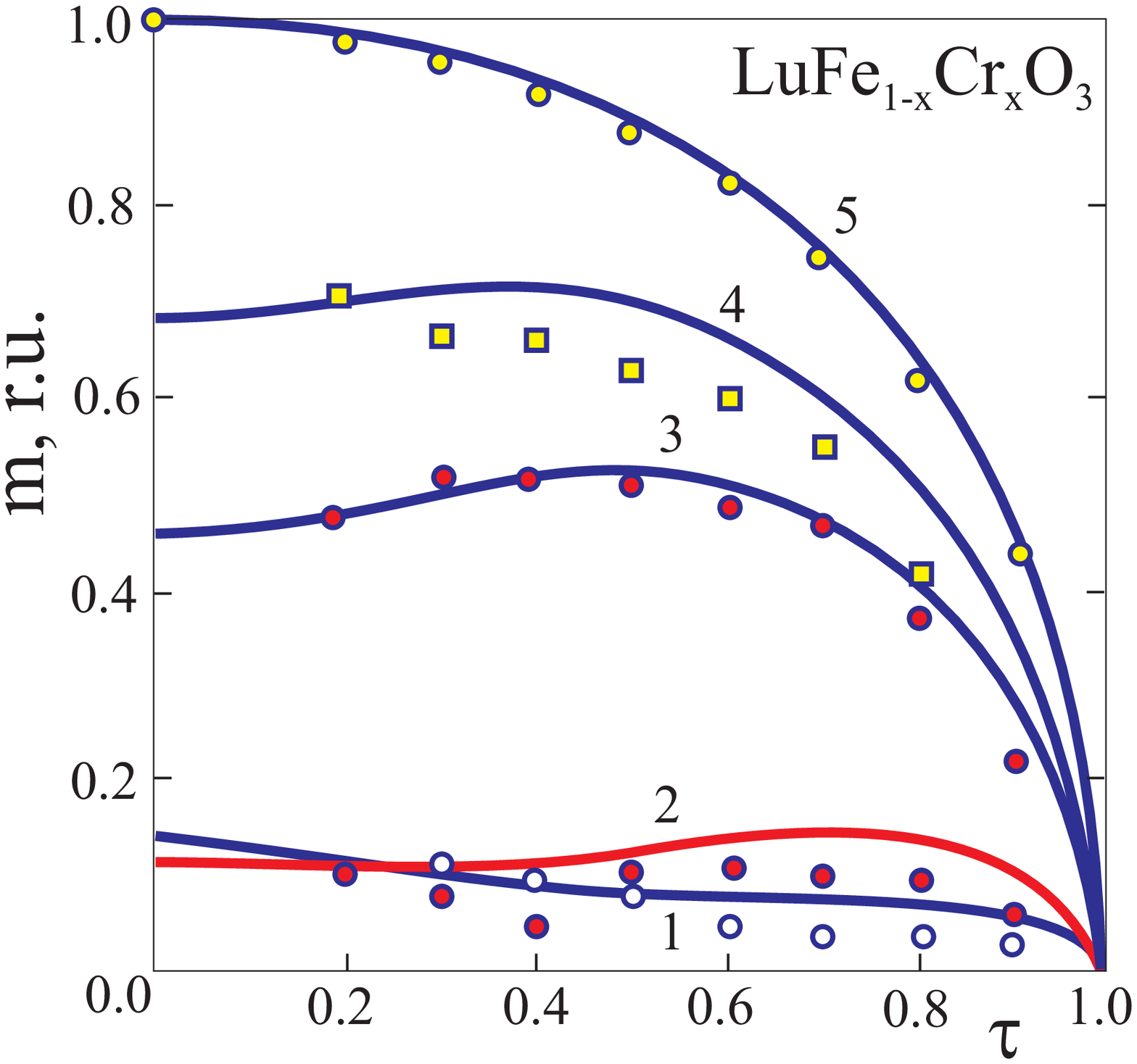} \\ b)}
\end{minipage}
\caption{a) Concentration dependence of the low-temperature ($T$\,=\,77\,K) magnetization in $LuFe_{1-x}Cr_xO_3$: experimental data (circles)\,\cite{LuFeO3}, the MFA calculations (solid curve) given $\delta$\,=\,-1.5. (b) Temperature dependence of magnetization in $LuFe_{1-x}Cr_xO_3$: circles -- experimental data\,\cite{LuFeO3} given $x$\,=\,0.6 (1), 0.5 (2), 0.2 (3), 0.1 (4), 0.0 (5),  solid curves -- the MFA calculations given $\delta$\,=\,-1.5.   }
\label{fig5}
\end{figure}

\begin{figure}[t]
\begin{minipage}[h]{0.30\linewidth}
\center{\includegraphics[width=0.5\linewidth]{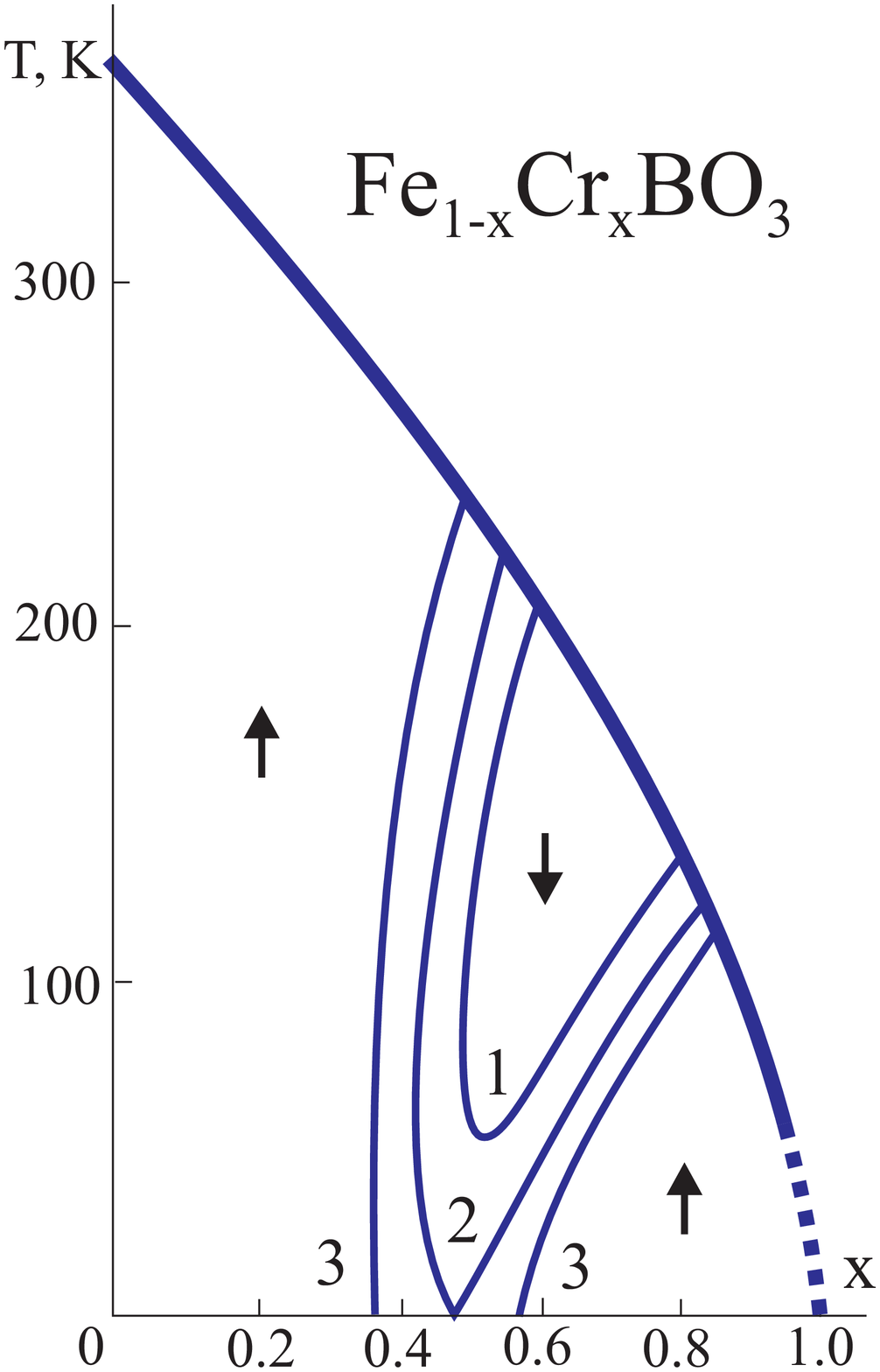} \\ a)}
\end{minipage}
\begin{minipage}[h]{0.35\linewidth}
\center{\includegraphics[width=0.5\linewidth]{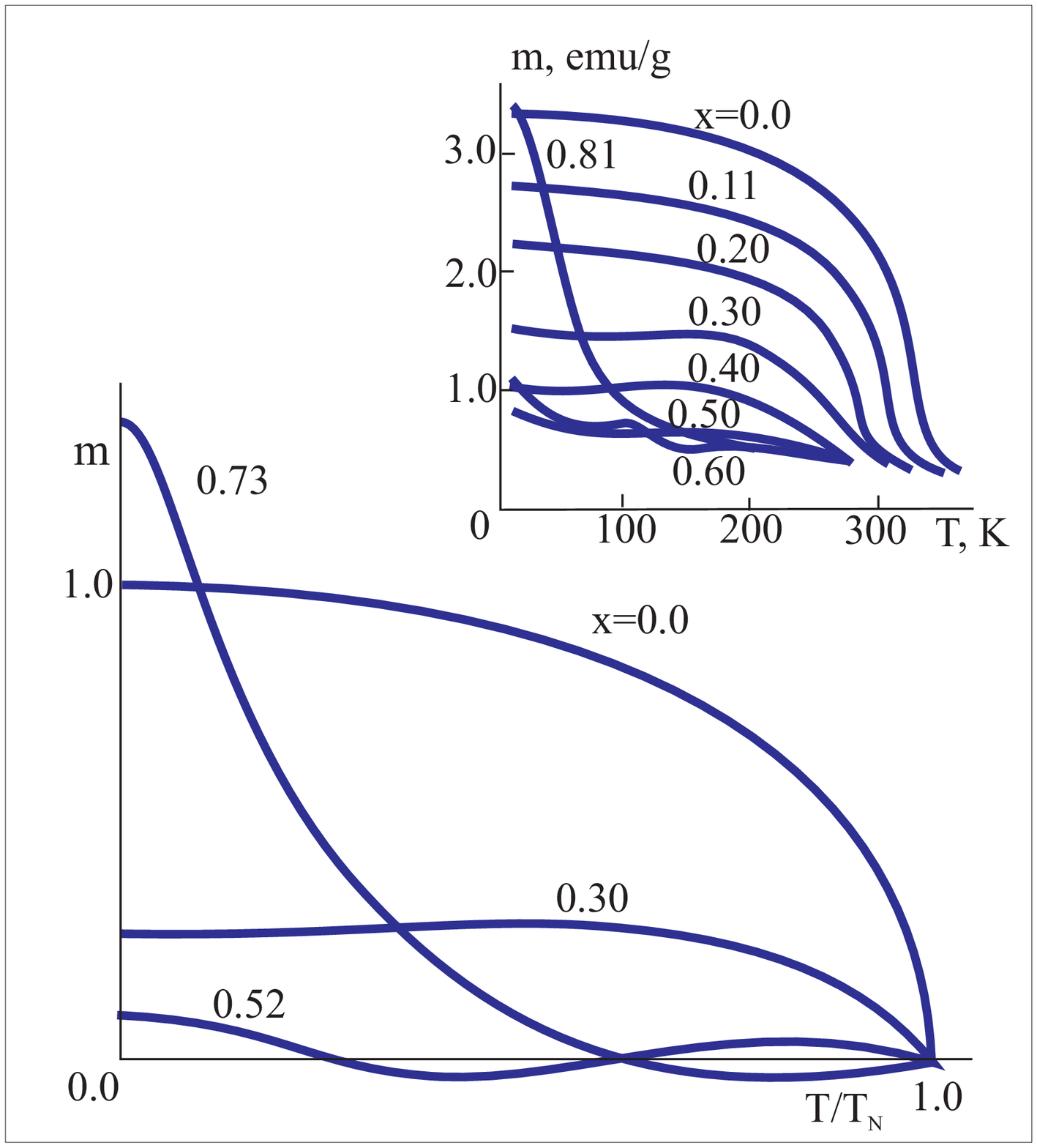} \\ b)}
\end{minipage}
\begin{minipage}[h]{0.30\linewidth}
\center{\includegraphics[width=0.5\linewidth]{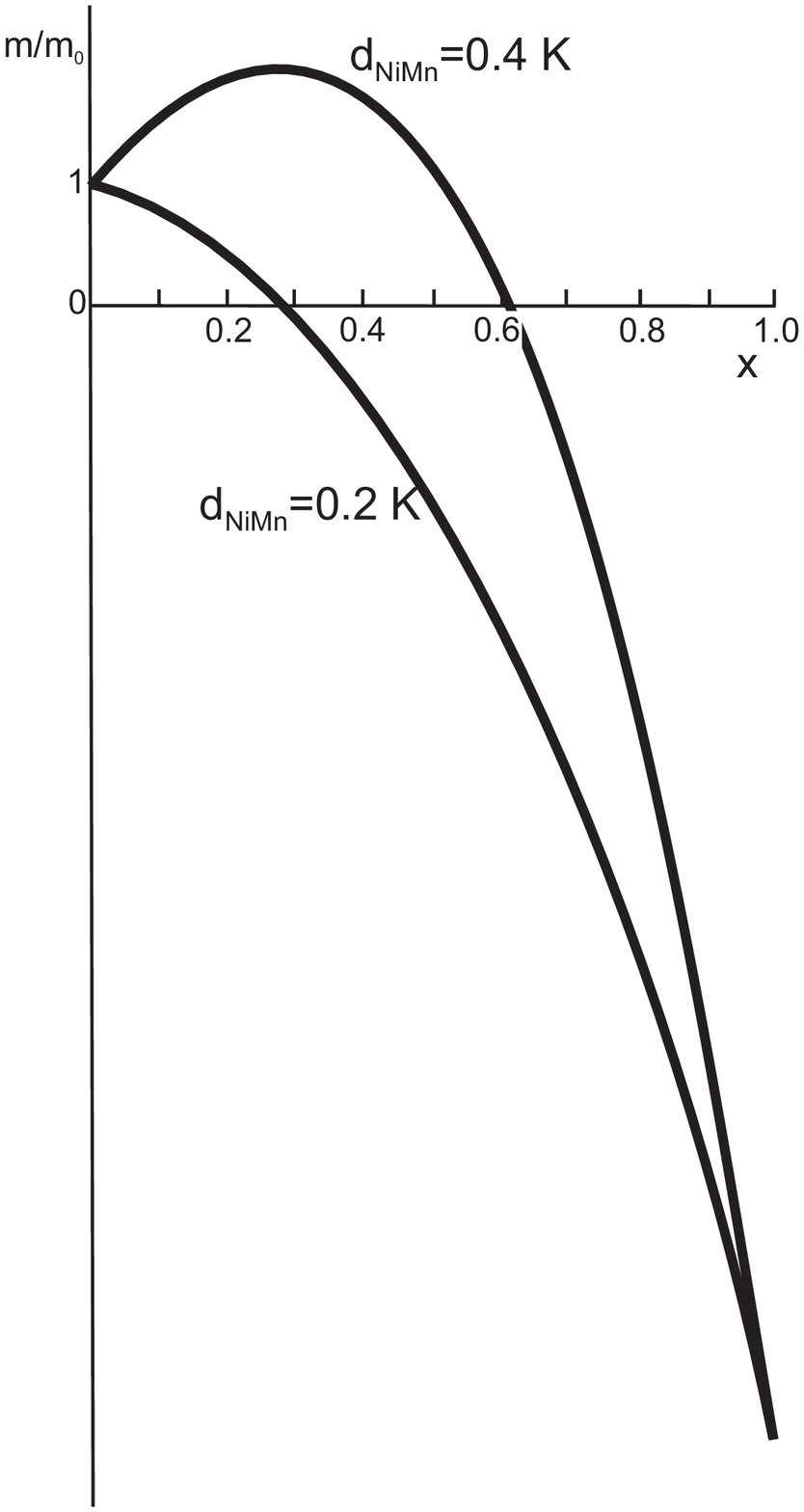} \\ c)}
\end{minipage}
\caption{a) The MFA simulation of the T-x-phase diagram of the weak ferro(ferri)magnet $Fe_{1-x}Cr_xBO_3$\,\cite{WFIM-3} given $I_{FeFe}=I_{FeCr}$=\,-20.3\,K, $I_{CrCr}$\,=\,2.0\,K, arrows point to orientation of the net weak ferromagnetic moment. Curves 1, 2, 3 mark the compensation points given $d_z(FeFe)$\,=\,$d_z(CrCr)$\,=\,0.67\,K, $d_z(FeCr)$\,=\,-0.67\,K (1), -0.75\,K (2), -0.90\,K (3), respectively. b) The MFA simulation of the temperature dependence of the net magnetization in $Fe_{1-x}Cr_xBO_3$\,\cite{WFIM-3} given $d_z(FeFe)$\,=\,$d_z(CrCr)$\,=\,-$d_z(FeCr)$\,=\,0.67\,K at different compositions, the insert shows experimental data from Ref.\,\cite{OHoro} taken at external magnetic field 1\,T. c) The MFA simulation of the concentration dependence of the low-temperature magnetization in $Mn_{1-x}Ni_xCO_3$\,\cite{WFIM-2} given $d_z(MnNi)>d_z^{(0)}(MnNi)$ and $d_z(MnNi)<d_z^{(0)}(MnNi)$, respectively.}
\label{fig6}
\end{figure}

First experimental studies of mixed orthoferrites-orthochromites YFe$_{1-x}$Cr$_x$O$_3$\,\cite{WFIM-1} performed in Moscow State University did confirm theoretical predictions regarding the signs of the Dzyaloshinskii vectors and revealed the {\it weak ferrimagnetic} behavior due to a competition of Fe-Fe, Cr-Cr, and Fe-Cr DM coupling with antiparallel orientation of the mean weak ferromagnetic moments of Fe and Cr subsystems in a wide concentration range. In other words, we have predicted a novel class of mixed 3d systems with competing signs of the Dzyaloshinskii vector, so-called weak ferrimagnets. Weak ferromagnetic moment of the $Cr^{3+}$ impurity ion in orthoferrite $YFeO_3$ can be evaluated as follows
\begin{equation}
{\bf m}_{Cr}=g\mu_BS_{Cr}(2\delta -1){\bf F}\, ,
\end{equation}
where dimensionless parameter
\begin{equation}
	\delta =\frac{d_{CrFe}}{d_{FeFe}}\frac{I_{FeFe}}{I_{CrFe}}
\end{equation}
does characterize a relative magnitude of the impurity-matrix DM coupling. By comparing $m_{Cr}$ with that of the matrix value ${\bf m}_{Fe}=g\mu_BS_{Fe}{\bf  F}$ we see that the weak ferromagnetic moment for the $Cr$  impurity is antiparallel to that of the $Fe$ matrix even for positive but small $\delta <1/2$. However, the effect is more pronounced for negative $\delta$, that is for different signs of $d_{CrFe}$ and $d_{FeFe}$.

Results of a simple mean-field calculation presented  in Figs.\,\ref{fig4}-\ref{fig6} show that the weak ferrimagnets such as RFe$_{1-x}$Cr$_x$O$_3$\,\cite{WFIM-1}, Mn$_{1-x}$Ni$_x$CO$_3$\,\cite{WFIM-2}, Fe$_{1-x}$Cr$_x$BO$_3$\,\cite{WFIM-3} do reveal very nontrivial concentration and temperature dependencies of magnetization, in particular, the compensation point(s).
In Fig.\,\ref{fig4}a,b,c we do present the MFA "weak ferrimagnetic"\, phase diagram,  concentration dependence of the low-temperature net magnetization and $Fe$, $Cr$ partial contributions  in $YFe_{1-x}Cr_xO_3$ calculated at constant value of  $\delta$\,=\,-4. Comparison with experimental data for the low-temperature net magnetization\,\cite{WFIM-1} and the MFA calculations with $\delta$\,=\,-2 (Fig.\,\ref{fig4}b) points to a reasonably well agreement everywhere except $x\sim$\,0.5, where $\delta$ parameter seems to be closer to $\delta$\,=\,-3.  
In Fig.\,\ref{fig4}d we compare first pioneering experimental data for the temperature dependence of magnetization $m(T)$ in weak ferrimagnet $YFe_{1-x}Cr_xO_3$ ($x$\,=\,0.38) (Kadomtseva {\it et al.}, 1977\,\cite{WFIM-1} -- curve 1) with recent  data for a close composition with $x$\,=\,0.4 (Dasari {\it et al.}, 2012\,\cite{Dasari} -- curve 2). It is worth noting that recent MFA calculations by Dasari {\it et al}.\,\cite{Dasari}  given $d_{FeCr}$\,=\,-0.39\,K provide very nice description of $m(T)$ at $x$\,=\,0.4. Note that the authors\,\cite{Dasari} found a rather strong dependence of the $d_{FeCr}$ parameter on the concentration $x$.
Interestingly, that concentration and temperature dependencies of magnetization in $LuFe_{1-x}Cr_xO_3$ are nicely described by a simple MFA model given constant value of $\delta$\,=\,-1.5 (Fig.\,\ref{fig5}a,b\,\cite{LuFeO3}). 

Fig.\,\ref{fig6}b shows a calculated phase diagram of the trigonal weak ferrimagnet Fe$_{1-x}$Cr$_x$BO$_3$\,\cite{WFIM-3}.
Temperature-dependent magnetization studies from 4.2 to 600\,K have been made for the solid solution system Fe$_{1-x}$Cr$_x$BO$_3$ where $0\leq x \leq $\,0.95\,\cite{OHoro}. A rapid decrease is observed in the saturation magnetization with increasing $x$ at 4.2\,K up to 0.40, after which a broad compositional minimum is found up to x\,=\,0.60. Compositions in the range of 0.40$\leq x \leq$0.60 display unusual magnetization behavior as a function of temperature in that maxima and minima are present in the curves below the Curie temperatures. Fig.\,\ref{fig6}b shows a nice agreement between experimental data\,\cite{OHoro} and our MFA calculations\,\cite{WFIM-3}.

At variance with the $d^5-d^3$ (Fe-Cr) mixed systems such as $YFe_{1-x}Cr_xO_3$ or $Fe_{1-x}Cr_xBO_3$ the manifestation of different DM couplings $Fe-Fe$, $Cr-Cr$, and $Fe-Cr$ in $(Fe_{1-x}Cr_x)_2O_3$  is all the more surprising because of different magnetic structures of the end compositions, $\alpha-Fe_{2}O_3$ and $Cr_{2}O_3$ and emergence of a nonzero DM coupling for the three-corner-shared $FeO_6$ and $CrO_6$ octahedra, "forbidden"\, for $Fe-Fe$ and $Cr-Cr$ bonding. All this makes magnetic properties of mixed compositions $(Fe_{1-x}Cr_x)_2O_3$ to be very unusual\,\cite{WFIM-4}.

It should be noted that in the Fe-Cr mixed systems  we are really dealing with the two concentration compensation points. 

At variance with the $d^5-d^3$ (Fe-Cr) mixed systems such as $YFe_{1-x}Cr_xO_3$ or $Fe_{1-x}Cr_xBO_3$, where the two concentration compensation points do occur given rather large $d_{FeCr}$ parameter, in the $d^5-d^8$ (Mn$^{2+}$-Ni$^{2+}$) systems we have the only concentration compensation point irrespective of the  $d_{MnNi}$ parameter. However, the character of the concentration dependence of the weak ferrimagnetic moment $m(x)$ depends strongly on its magnitude.
Given the increasing concentration the $m(x)$  is first rising or falling  with $x$ depending on whether $d_{MnNi}$ greater than, or less than $d_{MnNi}^{(0)}=(1+\frac{S_{Mn}}{S_{Ni}})\frac{I_{MnNi}}{2I_{MnMn}}d_{MnMn}$. Figure (\ref{fig6}c) does clearly demonstrate this feature. 

It should be noted that just recently Dmitrienko {\it et al.}\,\cite{Dmitrienko-EASTMAG} have first discovered experimentally that in accordance with our theory (see Table\,\ref{tablesign}) the sign of the Dzyaloshinskii vector in MnCO$_3$ ($d^5$-$d^5$) coincides with that of in FeBO$_3$ ($d^5$-$d^5$), whereas NiCO$_3$ ($d^8$-$d^8$)   demonstrates the opposite sign.

 The systems with  competing DM coupling were extensively investigated up to the end of 80ths including specific features of the DM coupling in some rare-earth ferrite-chromites, fluorine-substituted orthoferrites, disordered magnetic oxides\,\cite{WFIM+}.

 Recent renewal of interest to weak ferrimagnets as systems with the concentration and/or temperature compensation point was stimulated by  the perspectives of the applications in magnetic memory (see, e.g., Refs.\,\cite{Mao,Dasari} and references therein). 
Weak ferrimagnet YFe$_{0.5}$Cr$_{0.5}$O$_3$ exhibits magnetization reversal at low applied fields. Below a compensation temperature (T$_{comp}$), a tunable bipolar switching of magnetization is demonstrated by changing the magnitude of the field while keeping it in the same
direction. The compound also displays both normal and inverse magnetocaloric effects above and below 260\,K, respectively.
Recently the exchange bias (EB) effect was studied in LuFe$_{0.5}$Cr$_{0.5}$O$_3$ ferrite-chromite\,\cite{PRB_2018,JMMM_2018} which is a weak ferrimagnet below T$_N$\,=\,265\,K, exhibiting antiparallel orientation of the mean weak ferromagnetic moments (FM) of the Fe and Cr sublattices due to opposite sign of the  Fe-Cr Dzyaloshinskii vector as compared with that of Fe-Fe and Cr-Cr. 
Weak ferrimagnets can exhibit the tunable exchange bias effect\,\cite{Bora} and have potential applications in electromagnetic devices\,\cite{Mao}. Combining magnetization reversal effect with magnetoelectronics can exploit tremendous technological potential for device applications, for example, thermally assisted magnetic random access memories, thermomagnetic switches and other multifunctional devices, in a preselected and convenient manner.
Nowadays a large body of magnetic materials might be addressed as systems with competing antisymmetric exchange\,\cite{Kumar}, including novel class of mixed helimagnetic B20 alloys such as Mn$_{1-x}$Fe$_x$Ge where the helical nature of the main ferromagnetic spin structure is determined by a competition of the DM couplings Mn-Mn, Fe-Fe, and Mn-Fe. Interestingly, that the magnetic chirality in the mixed compound changes its sign at $x_{cr}\approx$\,0.75, probably due to different sign of the Dzyaloshinskii vectors for Mn-Mn and Fe-Fe pairs\,\cite{MnFeGe}.
 
 \section{Determination of the sign of the Dzyaloshinskii vector}


Determination of the "sign" of the Dzyloshinskii vector is of a fundamental importance from the standpoint of the microscopic theory of the DM coupling. For instance, this sign determines the handedness of spin helix in crystals with the noncentrosymmetric B20 structure.

How to measure the sign of the DM interaction in weak ferromagnets? According to Ref.\,\cite{Ozhogin}, an answer to this question can be given by determining experimentally the direction of rotation of the antiferromagnetism vector ${\bf l}$ around the magnetic field ${\bf H}$  in the geometry ${\bf H}\parallel {\bf d}\parallel$easy axis. However, as was pointed out later (see Ref.\,\cite{Chep}), the M\"{o}ssbauer experiment  on easy-axis hematite did not give an unambiguous result.

According to Dmitrienko {\it et al.}\,\cite{Dmitrienko}, first of all, a strong enough magnetic field should be applied to obtain the single domain state where the DM coupling pins antiferromagnetic ordering to the crystal lattice. Next, single crystal diffraction methods sensitive both to oxygen coordinates and to the phase of antiferromagnetic ordering should be used. In other words, one should observe those Bragg reflections $hkl$ where interference between magnetic scattering on Mn atoms and nonmagnetic scattering on oxygen atoms is significant. There are three suitable techniques: neutron diffraction, M\"{o}ssbauer $\gamma$-ray diffraction, and resonant x-ray scattering. The sign of the DM vector in weak ferromagnetic FeBO$_3$ was deduced from observed interference between resonant X-ray scattering and magnetic X-ray scattering\,\cite{Dmitrienko}.

The authors in Ref.\,\cite{Chep} claimed that the character of the field-induced transition from an antiferromagnetic phase to a canted phase in  cobalt fluoride CoF$_2$ is due to the "sign" of the Dzyaloshinskii interaction, and this affords an opportunity to determine experimentally the "sign" of the Dzyaloshinskii interaction. However, in fact they addressed a symmetric Dzyaloshinskii interaction that is magnetic anisotropy
$$
V_{sym}=-D(m_xl_y+m_yl_x)
$$
 rather than antisymmetric DM coupling.

\subsection{Ligand NMR in weak ferromagnets and first determination of the sign of the Dzyaloshinskii vector} 

As was firstly shown in our paper\,\cite{sign} reliable local information on the sign of the Dzyaloshinskii vector, or to be exact, that of the Dzyaloshinskii parameter $d_{12}$, can be extracted from the ligand NMR data in weak ferromagnets. The procedure was described in details for $^{19}$F NMR data in weak ferromagnet FeF$_3$\,\cite{sign}. 

\begin{figure}[t]
\centering
\includegraphics[width=8.5cm,angle=0]{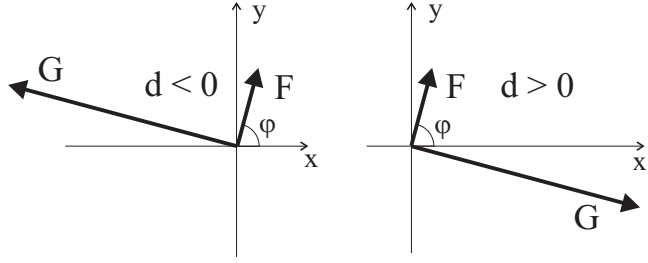}
\caption{Two mutual orientations of $\bf F$ and $\bf G$ vectors in a basal plane of FeF$_3$.}
\label{fig7}
\end{figure}

The $F^-$ ions in the unit cell of $FeF_3$ occupy six positions\,\cite{Hepworth}.
In a trigonal basis these are $\pm(x,1/2-x,1/4),\quad\pm(1/2-x,1/4,x),\quad
\pm(1/4,x,1/2-x)$, that correspond to i) $\;\pm(3p(x-1/4),\sqrt{3}p
(1/4-x),c/4),$ ii) $\;\pm(3p(1/4-x),\sqrt{3}p(1/4-x),c/4),$ 
and iii) $\;\pm(0,2\sqrt{3}p(x-1/4),c/4),$ in an orthogonal basis with $O_z\parallel C_3$ and $O_x\parallel C_2$. Each $F^-$ ion is surrounded by two $Fe^{3+}$ from different magnetic sublattices. Hereafter we introduce basic ferromagnetic {\bf F} and antiferromagnetic {\bf G} vectors:
\begin{equation}
\label{m6}
2S{\bf F}={\bf S_1}+{\bf S_2}, 2S{\bf G}={\bf S_1}-{\bf S_2},
{\bf F^2}+{\bf G^2}=1,
\end{equation}
where $Fe_1^{3+}$ and  $Fe^{3+}_2$ occupy positions (1/2,1/2,1/2) and (0,0,0), respectively. $FeF_3$ is an easy plane weak ferromagnet with {\bf F} and {\bf G} lying in (111) plane  with ${\bf F}\bot{\bf G}$. 
The two possible variants of the mutual orientation of the {\bf F} and {\bf G}
vectors in the basis plane,   tentatively called as "left" and "right", respectively,  are shown in Fig.\,\ref{fig7}. 
The DM energy per $Fe^{3+}-F^--Fe^{3+}$ bond can be written as follows
\begin{equation}
	E_{DM}=-2S^2d_z(12)(F_xG_y-F_yG_x)=-\frac{4\sqrt{3}}{l^2}p^2(x+\frac{1}{4})d(\theta )=+0.78S^2d(\theta )(F_xG_y-F_yG_x) \, .
\end{equation}
In other words, the "left" and "right" orientations of basic vectors are realized at $d(\theta )<0$ and $d(\theta )>0$, respectively.

Absolute magnitude of the ferromagnetic vector is numerically equals to an overt canting angle which can be found making use of familiar values of the Dzyaloshinskii field: $H_D=48.8$\,kOe and exchange field: $H_E=4.4\cdot
10^3$\,kOe\,\cite{Prozorova} as follows
\begin{equation}
F=H_D/2H_E\simeq 5.5\cdot 10^{-3}.
\end{equation}
If we know the Dzyaloshinskii field we can calculate the $d(\theta )$ parameter as follows
\begin{equation}
	H_D=\frac{6S}{g\mu_B}|d_z(12)|=\frac{6S}{g\mu_B}0.39|d(\theta )|=48.8\,kOe \, ,
\end{equation}
that yields $|d(\theta )|\cong$\,1.1\,K that is three times smaller than in YFeO$_3$.

The local field on the nucleus of the nonmagnetic $F^-$ anion in weak ferromagnet FeF$_3$, induced by neighboring magnetic S-type ion ($Fe^{3+}, Mn^{3+},\ldots$) can be written as follows\,\cite{Turov}
\begin{equation}
\label{m1}
{\bf H}=-\frac{1}{\gamma_n}\hat A{\bf S}
\end{equation}
($\gamma_n$ is a gyromagnetic ratio, $\gamma_n$=4.011 MHz/kOe, ${\bf S}$ is the spin moment of the magnetic ion), where the tensor of the transferred hyperfine interactions (HFI)
$\hat A$ consists of two terms: i) an isotropic contact term with $A_{ij}=A_s\delta_{ij}$
\begin{equation}
\label{m2}
A_s=\frac{f_s}{2S}A^{(0)}_s,\qquad A^{(0)}_s=\frac{16}{3}\pi\mu_B\gamma_n
|\varphi_{2s}(0)|^2\,  ;
\end{equation}
ii) anisotropic term with
\begin{equation}
A_{ij}=A_p(3n_in_j-\delta_{ij}),
\end{equation}
where ${\bf n}$ is a unit vector along the nucleus-magnetic ion bond and the 
 $A_p$ parameter includes the dipole and covalent contributions
\begin{equation}
\label{m4}
A_p=A_p^{cov}+A_d,
\end{equation}
\begin{equation}
\label{m5}
A_p^{cov}=\frac{(f_\sigma-f_\pi)}{2S}A_p^{(0)},
A_p^{(0)}=\frac{4}{5}\mu_B\gamma_n\langle\frac{1}{r^3}\rangle_{2p},
A_d=\frac{g_s\mu_B\gamma_n}{R^3}.
\end{equation}
Here $f_{s,\pi,\sigma}$ are parameters for the spin density transfer: magnetic ion - ligand along the proper $s-, \sigma-, \pi$-bond;
$|\varphi_{2s}(0)|^2$ is a probability density of the $2s$-electron on nucleus; $\langle \frac{1}{r^3}\rangle_{2p}$ is a radial average.

The transferred HFI for $\vphantom{F}^{19}F$ in fluorides are extensively studied by different techniques, NMR, ESR, and ENDOR\,\cite{Turov}. For $\vphantom{F}^{19}F$ one observes large values both of $A_s^{(0)}$ and $A_p^{(0)}$; $A_s^{(0)}=4.54\cdot~10^4$,
$A_p^{(0)}=1.28\cdot~10^3$ MHz\,\cite{Turov}, together with the 100\% abundance, nuclear spin $I=1/2$, and large   gyromagnetic ratio
this makes the study of the transferred HFI to be simple and available one.

Contribution of the isotropic and anisotropic transferred HFI to the local field on the $^{19}F$ can be written as follows
\begin{equation}
{\bf H}(iso)=-\frac{2S}{\gamma_n}A_s{\bf F}=a_F{\bf F},
{\bf H}(aniso)={\bf \stackrel{\leftrightarrow}{a}}{\bf G}, \hat a=-\frac{2S}{\gamma_n}({\bf \stackrel{\leftrightarrow}{A}}(1)-{\bf \stackrel{\leftrightarrow}{A}}(2)).
\end{equation}
The $A_s$ and $A_p$ parameters we need to calculate parameter $a_F$ and the HFI anisotropy tensor ${\bf \stackrel{\leftrightarrow}{a}}$ that is to calculate the "ferro-" and "antiferro-" contributions to $H$ one can find in the literature data for the pair $\vphantom{F}^{19}F-Fe^{3+}$. For instance, in $KMgF_3:Fe^{3+}$ ($R_{MgF}$\,=\,1.987\,\AA)\,\cite{Hall} $A_s=+72, A_p=+18$\,MHz, in $K_2NaFeF_6$ ($R_{FeF}$\,=\,1.91\,\AA),
 in $K_2NaAlF_6:Fe^{3+}$ $A_s$\,=\,+70.17, $A_p$\,=\,+20.34\,MHz\,\cite{Adam}. Thus, we expect in $FeF_3$ $|a_F|\sim 350\div
360$\,MHz $(a_F<0)$ and $H(iso)\simeq 2$\,MHz ($\simeq 0.5$\,kOe).

Calculated values of the components of the the local field anisotropy tensor $\hat a$ for different nuclei $\vphantom{F}^{19}F$ are listed in Table\,\ref{table3}. 
\begin{center}
\begin{table}
\caption{Values of the components of the the local field anisotropy tensor $\hat a$ for nuclei $^{19}F$ in positions 1, 2, 3 in $FeF_3$ ($a$\,=\,5.333\,\AA\,, $\alpha\,=\,57.72^{\circ}, p=(a/\sqrt{3})\sin\alpha$\,=\,1.486\,\AA\,,
$l$\,\,=1.914\,\AA\,, $x$\,=\,-\,0.157\,\cite{Jacobson})}
\centering
\begin{tabular}{|c|c|c|c|}
\hline
$a_{ij}$ & $^{19}F_1$ & $^{19}F_2$ &
$\vphantom{F}^{19}F_3$\\
\hline
$a_{xx}$ & 0 & $\frac{45p^2}{l^2}(x+\frac{1}{4})A_p$ & 
-$\frac{45p^2}{l^2}(x+\frac{1}{4})A_p$\\
&& $=2.53A_p$ & $=-2.53A_p$\\ \hline
$a_{yy}$ & 0 & $-\frac{45p^2}{l^2}(x+\frac{1}{4})A_p$ &
$\frac{45p^2}{l^2}(x+\frac{1}{4})A_p$\\
&& $=-2.53A_p$ & $=2.53A_p$\\ \hline
$a_{zz}$ & 0 & 0 & 0 \\ \hline
$a_{xy}$ & $\frac{30\sqrt{3}p^2}{l^2}(x+\frac{1}{4})A_p$ &
$-\frac{15\sqrt{3}p^2}{l^2}(x+\frac{1}{4})A_p$ &
$-\frac{15\sqrt{3}p^2}{l^2}(x+\frac{1}{4})A_p$\\
& $=2.92A_p$ & $=-1.46A_p$ & $-1.46A_p$\\ \hline
$a_{xz}$ & 0 &$\frac{15pc}{4l^2}(x+\frac{1}{4})A_p$  &
$-\frac{15pc}{4l^2}(x+\frac{1}{4})A_p$\\
&& $=1.89A_p$ & $-1.89A_p$\\ \hline
$a_{yz}$ & $-\frac{15\sqrt{3}pc}{2l^2}(x+\frac{1}{4})A_p$ &
$\frac{15\sqrt{3}pc}{4l^2}(x+\frac{1}{4})A_p$ &
$\frac{15\sqrt{3}pc}{4l^2}(x+\frac{1}{4})A_p$\\
& $=2.18A_p$ & $=1.09A_p$ & $=1.09A_p$\\ \hline
\end{tabular}
\label{table3}
\end{table}
\end{center}

In the absence of an external magnetic field the NMR frequencies for $\vphantom{F}^{19}F$ in positions 1, 2, 3 can be written as follows 
\begin{eqnarray}{c}
\label{m9}
\nu^2=\gamma_n^2[(\hat a{\bf G})^2+(a_f{\bf F})^2+2a_F{\bf F}\hat a{\bf G}]=\\
\gamma_n^2(a_{xy}^2+a_F^2F^2\pm 2a_Fa_{xy}F)+
\gamma_n^2(a_{yz}^2\mp 4a_Fa_{xy}F)\left\{
\parbox{3cm}{$\cos^2\varphi\nonumber\\
\cos^2(\varphi+60^o)\\ \cos^2(\varphi-60^o)$}\right.
\end{eqnarray}
where the $a_{xy},a_{yz}$ components are taken for $\vphantom{F}^{19}F_1$ in position 1; $\varphi$ is an azimuthal angle for ferromagnetic vector {\bf F} in the basis plane. The formula (\ref{m9}) and Table \ref{table3} do provide a direct linkage between the   $\vphantom{F}^{19}\!F$  NMR frequencies and parameters of the crystalline $(p,c,x,l)$ and magnetic $(F,\varphi,\pm)$ structures. As of particular importance one should note a specific dependence of the   $\vphantom{F}^{19}\!F$  NMR frequencies on mutual orientation of the ferro- and antiferromagnetic vectors or the sign of the Dzyaloshinskii vector: upper signs in  (\ref{m9}) correspond to "right orientation" while lower signs do to "left orientation" as shown in Fig.\ref{fig7}.

\begin{figure}[t]
\centering
\includegraphics[width=8.5cm,angle=0]{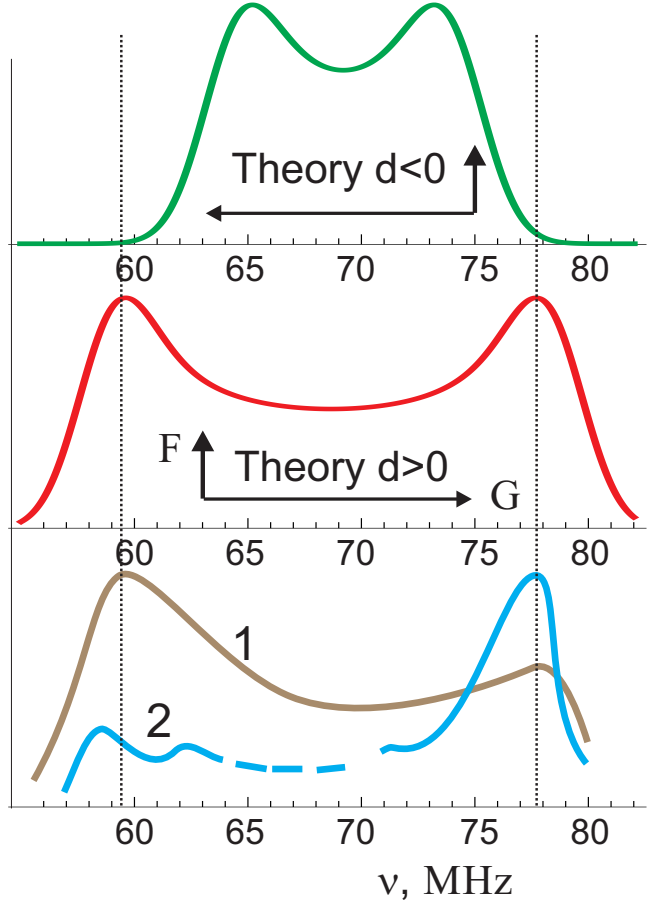}
\caption{Comparison of simulated (upper panels) and experimental (bottom panel) zero-field $^{19}F$  NMR spectra in $FeF_3$.}
\label{fig8}
\end{figure}	
	  
For minimal and maximal values of the $^{19}F$ NMR frequencies we have 
\begin{eqnarray}{}
\label{m10}
\nu^{\pm}_{min}=\gamma_n[a_{xy}^2+a_{F}^2\pm 2a_Fa_{xy}F]^{1/2},\nonumber\\
\nu^{\pm}_{max}=\gamma_n[a_{xy}^2+a_{yz}^2+a_F^2F^2\mp 2a_Fa_{xy}F]^{1/2}.
\end{eqnarray}
 
Taking into account smallness of isotropic HFI contribution, signs of $a_F$ and $A_{xy}$ we arrive at estimations
\begin{eqnarray}{}
\label{m11}
\nu^{\pm}_{min}\simeq\gamma_n(|a_{xy}|\mp |a_FF|)=
2.92A_p\mp|a_FF|,\nonumber\\   
\nu^{\pm}_{max}\simeq\gamma_n\left([a_{xy}^2+a_{yz}^2]^{1/2}\pm
\frac{|a_{xy}}{[a_{xy}^2+a_{yz}^2]^{1/2}}|a_FF|\right)=
3.65A_p\pm 0.8|a_FF|.
\end{eqnarray}
Thus
\begin{equation}
\label{m12}
(\nu_{max}-\nu_{min})^{\pm}=0.68A_p\pm 1.8|a_FF|.
\end{equation}

By using the $A_s$ and $A_p$ values, typical for $\vphantom{F}^{19}\!F
-Fe^{3+}$ bonds\,\cite{Hall,Adam} we get 
\begin{equation}
\label{m13}
\nu^+_{min}=57.6, \nu^+{max}=75.7, (\nu_{max}-\nu_{min})^+=18.1 \mbox{MHz}
\end{equation}
given "right orientation" of {\bf F} and {\bf G} (Fig.\ref{fig7})
\begin{equation}
\label{m14}
\nu^-_{min}=61.4, \nu^-_{max}=72.7, (\nu_{max}-\nu_{min})^-=11.3 \mbox{MHz}
\end{equation}
given "left orientation" of {\bf F} and {\bf G} (Fig.\ref{fig7}) .

 The zero-field $^{19}F$  NMR spectrum for single-crystalline samples of $FeF_3$ we simulated on assumption of negligibly small in-plane anisotropy\,\cite{Wolfe} is shown in Fig.\,\ref{fig8} for two different mutual orientations of ${\bf F}$ and ${\bf G}$ vectors. For a comparison in Fig.\,\ref{fig8} we adduce the experimental NMR spectra for polycrystalline samples of $FeF_3$\,\cite{Petrov,Zalesskii}, which are characterized by the same boundary frequencies despite rather varied shape. Obviously, the theoretically simulated NMR spectrum does nicely agree with the experimental ones only for "right" mutual orientations of ${\bf F}$ and ${\bf G}$ vectors, or $d(FeFe)>0$, in a full accordance with our theoretical sign predictions (see Table\,\ref{tablesign}). 

The same result, $d(FeFe)>0$ follows from the the magnetic $x$-ray scattering amplitude measurements in the weak ferromagnet FeBO$_3$\,\cite{Dmitrienko}.
 
\subsection{Sign of the Dzyaloshinskii vector in F$\mbox{e}$BO$_3$ and $\alpha$-F$\mbox{e}_2$O$_3$}
Making use of structural data for FeBO$_3$\,\cite{Diehl} we can calculate the $z$-component of the Dzyaloshinskii vector for Fe$_1$-O-Fe$_2$ pair, with Fe$_{1,2}$ in positions (1/2,1/2,1/2), (0,0,0), respectively, as follows:
\begin{equation}
	d_z(12)=d_{12}(\theta )\left[{\bf r}_1\times{\bf r}_2\right]_z=+\frac{1}{3}(\frac{1}{2}-x_h)\frac{ab}{l^2}d_{12}(\theta )\approx +0.61\,d_{12}(\theta ) \, ,
\end{equation}
where $a$\,=\,4.626\,\AA\,, $b$\,=\,8.012\,\AA\, are parameters of the orthohexagonal unit cell, $x_h$\,=\,0.2981 oxygen parameter, $l$\,=\,2.028\,\AA\, is a mean Fe-O separation\,\cite{Diehl}.

Similarly to FeF$_3$ the DM energy per $Fe^{3+}-O^{2-}-Fe^{3+}$ bond can be written as follows
\begin{equation}
	E_{DM}=d_z(12)\left[{\bf S}_1\times{\bf S}_2\right]_z = -2S^2d_z(12)(F_xG_y-F_yG_x)=+2\cdot 0.61\cdot S^2d_{12}(\theta )(F_xG_y-F_yG_x) \, .
\end{equation}
In other words, the "left" and "right" orientations of basic vectors are realized at $d(\theta )<0$ and $d(\theta )>0$, respectively.

Absolute magnitude of the ferromagnetic vector  equals numerically to an overt canting angle which can be found making use of familiar values of the Dzyaloshinskii field: $H_D\approx 100$\,kOe and exchange field: $H_E\approx 3.0\cdot
10^3$\,kOe\,\cite{Kotyuzhanskii,Diehl} as follows
\begin{equation}
F=H_D/2H_E\simeq 1.7\cdot 10^{-2}.
\end{equation}
If we know the Dzyaloshinskii field we can calculate the $d_{12}(\theta )$ parameter as follows
\begin{equation}
	H_D=\frac{6S}{g\mu_B}|d_z(12)|=\frac{6S}{g\mu_B}0.61|d(\theta )|=100\,kOe \, ,
\end{equation}
that yields $|d(\theta )|\cong$\,1.5\,K that is two times smaller than in YFeO$_3$. The difference can be easily explained, if one compares the superexchange bonding angles in FeBO$_3$ ($\theta \approx 125^{\circ}$) and YFeO$_3$ ($\theta \approx 145^{\circ}$), that is $cos\theta (FeBO_3)/cos\theta (YFeO_3)\approx$\,0.7, that makes the compensation effect of the $pd$- and $sd$-contributions to the $X$-factor (see Table\ref{tableXY}) more significant in borate than in orthoferrite. Interestingly that in their turn the structural factor $\left[{\bf r}_1\times{\bf r}_2\right]_z$ in FeBO$_3$ is 1.6 times larger than the mean value of the factor $\left[{\bf r}_1\times{\bf r}_2\right]_y$ in YFeO$_3$.  

The sign of the Dzyaloshinskii vector in FeBO$_3$ has been experimentally found recently due to making use of a new technique based on interference of the magnetic x-ray scattering with forbidden quadrupole resonant scattering\,\cite{Dmitrienko}. The authors found that the the magnetic twist follows the twist in the intermediate oxygen atoms in the planes between the iron planes, that is the DM coupling induces a small left-hand twist of opposing spins of atoms at (0,0,0) and (1/2,1/2,1/2). This means that in our notations the Dzyaloshinskii vector for Fe$_1$-O-Fe$_2$ pair is directed along $c$-axis, $d_z(12)>$\,0, that is $d_{12}(\theta )>$\,0 in a full agreement with theoretical predictions (see Table\,\ref{tablesign}).



\section{DM coupling in the three-center two-electron/hole system: cuprates}

\subsection{Effective Hamiltonian}

\begin{figure}[b]
\centering
\includegraphics[width=8.5cm,angle=0]{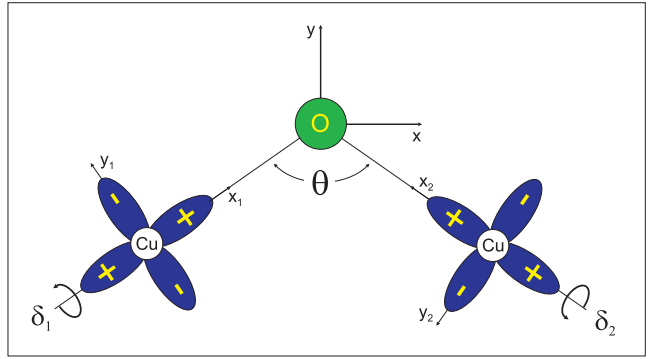}
\caption{Geometry of the three-center (Cu-O-Cu) two-hole system with ground Cu 3d$_{x^2-y^2}$ states.}
\label{fig9}
\end{figure}

As we pointed out above the Moriya approach to derivation of the effective DM coupling does not allow to uncover all the features of  this antisymmetric interaction, in particular, the structure of different contributions to the ${\bf D}_{12}$ vector, as well as the role of the intermediate ligands. For this and other features to elucidate we  address hereafter a typical for cuprates the three-center (Cu$^{2+}_1$-O-Cu$^{2+}_2$) two-hole system with tetragonal Cu on-site symmetry and ground Cu 3d$_{x^2-y^2}$ hole states (see Fig.\,\ref{fig9}) whose conventional bilinear effective spin Hamiltonian is written in terms of copper spins as follows\,\cite{JETP-2007,PRB-2007}
\begin{equation}
\hat H_s(12)=J_{12}(\hat{\bf s}_1\cdot \hat{\bf s}_2)+{\bf D}_{12}\cdot [\hat{\bf s}_1\times \hat{\bf s}_2]+\hat{\bf s}_1{\bf \stackrel{\leftrightarrow}{K}}_{12} \,\hat{\bf s}_2 \, ,\label{1}
\end{equation} 
 where $J_{12}>0$ is an exchange integral, ${\bf D}_{12}$ is the Dzyaloshinskii vector, ${\bf \stackrel{\leftrightarrow}{K}}_{12}$ is a symmetric second-rank tensor of the anisotropy constants. 
 In contrast with $J_{12}, {\bf \stackrel{\leftrightarrow}{K}}_{12}$, the Dzyaloshinskii vector ${\bf D}_{12}$
 is antisymmetric with regard to the site permutation: ${\bf D}_{12}=-{\bf D}_{21}$. Hereafter we will denote $J_{12}=J,{\bf \stackrel{\leftrightarrow}{K}}_{12}={\bf \stackrel{\leftrightarrow}{K}},{\bf D}_{12}={\bf D}$, respectively.
 It should be noted that making use of effective spin Hamiltonian (\ref{1}) implies a removal of orbital degree of freedom that calls for a caution with DM coupling as it changes both a spin multiplicity  and an orbital state.  
 
   It is clear that the applicability of such an operator as $\hat H_s(12)$ to describe all the "oxygen" effects is extremely limited. Moreover, the question arises in what concerns the composite structure and spatial distribution of what that be termed as the Dzyaloshinskii vector density. Usually this vector is assumed to be located somewhere on the bond connecting spins 1 and 2. 
  
  Strictly speaking, to within a constant the spin Hamiltonian $H_s(12)$ can be viewed as a result of the projection onto the purely ionic Cu$_1^{2+}(3d_{x^2-y^2})$-O$^{2-}(2p^6)$-Cu$_2^{2+}(3d_{x^2-y^2})$ ground state of the two-hole spin Hamiltonian
\begin{equation}
\hat	H_s=\sum_{i<j}I(i,j)(\hat{\bf s}(i)\cdot \hat{\bf s}(j))+\sum_{i<j}({\bf d}(i,j)\cdot [\hat{\bf s}(i)\times \hat{\bf s}(j)])+\sum_{i<j}\hat{\bf s}(i){\bf \stackrel{\leftrightarrow}{K}}(i,j)\, \hat{\bf s}(j) \, , \label{holes}
\end{equation}
where sum runs on the holes 1 and 2 rather than sites 1 and 2. This form   implies not only both copper and oxygen hole location, but allows to account for purely oxygen two-hole configurations. Moreover, such a form allows us to neatly separate both the one-center and two-center effects. 
 Two-hole spin Hamiltonian (\ref{holes}) can be projected onto three-center states incorporating the Cu-O charge transfer effects. 
 
\subsection{DM coupling}
	
We start with the construction of spin-singlet and spin-triplet wave functions for our three-center two-hole system taking account of the p-d hopping, on-site hole-hole repulsion, and crystal field effects for  excited configurations $\{n\}$\,=\,$\{n_1,n_0,n_2\}$ (011, 110, 020, 200, 002) with different hole occupation of Cu$_1$, O, and Cu$_2$ sites, respectively. In general, the on-site hole orbital basis sets include  the five 3d-functions on Cu$_1$ and Cu$_2$, and the three $p$-functions for the ligand site. Then we introduce a standard effective spin Hamiltonian operating in a fourfold spin-degenerated space of basic 101 configuration with two $d_{x^2-y^2}$ holes. 

The p-d hopping for Cu-O bond implies a conventional Hamiltonian
 \begin{equation}
 \hat H_{pd}=\sum_{\alpha \beta}t_{p\alpha d\beta}{\hat p}^{\dagger}_{\alpha}{\hat d}_{\beta}+h.c.\, ,
 \label{Hpd}
\end{equation}
 where ${\hat p}^{\dagger}_{\alpha }$ creates a hole in the $\alpha $ state on the ligand site, while  ${\hat d}_{\beta}$
annihilates a hole  in the $\beta $ state on the copper site; $t_{p\alpha d\beta}$ is a pd-transfer integral.
 
 For basic 101 configuration with two $d_{x^2-y^2}$ holes localized on its parent sites we arrive at a perturbed wave function as follows
 \begin{equation}
\Psi_{101;SM}=N_S(\Phi_{101;SM}+\sum_{\{n\}\Gamma}c_{\{n\}}({}^{2S+1}\Gamma)\Phi_{\{n\};\Gamma SM}) \, ,
	\label{Psi}
\end{equation}
where the summation runs both on different configurations and different orbital $\Gamma$ states, 
$$
N_S=[1+|\sum_{\{n\}\Gamma}c_{\{n\}}({}^{2S+1}\Gamma)|^2]^{-1/2}
$$
 is a normalization constant. It is important to highlight essentially different orbital functions for spin singlet and triplet states. 
The probability amplitudes $c_{\{011\}}, c_{\{110\}}\propto t_{pd}, c_{\{200\}}, c_{\{020\}}, c_{\{002\}}\propto t_{pd}^2$ can be easily calculated.

 
 For the microscopic expression for Dzyaloshinskii vector to derive Moriya\,\cite{Moriya} made use of the Anderson's formalism for the superexchange interaction\,\cite{PWA} with two main contributions of so-called kinetic and potential exchange, respectively. Then he took into account the spin-orbital corrections to the effective d-d transfer integral and  potential exchange. 
However, the Moriya's approach seems to be improper to account for purely ligand effects.
 In later papers (see, e.g. Refs.\,\cite{Shekhtman-93,Koshibae}) the authors made use of the Moriya scheme to account for spin-orbital corrections to p-d transfer integral, however, without any analysis of the ligand contribution.
It is worth noting that in both instances the spin-orbital renormalization of a single hole transfer integral leads immediately to a lot of problems with correct responsiveness of the on-site Coulomb hole-hole correlation effects.
 Anyway the effective DM spin-Hamiltonian evolves from the high-order perturbation effects that makes its analysis rather involved and leads to many misleading conclusions.

At variance with the Moriya approach  we consider the DM coupling 
\begin{equation}
\hat H_{DM}={\bf D}_{12}\cdot [\hat{\bf s}_1\times \hat{\bf s}_2]=\frac{1}{2}({\bf D}\cdot \hat{\bf T})	
\end{equation}
to be a result of a projection of the spin-orbital operator $\hat V_{so}=\hat V_{so}(Cu_1)+\hat V_{so}(O)+\hat V_{so}(Cu_2)$ on the ground state singlet-triplet manifold\,\cite{JETP-2007}. Then we calculate the singlet-triplet mixing amplitude due to the  three $local$ spin-orbital terms to find  the $local$ contributions to Dzyaloshinskii vector: 
 \begin{equation}
 {\bf D}={\bf D}^{(1)}+{\bf D}^{(0)}+{\bf D}^{(2)}.
 \end{equation}
Remarkably that the net Dzyaloshinsky vector ${\bf D}_{12}$ has a 
particularly local structure to be a superposition of $partial$ contributions of different ions ($i=1,0,2$) and ionic configurations $\{n\}=101,110,011,200,020,002$.
 Local spin-orbital coupling is taken as follows:  
\begin{equation}
V_{so}=\sum_i \xi_{nl}({\bf l}_i\cdot{\bf s}_i)=\frac{\xi _{nl}}{2}[({\bf \hat l}_1+{\bf \hat l}_2)\cdot {\bf \hat S}+({\bf \hat l}_1-{\bf \hat l}_2)\cdot {\bf \hat V}]
=\bm{\hat\Lambda}^S\cdot {\bf \hat S}+\bm{\hat\Lambda}^V\cdot {\bf \hat V}
\label{V_SO}
\end{equation}
 with a single particle constant $\xi_{nl}>0$ for electrons and $\xi_{nl}<0$ for holes. Here
\begin{equation}
	\hat{\bf S}=\hat{\bf s}_1+\hat{\bf s}_2\,\,\,; \hat{\bf V}=\hat{\bf s}_1-\hat{\bf s}_2\,.
\end{equation} 
 We make use of orbital matrix elements: for Cu 3d holes $\langle d_{x^2-y^2}|l_x|d_{yz}\rangle =\langle d_{x^2-y^2}|l_y|d_{xz}\rangle =i, \langle d_{x^2-y^2}|l_z|d_{xy}\rangle =-2i$, $\langle i|l_j|k\rangle =-i\epsilon _{ijk}$ with Cu 3d$_{yz}$=$|1\rangle$, 3d$_{xz}$=$|2\rangle$ 3d$_{xy}$=$|3\rangle$, and for the ligand np-holes $\langle p_{i}|l_j|p_{k}\rangle =i\epsilon _{ijk}$. 
 Free cuprous Cu$^{2+}$ ion is described by a large spin-orbital coupling with $|\xi _{3d}|\cong 0.1$ eV (see, e.g., Ref.\,\cite{Low}), though its value may be significantly reduced in oxides, chlorides... due to covalency effects.

Information regarding the $\xi _{np}$ value for the ligand $np$-orbitals is scant if any. Usually one considers the spin-orbital coupling on the oxygen in oxides to be much smaller than that on the copper, and therefore may be neglected\,\cite{Yildirim,Sabine}. However, even for a free oxygen atom the electron spin orbital coupling turns out to reach of appreciable magnitude: $\xi _{2p}\cong 0.02$ eV\,\cite{Herzberg} while for the oxygen O$^{2-}$ ion in oxides one expects the visible enhancement of spin-orbital coupling due to a larger compactness of 2p wave function\,\cite{Meier}. If to account for $\xi _{nl}\propto \langle r^{-3}\rangle _{nl}$ and compare these quantities for the copper ($\langle r^{-3}\rangle _{3d}\approx 6-8$ a.u. \cite{Meier}) and the oxygen ($\langle r^{-3}\rangle _{2p}\approx 4$ a.u.\cite{Walstedt,Meier}) we arrive at a maximum factor two difference in $\xi _{3d}$ and $\xi _{2p}$. However, for other ligands the spin-orbital effects can be of comparable value with that of $Cu^{2+}$.
For a free chlorine atom the electron spin-orbital coupling turns out to reach of appreciable magnitude: $\xi _{3p}\cong 0.07$ eV\,\cite{Herzberg} close to $\xi _{3d}$ while for the chlorine Cl$^{-}$ ion in chlorides one expects the visible enhancement of spin-orbital coupling due to a larger compactness of 3p wave function. 

As for the DM interaction we deal with two competing contributions\,\cite{JETP-2007,PRB-2007}. The first one is determined by the inter-configurational mixing effect and is derived as a first order contribution which does not take account of Cu$_{1,2}$ 3d-orbital fluctuations for a ground state 101 configuration. Projecting the spin-orbital coupling (\ref{V_SO}) onto  states (\ref{Psi}) we see that $\bm{\hat\Lambda}^V\cdot {\bf \hat V}$  term is equivalent to a spin DM coupling with local contributions to Dzyaloshinskii vector
\begin{equation}
D^{(m)}_{\alpha}=-2i\langle 00|V_{so}(m)|1\alpha\rangle=
-2i\sum_{\{n\}\Gamma_1,\Gamma_2}c_{\{n\}}^*({}^{1}\Gamma_1)c_{\{n\}}({}^{3}\Gamma_2)\langle \Phi_{\{n\};\Gamma _{1} 00} |\Lambda^V_i|\Phi_{\{n\};\Gamma_21\alpha}\rangle \label{DM2}\, ,
\end{equation}
where $m$\,=\,Cu$_1$, O, Cu$_2$, $\alpha =x,y,z$.
In all the instances the nonzero contribution to the local Dzyaloshinskii vector is determined solely by the spin-orbital singlet-triplet mixing for the on-site 200, 020, 002 and two-site 110, 011 two-hole configurations, respectively. For on-site two-hole configurations we have ${\bf D}^{(200)}={\bf D}^{(1)}$, ${\bf D}^{(020)}={\bf D}^{(0)}$, ${\bf D}^{(002)}={\bf D}^{(2)}$.

The second, "Moriya"-type, contribution,  associated with Cu$_{1,2}$ 3d-orbital fluctuations  within a ground state 101 configuration, is more familiar one and evolves from a second order combined effect of Cu$_{1,2}$ spin-orbital $V_{so}(Cu_{1,2})$ and effective orbitally anisotropic Cu$_1$-Cu$_2$ exchange coupling 
\begin{eqnarray}
D^{(m)}_{\alpha}=-2i\langle 00|V_{so}(m)|1\alpha\rangle=
\nonumber \\
-2i\sum_{\Gamma}\frac{\langle \{101\};\Gamma _{s} 00 |\hat\Lambda^V_{\alpha}|\{101\};\Gamma 1\alpha\rangle \langle \{101\};\Gamma 1\alpha |\hat H_{ex}|\{101\};\Gamma_t1\alpha\rangle}{E_{^3\Gamma _t}(\{101\})-E_{^3\Gamma}(\{101\})} 
 \nonumber \\
 -2i\sum_{\Gamma}\frac{\langle \{101\};\Gamma _{s} 00 |\hat H_{ex}|\{101\};\Gamma 00\rangle \langle \{101\};\Gamma 00 |\hat\Lambda^V_{\alpha}|\{101\};\Gamma_t1\alpha\rangle}{E_{^1\Gamma _s}(\{101\})-E_{^1\Gamma}(\{101\})} \, .
\end{eqnarray}
It should be noted that at variance with the original Moriya approach\,\cite{Moriya} both spinless and spin-dependent parts of exchange Hamiltonian contribute additively and  comparably to DM coupling, if one takes account of the same magnitude and opposite sign of the singlet-singlet and triplet-triplet exchange matrix elements  on the one hand and  orbital antisymmetry of spin-orbital matrix elements on the other hand.

It is easy to see that the contributions of 002 and 200 configurations to Dzyaloshinskii vector  bear a similarity to 
the respective second type ($\propto V_{so}H_{ex}$) contributions, however, in the former  we deal with spin-orbital coupling for two-hole Cu$_{1,2}$ configurations, while in the latter with that of one-hole  Cu$_{1,2}$ configurations.

\subsubsection{Copper contribution}
First we address a relatively simple example of strong rhombic crystal field for the intermediate ligand ion with the crystal field axes oriented along global coordinate $x,y,z$-axes, respectively. It is worth noting that in such a case the ligand np$_z$ orbital does not play an active role both in symmetric and antisymmetric (DM) exchange interaction as well as Cu 3d$_{yz}$ orbital appears to be inactive in DM coupling due to a zero overlap/transfer with ligand $np$ orbitals.

For illustration, hereafter we address the first contribution (\ref{DM2}) of two-hole on-site 200, 002 $d^2_{x^2-y^2}$, $d_{x^2-y^2}d_{xy}$, and $d_{x^2-y^2}d_{xz}$ configurations, which do covalently mix with ground state configuration\,\cite{JETP-2007,PRB-2007}.
Calculating the singlet-triplet mixing matrix elements in the global coordinate system we find all the  components of the local Dzyaloshinskii vectors. The Cu$_1$ contribution turns out to be nonzero only for 200  configuration, and may be written as a sum of several terms. First we present the contribution of the singlet $(d^2_{x^2-y^2})^1A_{1g}$ state:
$$
D^{(1)}_x=-2i\langle 00|V_{so}(Cu_1)|1x\rangle =D^{(1)}(\theta ,\delta _{1})\cos\frac{\theta}{2};\,\,
D^{(1)}_y=-2i\langle 00|V_{so}(Cu_1)|1y\rangle =-D^{(1)}(\theta ,\delta _{1})\sin\frac{\theta}{2};
$$
\begin{equation}
 D^{(1)}_z=-2i\langle 00|V_{so}(Cu_1)|1z\rangle =-{\sqrt{2}}\xi _{3d}\,c_{200}(^1A_{1g})[ c_{200}(^3E_{g})\sin\delta _1 -2c_{200}(^3A_{2g})\cos\delta _1 ]\, .	
\label{Cu}
\end{equation}
where
\begin{equation}
	D^{(1)}(\theta ,\delta _{1})={\sqrt{2}}\xi _{3d}\,c_{200}(^1A_{1g})[c_{200}(^3E_{g})\cos\delta _1 -2c_{200}(^3A_{2g})\sin\delta _1 ]\propto \left[\frac{\sin^2\frac{\theta}{2}}{\epsilon _x} -\frac{\cos^2\frac{\theta}{2}}{\epsilon _y}\right]\sin\theta \,\sin2\delta _{1}\, , 
\end{equation}
where $\epsilon _{x,y}$ are the ligand $p_{x,y}$-hole energies.
  In a vector form we arrive at 
\begin{equation}
	{\bf D}^{(1)}(\theta ,\delta _{1})=D^{(1)}(\theta ,\delta _{1})[{\bf r}_1\times {\bf z}]+D^{(1)}_z(\theta ,\delta _{1}){\bf z} \, ,
\end{equation}
where ${\bf r}_1$ is an unit vector directed along Cu$_1$-O bond, ${\bf z}$ is the unit vector in $xyz$ system.
 Taking into account that $c_{002}(^1A_{1g})=c_{200}(^1A_{1g})$, $c_{002}(^3A_{2g})=c_{200}(^3A_{2g})$, $c_{002}(^3E_{2g})=c_{200}(^3E_{g})$ \cite{remark} we see that the Cu$_2$ contribution to the Dzyaloshinskii vector can be obtained from Exps. (\ref{Cu}), if  $\theta ,\delta _1$  replace by $-\theta ,\delta _2$, respectively.


 	Both collinear ($\theta = \pi$) and rectangular
 ($\theta = \pi /2$) superexchange geometries appear to be unfavorable for copper contribution to antisymmetric exchange, though in the latter  the result  depends strongly on the relation between the energies of the ligand np$_x$ and np$_y$ orbitals.
  Contribution of singlet $(d_{x^2-y^2}d_{xy})^{1}A_{2g}$ and $(d_{x^2-y^2}d_{xz})^{1}E_{g}$ states to the Dzyaloshinskii vector yields  
$$
d^{(1)}_x=d^{(1)}\sin\frac{\theta}{2}\, ,\,\,
d^{(1)}_y=d^{(1)}\cos\frac{\theta}{2}\, , \,\,d^{(1)}_z=0\, ,
$$
where
\begin{equation}
	d^{(1)}=\xi _{3d}(c_{200}(^1A_{2g})c_{200}(^3E_{g})-c_{200}(^1E_{g})c_{200}(^3A_{2g}))\propto  \sin^2\theta \sin2\delta _1\, .
\end{equation}
Here we deal with a vector
\begin{equation}
	{\bf d}^{(1)}=d^{(1)}{\bf r}_1
\end{equation}
directed along the Cu$_1$-O bond whose   magnitude is determined by a partial cancellation of two terms.

It is easy to see that the copper $V_{so}(1)$ contribution to the Dzyaloshinskii vector for two-site 110 and 011 configurations is determined by a $dp$-exchange.
 

\subsubsection{Ligand contribution}

In frames of the same assumption regarding the  orientation of rhombic crystal field axes for the intermediate ion the local ligand contribution to the Dzyaloshinskii vector for one-site 020 configuration  appears to be oriented along local O$_z$ axis and  may be written as follows\,\cite{JETP-2007,PRB-2007}
\begin{equation}
D^{(0)}_z=-2i\langle 00|V_{so}(O)|1z\rangle ={\sqrt{2}}\xi _{2p}\,c_t(p_xp_y)[c(p_x^2)+c(p_y^2)]\, .	
\label{O}
\end{equation} 
  This vector  can be written as 
 \begin{equation}
	{\bf D}^{(0)}=D_O(\theta)[{\bf r}_1\times {\bf r}_2]\, ,
	\label{x}
\end{equation}
 where ${\bf r}_{1,2}$ are unit radius-vectors along Cu$_1$-O, Cu$_2$-O bonds, respectively, and
	\begin{equation}
	D_O(\theta)=\frac{9\xi _{2p}t_{pd\sigma}^4}{16}\frac{1}{E_t(p_xp_y)}\left(\frac{1}{\epsilon _x} +\frac{1}{\epsilon _y}\right)\left[\frac{\cos^2\frac{\theta}{2}}{\epsilon _xE_s(p_x^2)}-\frac{\sin^2\frac{\theta}{2}}{\epsilon _yE_s(p_y^2)}\right] \, ,
	\label{OO}
\end{equation}
where $E_s(p_{x,y}^2)$ are the two-hole singlet energies.
It is worth noting that ${\bf D}^{(0)}$ does not depend on the $\delta _1, \delta _2$ angles. The  $D_O(\theta)$ dependence is expected to be rather  smooth without any singularities for collinear  and rectangular superexchange geometries. 

The local ligand contribution to the Dzyaloshinskii vector for the two-site 110 and 011 configurations  is determined by a direct $dp$-exchange and may be written similarly to (\ref{x}) with
\begin{equation}
	D_O(\theta)=\frac{3\xi _{2p}t_{pd\sigma}^2}{8}\frac{1}{\epsilon _x\epsilon _y}\left(\frac{I_{dpx}}{\epsilon _x} -\frac{I_{dpy}}{\epsilon _y}\right)\approx \frac{3\xi _{2p}t_{pd\sigma}^2}{8}\frac{1}{\epsilon _x\epsilon _y}\left(\frac{\sin^2\frac{\theta}{2}}{\epsilon _x} -\frac{\cos^2\frac{\theta}{2}}{\epsilon _y}\right)I_{dp\sigma},\label{OO11}
\end{equation}
where we take account only of the $dp\sigma$ exchange ($I_{dp\sigma}\propto t^2_{pd\sigma}$).

\subsubsection{DM coupling in La$_2$CuO$_4$ and related cuprates}
The DM coupling and magnetic anisotropy in La$_2$CuO$_4$ and related copper oxides
has attracted considerable interest in 90-ths (see, e.g., Refs.\,\cite{Coffey,Koshibae,Shekhtman,debate}), and is still debated in the literature\,\cite{Tsukada,Kataev}. 
In the low-temperature tetragonal (LTT) and orthorhombic (LTO) phases of La$_2$CuO$_4$, the
oxygen octahedra surrounding each copper ion rotate by a
small tilting angle ($\delta _{LTT}\approx 3^0,\delta _{LTO}\approx 5^0$) relative to their location in the high-temperature tetragonal (HTT) phase.
The structural distortion  allows for the appearance of the antisymmetric DM coupling.
In terms of our choice for structural parameters to describe the Cu$_1$-O-Cu$_2$ bond we have for LTT phase:
$ \theta =\pi; \delta _1=\delta _2=\frac{\pi}{2}\pm \delta _{LTT}$ for bonds oriented perpendicular to the tilting plane, and $ \theta =\pm(\pi -2\delta _{LTT}); \delta _1=\delta _2=\frac{\pi}{2}$ for bonds oriented parallel to the tilting plane. It means that all the local Dzyaloshinskii vectors turn into zero for the former bonds, and turn out to be perpendicular to the tilting plane for the latter bonds. For LTO phase:$ \theta =\pm(\pi -\sqrt{2}\delta _{LTO}); \delta _1=\delta _2=\frac{\pi}{2}\pm \delta _{LTO}$. The largest ($\propto \delta _{LTO}$) component of the local Dzyaloshinskii vectors (z-component in our notation) turns out to be oriented perpendicular to the Cu$_1$-O-Cu$_2$ bond plane. Other two components of the local Dzyaloshinskii vectors are fairly small: that of perpendicular to CuO$_2$ plane (y-component in our notation) is of the order of $\delta _{LTO}^2$, while that of oriented along Cu$_1$-Cu$_2$ bond  axis (x-components in our notation) is of the order of $\delta _{LTO}^3$.

Comparative analysis of Exps. (\ref{Cu}), (\ref{OO}), and (\ref{OO11}) given estimations for different parameters typical for cuprates\,\cite{Eskes} ($t_{pd\sigma}\approx 1.5$ eV, $t_{pd\pi}\approx 0.7$ eV, $A=6.5$ eV, $B=0.15$ eV, $C=0.58$ eV, $F_0=5$ eV, $F_2=6$ eV) evidences that copper and oxygen Dzyaloshinskii vectors can be of a comparable magnitude, however, in fact it strongly depends on the Cu$_1$-O-Cu$_2$ bond geometry, correlation energies, and crystal field effects. The latter determines the single hole energies both for O 2p- and Cu 3d-holes such as $\epsilon _{x,y}$ and $\epsilon _{xy,xz}$, whose values are usually of the order of 1\,eV and 1-3\,eV, respectively.  It is worth noting that for two limiting bond geometries: $\theta \sim \pi$ and  $\theta \sim \pi /2$ (near collinear and near rectangular bonding, respectively) we deal with a strong "geometry reduction" of the DM coupling due to the $\sin\theta$ factor for the former and the factor like $\left[\frac{\sin^2\frac{\theta}{2}}{\epsilon _x} -\frac{\cos^2\frac{\theta}{2}}{\epsilon _y}\right]$ for the latter. Really, the resulting effect for the near rectangular Cu$_1$-O-Cu$_2$ bonding appears to be very sensitive to the local oxygen crystal field. A critical angle $\theta _{Cu}$ to turn the Cu-contribution to Dzyaloshinskii vector into zero is defined as follows: 
 $$
 \tan^2\frac{\theta _{Cu}}{2}=\epsilon _x/\epsilon _y \, ,
 $$
while for the oxygen contribution (\ref{OO}) we arrive at another critical angle: 
$$
\tan^2\frac{\theta _{O}}{2}=\epsilon _yE_s(p_y^2)/\epsilon _xE_s(p_x^2).
$$ 
 Maximal value of the scalar parameter $D_O(\theta)$ which determines the oxygen contribution to Dzyaloshinskii vector can be estimated to be of $\leq$1\,meV given the above mentioned typical parameters, however, the unfavorable geometry of the Cu-O-Cu bonds in the corner-shared cuprates  leads to a small value of the resulting Dzyaloshinskii vector and canting angles\,\cite{Thio}.
 As a whole, our model microscopic theory is believed to provide a reasonable estimation of the direction, sense, and numerical value of the Dzyaloshinskii vectors. Seemingly  more important result concerns the elucidation of the role played by Cu$_1$-O-Cu$_2$ bond geometry, crystal field, and correlation effects.

\subsection{DM coupled Cu$_1$-O-Cu$_2$ bond in external fields}

 Application of an uniform external magnetic field ${\bf h}_S$ will produce a net staggered spin polarization in the antiferromagnetically coupled Cu$_1$-Cu$_2$ pair
\begin{equation}
	\langle {\bf V}_{12}\rangle ={\bf L}=-\frac{1}{J_{12}^2}[\sum_i{\bf D}_{12}^{(i)}\times {\bf h}^S]=\bm{\chi}^{VS}{\bf h}^S
\label{V}
\end{equation}
with antisymmetric $VS$-susceptibility tensor: $\chi _{\alpha\beta}^{VS}=-\chi _{\beta\alpha}^{VS}$. It is worth noting that only in a classical representation the net contribution of the three local spin-orbital couplings does reduce to  a conventional antiferromagnetic spin order: 
$$
\langle {\bf V}_{12}\rangle ={\bf L}_{12}= {\bf S}_{1}-{\bf S}_{2}\, ,
$$
while in quantum representation one should say about emergence of some nonequivalence of spins for holes formally numbered as 1 and 2 on different sites. Puzzlingly, we arrive at a very unusual effect of the on-site staggered spin order to be a result of the on-site spin-orbital coupling and the the cation-ligand spin density transfer.
One sees that the sense of a staggered spin polarization, or antiferromagnetic vector, depends on that of  Dzyaloshinskii vector.  The $VS$ coupling
results in many interesting effects for the DM systems, in particular, the "field-induced gap" phenomena in 1D s=1/2 antiferromagnetic Heisenberg system with alternating DM coupling\,\cite{Affleck}. Approximately, the phenomenon is described by a so-called $staggered$  s=1/2 antiferromagnetic Heisenberg model with the Hamiltonian
\begin{equation}
\hat H=J\sum_i (\hat{\bf s}_i\cdot \hat{\bf s}_{i+1})-h_u\hat s_{iz}-(-1)^ih_s\hat s_{ix} \, ,\label{staggered}
\end{equation}  
which includes the effective uniform field $h_u$ and the induced staggered field $h_s\propto h_u$ perpendicular both to the applied uniform magnetic field and Dzyaloshinskii vector.

The DM copling for ferromagnetically coupled Cu$_1$-Cu$_2$ pair does also produce a net staggered spin polarization 
\begin{equation}
	\langle {\bf V}_{12}\rangle =\frac{1}{2J_{12}}[\sum_i{\bf D}_{12}^{(i)}\times {\bf S}] \, ,
\label{VFM}
\end{equation}
oriented perpendicular both to the net magnetic moment and Dzyaloshinskii vector. It should be noted that all the partial contributions to the net staggered spin polarization  can, in general, have distinct orientations.


 Application of a staggered field ${\bf h}^V$ for an antiferromagnetically coupled Cu$_1$-Cu$_2$ pair will produce a local spin polarization both on copper and oxygen sites 
\begin{equation}
	\langle {\bf S}_{i}\rangle=\frac{1}{J_{12}^2}[{\bf D}_{12}^{(i)}\times {\bf h}^V]=\bm{\chi}^{SV}(i){\bf h}^V,
\label{S}
\end{equation}
 which can be detected by different site-sensitive methods including neutron diffraction and, mainly, by nuclear magnetic resonance. It should be noted that $SV$-susceptibility tensor is the antisymmetric one: $\chi _{\alpha\beta}^{SV}=-\chi _{\beta\alpha}^{SV}$. 
 Strictly speaking, the both formulas (\ref{V}) and (\ref{S}) work well only in a paramagnetic regime and for relatively weak external fields. 
 
 Above we addressed a single Cu$_1$-O-Cu$_2$ bond, where, despite a site location, the direction and magnitude of Dzyaloshinskii vector  depends strongly on the bond strength and geometry. 
 It is clear that a site rather than a bond location of DM vectors would result in a revisit of conventional symmetry considerations and of magnetic structure in weak ferro- and antiferromagnets. Interestingly that the expression (\ref{S}) predicts the effects of a constructive or destructive (frustration) interference of copper spin polarizations in 1D, 2D, and 3D lattices depending on the relative sign of Dzyaloshinskii vectors and staggered fields for nearest neighbours. It should be noted that with the destructive interference the local copper spin polarization may turn into zero and DM coupling will manifest itself only through the oxygen spin polarization.  
\begin{figure}[h]
\centering
\includegraphics[width=8.5cm,angle=0]{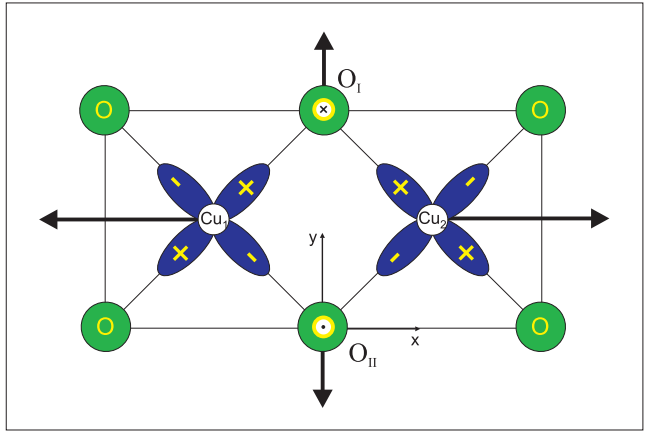}
\caption{The fragment of a typical edge-shared CuO$_2$ chain with copper and oxygen spin orientation under staggered field applied along x-direction. Note the antiparallel orientation of oxygen Dzyaloshinskii vectors.}
\label{fig10}
\end{figure} 
 Another interesting manifestation of the ligand DM antisymmetric exchange coupling concerns the edge-shared CuO$_2$ chains (see Fig.\,\ref{fig10}), ubiquitous for many cuprates, where we deal with an exact compensation of copper contributions to Dzyaloshinskii vectors and the unique possibility to observe the effects of uncompensated though oppositely directed  local oxygen contributions. It is worth noting that for purely antiferromagnetic in-chain ordering the oxygen spin polarization induced due to the  $dp$-covalency by two neighboring Cu ions is really compensated. In other words, the oxygen ions are expected to be nonmagnetic. However, the situation varies, if one accounts for a nonzero oxygen DM coupling. Indeed,  applying the staggered field, for instance, along chain direction (O$_x$) we arrive in accordance with Exp.(\ref{S}) at a staggered spin polarization of oxygen ions in an orthogonal O$_y$ direction  whose magnitude is expected to be strongly enhanced due to usually small magnitudes of 90$^{\circ}$ symmetric superexchange. In general, the direction of the oxygen staggered spin polarization is to be perpendicular both to the main chain antiferromagnetic vector and the CuO$_2$ chain normal. 
 
 It should be emphasized that the net in-chain Dzyaloshinskii vector ${\bf D}={\bf D}^{(1)}+{\bf D}^{(O_I)}+{\bf D}^{(O_{II})}+{\bf D}^{(2)}$ turns into zero  hence in terms of a conventional approach to DM theory we miss the anomalous oxygen spin polarization effect. In this connection it is worth noting the neutron diffraction data by Chung {\it et al.}\cite{Chung}  which unambiguously evidence the oxygen momentum formation and canting in  edge shared CuO$_2$ chain cuprate Li$_2$CuO$_2$. Anyhow, we predict an interesting possibility to find out the purely oxygen contribution to the DM coupling.

 \section{$^{17}$O NMR in L$\mbox{a}_2$C$\mbox{u}$O$_4$: Field-induced staggered magnetization}
Effect of the field-induced staggered magnetization was firstly discussed by Ozhogin {\it et al.}\,\cite{Ozhogin-1975} by means of the $^{57}Fe$ M\"{o}ssbauer measurements in the orthoferrite YFeO$_3$ in a paramagnetic region near $T_N$. Earlier on  we pointed to the ligand NMR as, probably, the only experimental technique to measure both staggered spin polarization, or antiferromagnetic vector in weak 3d ferromagnets  and the value, direction, and the sense of Dzyaloshinskii vector. The latter possibility was realized with $^{19}$F NMR for  weak ferromagnet FeF$_3$\,\cite{sign}. 
Hereafter we address the problem for generic  weak ferromagnetic cuprate La$_2$CuO$_4$ making use of ligand $^{17}$O NMR as an unique local probe to study the charge and spin densities on oxygen sites.


Detailed study of the ligand  $^{17}$O hyperfine couplings in weak ferromagnetic La$_2$CuO$_4$ for temperatures ranging from 285 to 800\,K  undertaken by R. Walstedt {\it et al}.\,\cite{Walstedt} has uncovered puzzling anomalies of the $^{17}$O Knight shift. With  approaching  to the ordered magnetic phase,  the authors observed anomalously large negative $^{17}$O Knight shift for planar oxygens whose anisotropy resembled that of weak ferromagnetism in this cuprate. The giant shift was observed only when external field was parallel to the local Cu-O-Cu bond axis (PL1 lines) or perpendicular to CuO$_2$ plane. The effect was not observed  for PL2 lines which correspond to oxygens in the local Cu-O-Cu bonds whose axis is perpendicular to the in-plane external field.  
It is worth noting once more, that the most part of experimental data was collected in paramagnetic state for temperatures well above T$_N$ where there is no frozen moments!
The data were first interpreted as an indication of a direct oxygen spin polarization due to a local DM antisymmetric exchange coupling. However, it demands unphysically large values for such a polarization, hence the puzzle remained to be unsolved\,\cite{Walstedt}. 

Our interpretation of the ligand NMR data in low-symmetry systems as La$_2$CuO$_4$ implies a thorough analysis both of the spin canting effects and of the transferred hyperfine interactions with a revisit of some textbook results  being typical for the model high-symmetry systems\,\cite{PRB-2007}. First, we start with spin-dipole hyperfine interactions for O 2p-holes which are main participants of Cu$_1$-O-Cu$_2$ bonding. 
 Making use of a conventional formula for a spin-dipole contribution to local field
 and calculating  appropriate  matrix elements on oxygen 2p-functions
we present the local field on the $^{17}$O nucleus in Cu$_1$-O-Cu$_2$ system as a sum of ferro- and antiferromagnetic contributions as follows\cite{PRB-2007} 
 \begin{equation}
{\bf \cal H}_n={\bf \stackrel{\leftrightarrow}{A}}^S\cdot \langle{\bf \hat S}\rangle+{\bf \stackrel{\leftrightarrow}{A}}^V\cdot \langle{\bf \hat V}\rangle
\label{Hn}
\end{equation}
where along with a conventional textbook  ferromagnetic ($\propto \langle{\bf \hat S}\rangle$) transferred hyperfine contribution to local field which simply mirrors a sum total of two Cu-O bonds,  we arrive at an additional unconventional antiferromagnetic ($\propto \langle{\bf \hat V}\rangle$) contribution whose symmetry and  magnitude strongly depend on the orientation of the oxygen crystal field axes and  Cu$_1$-O-Cu$_2$ bonding angle. 
In the case of Cu$_1$-O-Cu$_2$ geometry shown in Fig.\ref{fig9} we arrive at a diagonal ${\bf \stackrel{\leftrightarrow}{A}}^S$ tensor: 
$$
A^S_{xx}=2A_{p}(3\sin^2\frac{\theta}{2} -1);\,A^S_{yy}=2A_{p}(3\cos^2\frac{\theta}{2} -1);
$$
\begin{equation}
A^S_{zz}=-2A_{p},	
\end{equation}
and the only nonzero components of ${\bf \stackrel{\leftrightarrow}{A}}^V$ tensor: 
\begin{equation}
A^V_{xy}=A^V_{yx}=3A_{p}\sin\theta 	
\end{equation}
with
\begin{equation}
A_{p}=\frac{3}{4}\left(\frac{t_{dp\sigma}}{\epsilon_{p}}\right)^2A_{p}^{0}=f_{\sigma}A_{p}^{0},	
\end{equation}
where $f_{\sigma}$ is the parameter of a transferred spin density and we made use of a simple approximation $E_{s,t}(dp_{x,y})\approx \epsilon_{p}$. 
Thus, the ligand $^{17}$O NMR provides an effective tool to inspect the spin canting effects in oxides with DM coupling both in paramagnetic and ordered phases.

The two-term structure of the oxygen local field (\ref{Hn}) implies a two-term  S-V structure of the $^{17}$O Knight shift
\begin{equation}
^{17}K={\bf \stackrel{\leftrightarrow}{A}}^S\bm{\chi}^{SS} +
{\bf \stackrel{\leftrightarrow}{A}}^V\bm{\chi}^{VS}	
\end{equation}
that points to Knight shift as an effective tool to inspect both uniform and staggered spin polarization. 
 The existence of antiferromagnetic term in oxygen hyperfine interactions yields a rather simple explanation of the $^{17}$O Knight shift anomalies in La$_2$CuO$_4$\,\cite{Walstedt} as a result of the external field induced  staggered spin polarization
$\langle{\bf \hat V}\rangle$\,=\,$\bf L$\,=\,$\bm{\chi}^{VS}$${\bf \cal H}_{ext}$. Indeed, "our" local $y$ axis for Cu$_1$-O-Cu$_2$ bond corresponds to  the crystal tetragonal $c$-axis oriented perpendicular to CuO$_2$ planes both in LTO and LTT phases of La$_2$CuO$_4$ while $x$-axis does to local Cu-O-Cu bond axis. It means that for the geometry of the experiment by Walstedt {\it et al.}\cite{Walstedt} (the crystal is oriented so that the external uniform field is either $\parallel$ or $\perp$ to the local Cu-O-Cu bond axis) the antiferromagnetic contribution to $^{17}$O Knight shift will be observed  only a) for oxygens in the Cu$_1$-O-Cu$_2$ bonds oriented  along external field or b) for external field along the tetragonal $c$-axis. Experimental data\,\cite{Walstedt} agree with staggered magnetization along the tetragonal $c$-axis in the former and along the rhombic $c$-axis (tetragonal $ab$-axis) in the latter.
 Given $L=1$, $A_{p}^{(0)}\approx 100$ kG/spin\,\cite{Walstedt}, $|\sin\theta | \approx 0.1$, and $f_{\sigma}\approx 20\%$ we obtain $\approx 6$ kG as a maximal value of a low-temperature antiferromagnetic contribution to hyperfine field which is equivalent to a giant   $^{17}$O Knight shift of the order of almost $\sim$10\%. Nevertheless, this value agrees with a low-temperature extrapolation of high-temperature experimental data by Walstedt {\it et al.}\,\cite{Walstedt}. Interestingly, the sizeable effect of anomalous negative contribution to $^{17}$O Knight shift has been observed in  La$_2$CuO$_4$ well inside the paramagnetic state for temperatures $T\sim 500$ K that is essentially higher than $T_N \approx 300$\,K. It points to the close relation between the magnitude of field-induced staggered magnetization and spin-correlation length which goes up as one approaches T$_N$. 
 
The ferro-antiferromagnetic S-V structure  of the local field on the nucleus of an intermediate oxygen ion in the  Cu$_1$-O-Cu$_2$ triad points to the $^{17}$O NMR as, probably, the only experimental technique to measure both the value, direction, and the sense of Dzyaloshinskii vector. 
For instance, the negative sign of $^{17}$O Knight shift in La$_2$CuO$_4$\,\cite{Walstedt}  points to a negative sign of $\bm{\chi}^{VS}$ for Cu$_1$-O-Cu$_2$ triad with $A_{xy}^V>0$, hence to a positive sense of $z$-component of  the net Dzyaloshinskii vector in Cu$_1$-O-Cu$_2$ triad with geometry shown in Fig.\ref{fig9} given $\theta \leq \pi$, $\delta_1=\delta_2\approx \pi /2$.
It should be emphasized that the above effect is determined by the net  Dzyaloshinskii vector in Cu$_1$-O-Cu$_2$ triad rather than by a local oxygen "weak-ferromagnetic" polarization as it was firstly proposed by Walstedt {\it et al.}\,\cite{Walstedt}. 

Similar effect of anomalous ligand $^{13}$C Knight shift has recently been observed in copper pyrimidine dinitrate  [CuPM(NO$_3$)$_2$(H$_2$O)$_2$]$_n$, a one-dimensional S=1/2 antiferromagnet with alternating local symmetry, and was also interpreted in terms of the field-induced staggered magnetization\,\cite{CuPM}.
However, the authors did take into account  only the inter-site  magneto-dipole contribution to ${\bf \stackrel{\leftrightarrow}{A}}^V$ tensor that  questions their quantitative conclusions regarding the "giant" spin canting. 


\section{DM coupling in helimagnetic C$\mbox{s}$C$\mbox{u}$C$\mbox{l}$$_3$}

 All the systems  described above were somehow or other connected with weak ferromagnets where DM coupling manifests itself in the canting of a basic antiferromagnetic structure. 
Caesium cupric chloride, CsCuCl$_3$ is a unique screw antiferroelectric crystal with a low-temperature helimagnetically distorted ferromagnetic order.
CsCuCl$_3$ possesses the hexagonal CsNiCl$_3$-type D$^4_{6h}$($P6_3/mmc$)
structure\,\cite{Achiwa} above the transition temperature T$_c$ ($\approx$\,423\,K)
and is distorted through a first-order phase transition to low symmetry by a cooperative Jahn-Teller effect below T$_c$\,\cite{JT}. In the high-temperature phase CuCl$_6$ octahedra are linked together
by sharing faces, thus forming a one-dimensional chain structure along the $c$ axis. The octahedra are not regular
but trigonally compressed along the $c$ axis, with all Cu-Cl distances remaining equal.
At T$_c$ all of the constituent atoms are displaced from the normal position along the c-axis to form a helix whose period is three
times the lattice constant $c$ of the high-temperature phase.
The room-temperature structure was determined by x-ray diffraction\,\cite{Xray}. The space group is one of two
enantiomorphous groups D$^2_{6}$($P6_122$) or D$^3_{6}$($P6_522$) without a center of symmetry, corresponding to the right and left helixes, respectively, with six formula units in a unit cell.
Deformation of each CuCl$_6$ octahedron associated with the transition at T$_c$ is the Cu$^{2+}$ displacement and  tetragonal elongation  with the directions of their longest axes alternating by 120$^\circ$ in adjacent octahedra lying along the chain.



 In addition, CsCuCl$_3$ has a peculiar magnetic property. It is a quantum frustrated magnetic system
with a triangular lattice of antiferromagnetically coupled s=1/2  spins of Cu$^{2+}$ in the $ab$
plane.
In the magnetically ordered state, below T$_N$ (10.5-10.7\,K) spins lie in the basal plane and form
the 120$^\circ$-structure, while along the $c$-direction, a long period (about 71 triangular layers) helical incommensurate
arrangement (Dzyaloshinskii helix\,\cite{Plakhty}) with a slow spin spiraling (pitch angle of about 5$^\circ$) is realized\,\cite{Adachi}, due to the competition between the dominant ferromagnetic interaction and the additional DM coupling along the chain. The DM coupling forces the spins to lie almost flat in the $ab$ plane, so this spin system is approximately an
XY-system. In fact, from the structure determination, the spins are known to be slightly canted out of the
$ab$ plane\,\cite{Adachi}. CsCuCl$_3$ is the first example having a helical magnetic structure due to the antisymmetric exchange interaction.
The reduction (to 0.58\,$\mu_B$) of the ordered moment of the s\,=\,1/2 spin of the Cu$^{2+}$ ion is not uncommon in frustrated triangular-lattice antiferromagnetic systems. It is worth noting that Plakhty et al.\,\cite{Plakhty} have  revealed a modulation of the CsCuCl$_3$ crystal structure with the periodicity of the incommensurate long-period Dzyaloshinskii helix.

 Despite numerous experimental and theoretical studies many details of spin structure in CsCuCl$_3$ remain to be answered. 
The  NMR data do not support incommensurability in CsCuCl$_3$, the $^{63,65}$Cu NMR spectra  clearly indicate that the Cu$^{2+}$ moments refer regularly to a local symmetry axis rather than to a spin spiral arrangement\,\cite{NMR-76}, or the in-plane  spin projection forms a commensurate spiral with the pitch angle 60$^{\circ}$\,\cite{NMR-77}.
According to Ref.\,\cite{Adachi} the Dzyaloshinskii vector appears to be parallel to the vector between the nearest along $c$ helically displaced Cu$^{2+}$ ions. However, Plakhty et al.\,\cite{Plakhty} argued that the vector should be directed perpendicular to a plane formed by the Cu-Cl-Cu  triad. 
Another point of a great importance for a detailed spin structure determination in CsCuCl$_3$ is a nonzero chlorine spin polarization whose accounting can strongly influence the interpretation of magnetic, neutron, and NMR data. To the best of our knowledge the  
neutron diffraction data for chain cuprate Li$_2$CuO$_2$ by Chung {\it et al.}\,\cite{Chung}  provided the first unambiguous evidence of the ligand (oxygen) magnetic momentum formation and canting.

The change of the CsCuCl$_3$ crystal structure  occurring at the transition at T$_c$ due to a cooperative Jahn-Teller (JT) effect\,\cite{JT} consists of helical displacements of all the constituent atoms.  The period of the helices is three times the lattice constant $c$ of the high-temperature phase, thus the
$c$ axis is tripled below T$_c$. The Cu atoms displace by $u$\,=0.06136 from the $c$ axis with the directions of the displacements alternating by 60$^{\circ}$ in adjacent Cu-Cu pairs lying along the chain. The three equivalent Cl ions forming a regular triangle in the $c$ plane in the high-temperature phase form two types of chlorine atoms below T$_c$ labelled as Cl(1) and Cl(2). The Cl(1) atom moves along the symmetrical line Y = 2X, while two Cl(2) atoms move into general positions.
Main local deformation of each CuCl$_6$ octahedron associated with the JT transition is a tetragonal elongation along the Cu-Cl(2) direction. When viewed along the $c$ axis, the directions of the elongated axes alternate by 120$^{\circ}$ in adjacent octahedra lying along the chain. The Cu ion  displaces from the tetragonal axis along the edge of nearly square CuCl$_2$(1)Cl$_2$(2) plaquette with two Cu-Cl separations of 2.3525\,\AA\, and two of 2.2837\,\AA\, which are sizeably shorter than the  Cu-Cl(2) separations of 2.7758\,\AA\, along the tetragonal axis. 
The JT effect  results in a strong distortion of the CuCl$_6$ edge sharing. Instead of three equivalent Cu-Cl-Cu bonds we arrive at   two Cu-Cl(2)-Cu bonds with bonding angle of 73.74$^{\circ}$ and  with 
the shortest (2.2837\,\AA\,) and longest (2.7758\,\AA\,) Cu-Cl(2) separations, respectively, and the Cu-Cl(1)-Cu bond with bonding angle of 81.17$^{\circ}$ and equal 
Cu-Cl(1) separations of 2.3525\,\AA\,. Namely the latter Cu-Cl(1)-Cu bond should be addressed to be a main contributor to Cu-Cu exchange coupling. Indeed, 
the longest (2.7758\,\AA\,) Cu-Cl(2) separation is too long to provide a meaningful exchange coupling and will take into account only the Cu-Cl(1)-Cu bond. From the other hand, the anticipated  $d_{x^2-y^2}$-type hole ground state of the  CuCl$_4$ plaquette  typical for Cu$^{2+}$ squarely coordinated with four ligands cannot provide meaningful Cu-Cl(2) coupling along tetragonal axis which is required for enabling the efficient Cu-Cl(2)-Cu coupling. The almost rectangular Cu-Cl(1)-Cu  superexchange with rather long Cu-Cl(1) separation of 2.3525\,\AA\, ought to be small ferromagnetic that explains rather low temperature of magnetic ordering.  

To make a semiquantitative analysis of the Cu-Cl(1)-Cu DM coupling, hereafter we assume a tetragonal symmetry at Cu sites with local coordinate systems as shown in Fig.\,\ref{fig9}. 
The net Dzyaloshinskii vector ${\bf D}$ for the Cu$_1$-Cl(1)-Cu$_2$  superexchange is a superposition of three contributions ${\bf D}={\bf D}^{(1)}+{\bf D}^{(O)}+{\bf D}^{(2)}$ attached to the respective sites. In general, all the vectors are oriented differently. In other words, the direction of the net Dzyaloshinskii vector ${\bf D}_{nn+1}$ seems to be more complicated that it is suggested in Refs.\,\cite{Plakhty,Adachi}. Interestingly, the $x$-component of the Dzyaloshinskii vector, or its projection onto the Cu$_n$-Cu$_{n+1}$ direction gives rise to a helical spin ordering along $c$-axis with spins in $ab$-plane, while $y$ and $z$ components compete  for the spin canting upward and downward from the $ab$-plane with a periodicity of six Cu$^{2+}$ ion spacings along the $c$-axis.   

Comparative analysis of Exps. (\ref{Cu}), (\ref{OO}), and (\ref{OO11}) given estimations for different parameters typical for cuprates\,\cite{Eskes} evidences that copper and chlorine Dzyaloshinskii vectors can be of a comparable magnitude, however, in fact it strongly depends on the Cu$_1$-Cl-Cu$_2$ bond geometry, correlation energies, and crystal field effects. 
 Maximal value of the scalar parameter $D_O(\theta)$ which determines the chlorine contribution to Dzyaloshinskii vector can be estimated to be of $\sim$\,1\,meV given the above mentioned typical parameters.
 As a whole, our model microscopic theory is believed to provide a reasonable estimation of the direction, sense, and numerical value of the Dzyaloshinskii vectors and the role of the Cu$_1$-Cl-Cu$_2$ bond geometry, crystal field, and correlation effects.

\section{Effective two-ion symmetric spin anisotropy due to DM coupling}
When one says about an effective  spin anisotropy due to DM coupling one usually addresses a simple classical two-sublattice weak ferromagnet where the free energy has a minimum when both ferro- ($\propto \langle \hat {\bf S}\rangle $, $\hat {\bf S}=\hat {\bf S}_1+\hat {\bf S}_2$) and antiferromagnetic ($\propto \langle \hat {\bf V}\rangle $,  $\hat {\bf V}=\hat {\bf S}_1-\hat {\bf S}_2$) vectors, being perpendicular to each other, lie in the plane perpendicular to the Dzyaloshinskii vector ${\bf D}$. However, the issue is rather involved and appeared for a long time to be hotly debated. In our opinion, first of all we should define what the spin anisotropy is. Indeed, the description of any spin system implies  the free energy $\Phi$ depends on a set of vector order parameters (e.g., $\langle \hat {\bf S}\rangle ,\langle \hat {\bf V}\rangle $) under  kinematic constraint, rather than a single magnetic moment as in a simple ferromagnet, that can make the orientational dependence of the free energy $\Phi$ extremely involved. Such a situation needs in a careful analysis of respective spin Hamiltonian with a choice of proper approximations.

Effective symmetric spin anisotropy due to DM interaction can be easily derived as a second order perturbation correction due to DM coupling.  For antiferromagnetically coupled spin 1/2 pair $\hat H_{an}^{DM}$ may be written as follows\,\cite{JETP-2007,PRB-2007}:
$$
\hat H_{an}^{DM}=\sum_{ij}\Delta K^V_{ij} {\hat V}_i{\hat V}_j
$$
with
$\Delta K^V_{ij} =\frac{1}{8J} D_{i}D_{j}$ provided $|{\bf D}|\ll J$. We see that in frames of a simple MFA approach this anisotropy stabilizes a N\'{e}el state with $\langle \hat {\bf V}\rangle \perp {\bf D}$. However, in fact we deal with a MFA artefact. Indeed, let examine the second order perturbation correction to the ground state energy of an antiferromagnetically coupled spin 1/2 pair in  a N\'{e}el-like staggered field ${\bf h}^V\parallel {\bf n}$ ($\Psi _{\alpha ,0} =\cos\alpha |00\rangle +\sin\alpha |1n\rangle $, $\tan2\alpha=\frac{2h^V}{J}$): 
\begin{equation}
E_{an}^{DM}=-\frac{|{\bf D}\cdot {\bf n}|^2}{4(E_{\parallel}-E_g)}-\frac{|{\bf D}\times {\bf n}|^2}{4(E_{\perp}-E_g)}\cos^2\alpha \, ,
\label{an}
\end{equation}
where $E_{\perp}=J; E_{\parallel}=J\,\cos^2\alpha +h^V\,\sin2\alpha; E_{g}=J\,\sin^2\alpha -h^V\,\sin2\alpha$. First term in (\ref{an}) stabilizes ${\bf n}\parallel{\bf D}$ configuration while the second one does the ${\bf n}\perp{\bf D}$
configuration. Interestingly that $(E_{\parallel}-E_g)\cos^2\alpha = (E_{\perp}-E_g)$, that is for any  staggered field $E_{an}^{DM}$  does not  depend on its orientation, if to account for: $|{\bf D}\cdot {\bf n}|^2+|{\bf D}\times {\bf n}|^2=|{\bf D}|^2$. In other words, at variance with a simple MFA approach, the DM contribution to the energy of anisotropy for an exchange coupled spin 1/2 pair in a staggered field turns into zero. Anyway, the $\hat H_{an}^{DM}$ term has not to be included into an effective spin anisotropy Hamiltonian for quantum 1/2 spins. However, for large spins $S\gg$\,1/2 the MFA, or classical approach to anisotropy induced by the DM coupling can be more justified.   

\section{"First-principles" calculations of the DM coupling}

The electronic states in strongly correlated 3$d$ oxides manifest both significant localization and dispersional features. One strategy to deal with this dilemma is to restrict oneself to small many-electron clusters embedded to a whole crystal, then creating model effective lattice Hamiltonians whose spectra may reasonably well represent the energy and dispersion of the important excitations of the full problem. Despite some shortcomings the method did provide a clear physical picture of the complex electronic structure and the energy spectrum, as well as the possibility of a quantitative modeling. 

However, last decades the condensed matter community faced an expanding flurry of papers with {\it ab initio} calculations of electronic structure and physical properties for strongly correlated systems such as 3$d$ compounds  based on density functional theory (DFT). The modern formulation of the DFT originated in the work of Hohenberg and Kohn\,\cite{HK}, on which based the other classic work in this field by Kohn and Sham\,\cite{KS}. The Kohn-Sham equation, has become a basic mathematical model of much of present-day methods for treating electrons in atoms, molecules, condensed matter, solid surfaces, nanomaterials, and man-made structures\,\cite{Kryachko}. 

However, DFT still remains, in some sense, ill-defined: many of the DFT statements were ill-posed or not rigorously proved.
Most widely used DFT computational schemes start with a "metallic-like" approaches making use of approximate energy functionals, firstly LDA (local density approximation) scheme, which are constructed as expansions around the homogeneous electron gas limit and fail quite dramatically in capturing the properties of strongly correlated systems. 
         The LDA+U and LDA+DMFT (DMFT, dynamical mean-field theory)\,\cite{LDA+U+DMFT} methods are believed to correct the inaccuracies of approximate DFT exchange correlation functionals, however, these preserve many shortcomings of the DFT-LDA approach.
All efforts to account for the correlations beyond LDA encounter an insoluble problem of double counting (DC) of interaction terms which had just included into Kohn-Sham single-particle potential.
In a certain sense the cluster based calculations seem to provide a better description of the overall electronic structure of insulating 3$d$ oxides and its optical response than the DFT based band structure calculations, mainly due to a clear physics and a better account for correlation effects (see, e.g., Refs.\,\cite{Eskes,Ghijsen}).

Basic drawback of the spin-polarized DFT approaches is that these start with a  local density functional in the form 
(see, e.g. Ref.\,\cite{Sandratskii})
\begin{equation}
	{\bf v}({\bf r})=v_0[n({\bf r})]+\Delta v[n({\bf r}),{\bf m}({\bf r})](\bm{\sigma}\cdot \frac{{\bf m}({\bf r})}{|{\bf m}({\bf r})|})\, ,
\label{xc}
\end{equation}
where $n({\bf r}),{\bf m}({\bf r})$ are the electron and spin magnetic density, respectively, $\bm{\sigma}$ is the Pauli matrix, that is these approaches imply  presence of
a large fictious local {\it one-electron} spin-magnetic field $\propto (v^{\uparrow}-v^{\downarrow})$, where $v^{\uparrow ,\downarrow}$ are the on-site LSDA spin-up and spin-down potentials. Magnitude of the field is considered to be  governed by the intra-atomic Hund exchange, while its orientation does by the effective molecular, or inter-atomic exchange fields. Despite the supposedly spin nature of the field it produces an unphysically giant spin-dependent rearrangement of  the charge density that cannot be reproduced within any conventional technique operating with spin Hamiltonians. Furthermore, a  direct link with the orientation of the field makes the effect of the spin configuration onto the charge distribution to be unphysically large. However, magnetic long-range order has no significant influence on the redistribution of the charge density. The DFT-LSDA community needed many years to understand such a physically clear point. 

In general, the  LSDA method to handle a spin degree of freedom is absolutely incompatible with a conventional approach based on  the spin Hamiltonian concept. There are some intractable problems with a match making between the conventional formalism of a spin Hamiltonian and LSDA approach to the exchange and exchange-relativistic effects.
Visibly plausible numerical results for different exchange and exchange-relativistic  parameters reported in many LSDA investigations (see, e.g., Refs.\,\cite{Mazurenko})  evidence only a potential capacity of the LSDA based models for semiquantitative estimations, rather than for reliable  quantitative data.
It is worth noting that for all of these "advantageous" instances the matter concerns the handling of certain classical N\'eel-like  spin configurations (ferro-, antiferro-, spiral,...) and search for a compatibility with a mapping made with a  conventional quantum spin Hamiltonian. It's quite another matter when one addresses the search of the charge density redistribution induced by a spin configuration as, for instance, in multiferroics. In such a case the straightforward application of the LSDA scheme can lead to an unphysical overestimation of the effects or even to qualitatively incorrect results due to an unphysically strong effect of a breaking of spatial symmetry induced by a spin configuration (see, e.g. Refs.\,\cite{multiferro} and references therein).



As a typical starting point for the "first-principles"\, calculation of the exchange interactions and DM coupling one makes use of a predetermined classical spin configuration and classical Hamiltonian as follows
\begin{equation}
	H=H_{ex}+H_{DM}=\sum_{i\not= j}J_{ij}({\bf e}_i\cdot {\bf e}_j)+\sum_{i\not= j}{\bf D}_{ij}\cdot [{\bf e}_i\times {\bf e}_j] \, ,
\label{class}
\end{equation}
where ${\bf e}_i$ is a unit vector in the direction of the $i$th site magnetization, $J_{ij}$ is the exchange interaction, and ${\bf D}_{ij}$ is the Dzyaloshinskii vector. 
It should be noted that this oversimplification together with an exceptionally one-particle nature of the LDA approach bounds all the efforts to account of intricate quantum effects that perturbatively define the DM coupling for many-electron ions, though  keeps a possibility of a plausible estimation.  We'd like to remind that classical approximation for the singlet-triplet exchange splitting in the pair of quantum s=1/2 spins yields the three times smaller value than the quantum result.

Obviously, the LDA based approaches cannot provide a comprehensive description of the DM coupling and other anisotropic interactions that are derived from the higher than the isotropic exchange perturbation orders and imply an intricate interplay of different many-electron quantum fluctuations. It is worth noting that at variance with isotropic exchange the DM coupling does mix spin multiplicity that cannot be distinctly reproduced in classical approach.
The so-called LDA+U+SO approach that attempts (and fails) to incorporate spin-orbit coupling within LDA+U scheme leads to unphysical results such as an "intra-atomic noncollinear magnetic ordering"\, when the spins of different orbitals appear to be noncollinear to each other or an appearance of the single-ion anisotropy for s=1/2 ions ($Cu^{2+}$)\,\cite{Whangbo}. 
The LDA+U+SO calculations\,\cite{Mazurenko} show appearance of unphysical on-site contribution in the magnetic torque and DM coupling, moreover this false term gives the main contribution to Dzyaloshinskii vector (!?). 
Recently a distinct approach for calculations of DM coupling and other anisotropic interactions in molecules and crystals has been
proposed\,\cite{Kats-2010,Secchi}. 
The authors derive a set of equations expressing the parameters of the magnetic interactions characterizing a
strongly correlated electronic system in terms of single-electron Green's functions and self-energies. This
allows to establish a mapping between the initial electronic system and a classical spin model (\ref{class}) including up to quadratic interactions between the effective spins, with a general interaction (exchange) tensor that accounts for DM coupling, single- and two-ion anisotropy. 
As they argue, the scheme leads to a simple and transparent analytical expression for the Dzyaloshinskii vector with a natural separation into spin and orbital contributions, though they do not present physical explanation for such a separation. However, the mere possibility of such a mapping seems to be unacceptable, as any ions with a bare spin and orbital degeneracy are characterized by a number of multicomponent spin and orbital order parameters that cannot be reduced to the only vector order parameter. The application of  inappropriate techniques makes it often hard to compare results obtained by different "first-principles" calculations even for the same  weak ferromagnet. For instance, for the spin canting angle in $La_2CuO_4$ one obtains 0.7$\cdot 10^{-3}$\,\cite{Mazurenko} and 5$\cdot 10^{-3}$\,\cite{Kats-2010} as compared with experimental value of (2-3)$\cdot 10^{-3}$. 
  



In our opinion, any comprehensive physically valid description of the exchange and exchange-relativistic effects for strongly correlated systems  should  combine simple physically clear cluster ligand-field analysis with a numerical calculation technique such as LDA+MLFT\,\cite{Haverkort} with a regular appeal to experimental data.

\section{Conclusion}

We performed an  overview of the microscopic theory of the Dzyaloshinskii-Moriya coupling in strongly correlated 3d compounds. 
Most attention in the paper focused on the derivation of the Dzyaloshinskii vector, its value, orientation, and sense both for different types of 3d ions and  under different types of the (super)exchange interaction and crystal field.

We considered both the Moriya mechanism of the antisymmetric interaction and novel contributions, in particular, that of spin-orbital coupling on the intermediate ligand ions.  Microscopically derived expression for the dependence of the Dzyaloshinskii vector on the superexchange geometry allows to find all the overt and hidden canting angles in orthoferrites RFeO$_3$. Being based on the theoretical predictions regarding the sign of the Dzyaloshinskii vector we have predicted a novel magnetic phenomenon, {\it weak ferrimagnetism} in mixed weak ferromagnets with competing signs of the Dzyaloshinskii vectors. Weak ferrimagnets can exhibit the tunable exchange bias effect.
 We revisited the problem of  the DM antisymmetric exchange coupling for a single bond in cuprates specifying the local spin-orbital contributions to Dzyaloshinskii vector focusing on the oxygen term. The Dzyaloshinskii vector and respective weak ferromagnetic moment is shown to be a superposition of comparable and, sometimes, competing local Cu and O contributions. We predict a novel puzzling effect of the on-site staggered spin polarization to be a result of the on-site spin-orbital coupling and the the cation-ligand spin density transfer. The ligand NMR measurements are shown to be an effective tool to inspect the effects of the DM coupling in an external magnetic field. We predict the effect of $strong$ oxygen weak antiferromagnetism in edge-shared CuO$_2$ chains due to uncompensated oxygen Dzyaloshinskii vectors. We revisited the effects of symmetric spin anisotropy directly induced by the DM coupling and demonstrated the specific feature of the quantum s=1/2 magnets.  
  Theoretical results are applied to different classes of 3d
 compounds from conventional weak ferromagnets (FeF$_3$, $\alpha$-Fe$_2$O$_3$, RFeO$_3$, RCrO$_3$,.. ) to unconventional systems such as weak ferrimagnets (RFe$_{1-x}$Cr$_x$O$_3$, Fe$_{1-x}$Cr$_x$BO$_3$, Mn$_{1-x}$Ni$_x$CO$_3$), helimagnets (CsCuCl$_3$) and parent cuprates (La$_2$CuO$_4$,...).

In all cases, the magnitude of the Dzyaloshinskii vector ${\bf d}_{12}$ is anticorrelated with the magnitude of the superexchange integral $J_{12}$ in the sense that the superexchange geometry, favorable for the former, is unfavorable for the latter. As a typical example, parent cuprates can be cited, where the small value of the Dzyaloshinsky vector is determined by only a small tilting of the CuO$_6$ octahedra from the CuO$_2$ planes, which practically does not affect the large value of the exchange integral, $J_{12}\geq$\,0.1\,eV\,\cite{Thio}. The specific supersensitivity of the DM coupling to the superexchange geometry and the energy of orbital excitations for Cu and O ions allows us to consider this interaction, first of all, the value and orientation of the Dzyaloshinskii vector, as one of the most important indicators determining the role of structural factors, in particular, the
tilts and bond disproportions in the CuO$_2$ lattice network  associated with "lattice strain"\,\cite{Agrestini,Gav,Ivashko,Liao,Campi}, and different orbital excitations\,\cite{Seino,Oles} in the formation of an unusual electronic structure of the normal and superconducting state of HTS cuprates.

 The work clearly shows advantages of the cluster method as compared with the DFT-based technique to provide an adequate description of the DM coupling and related exchange-relativistic effects in strongly correlated 3d compounds such as exchange anisotropy\,\cite{TIA}, spin-other-orbit interaction\,\cite{MO-Pisarev,MO,PSS_2019}, antisymmetric magnetoelectric coupling\,\cite{multiferro}, and electron-nuclear antisymmetric supertransferred hyperfine  interactions\,\cite{Moskvin-ASTHF,Rokeah}.
However, it should be noted that the DFT with functionals more advanced than LDA  can be effective in calculating correctly the sign and strength of the DM coupling in non-correlated materials\,\cite{Yang1,Yang2,Jadaun}.


\acknowledgments{I thank E.V. Sinitsyn and I.G. Bostrem for very fruitful multi-year collaboration, I.E. Dzyaloshinskii, V.I. Ozhogin, R.E. Walstedt, S.V. Maleev, B.Z. Malkin, B.S. Tsukerblatt, S.-L. Drechsler, and V.E. Dmitrienko for stimulating and encouraging discussions. 
Supported by Act 211 Government of the Russian Federation, agreement No. 02.A03.21.0006 and by the Ministry of Education and Science, projects No. 2277 and No. 5719.}



\begin{thebibliography}{999}
\bibitem{remark}
This is true to within a replacement $\delta _1 \leftrightarrow \delta _2$.
\end{thebibliography}
\end{document}